\documentclass[10pt,journal,twocolumn]{IEEEtran}

\usepackage[english]{babel}
\usepackage[latin1]{inputenc}
\usepackage[T1]{fontenc}
\usepackage{url}
\usepackage{graphicx}
\usepackage{subfigure}
\usepackage{color}
\usepackage[cmex10]{amsmath}
\usepackage{amsthm,amsfonts,amssymb}
\usepackage{colortbl}                   
\usepackage{url}
\usepackage{booktabs}
\usepackage{epsfig}
\usepackage{pstricks}
\usepackage{pst-node}
\usepackage{pst-grad}
\usepackage{pst-plot}
\usepackage{pstricks-add}
\usepackage{ifthen}
\usepackage{algorithm}
\usepackage{algpseudocode}
\usepackage{float}
\usepackage{fp}
\usepackage{cite}

\newfloat{algorithm}{t}{lop}

\newboolean{draft}
\newcommand{\isdraft}[2]{\ifthenelse{\boolean{draft}}{#1}{#2}}

\isdraft{\usepackage{setspace}}{}                              
\isdraft{\usepackage[footnotesize]{caption}}{}                 
\isdraft{\usepackage{paralist}}{}                              

\newboolean{ShorterVersion}
\setboolean{ShorterVersion}{true}
\newcommand{\Shorter}[2]{\ifthenelse{\boolean{ShorterVersion}}{#1}{#2}}

\def\today{\ifcase\month\or
  January\or February\or March\or April\or May\or June\or
  July\or August\or September\or October\or November\or December\fi \space \number\year}

\newcommand{\fref}[1]{Fig.~\ref{#1}}

\newcommand{\cref}[1]{Chapter~\ref{#1}}
\newcommand{\sref}[1]{Section~\ref{#1}}
\newcommand{\ssref}[1]{Subsection~\ref{#1}}
\newcommand{\aref}[1]{Appendix~\ref{#1}}

\newcommand{\pref}[1]{Proposition~\ref{#1}}
\newcommand{\thref}[1]{Theorem~\ref{#1}}
\newcommand{\dref}[1]{Definition~\ref{#1}}
\newcommand{\lref}[1]{Lemma~\ref{#1}}

\newcommand{\Arxiv}[1]{preprint: \url{#1}}
\newcommand{\OptimizationOnline}[1]{preprint: \url{#1}}

\theoremstyle{plain}
\newtheorem{Theorem}{Theorem}
\newtheorem{Definition}{Definition}

\newtheorem{Proposition}{Proposition}
\newtheorem{Lemma}{Lemma}

\newcommand{\mypar}[1]{{\bf #1.}}

\isdraft{}{}

\title{Compressed Sensing with Prior Information: \\	Optimal Strategies, Geometry, and Bounds}

\author{Jo\~ao F.~C.~Mota, Nikos Deligiannis, and Miguel R.~D.~Rodrigues
\IEEEcompsocitemizethanks{
	\IEEEcompsocthanksitem All authors are with the Electronic \& Electrical Engineering Department at University College London, UK.	
}%
}

\begin{document}

\maketitle

\begin{abstract}
	We address the problem of compressed sensing (CS) with prior information: \textit{reconstruct a target CS signal with the aid of a similar signal that is known beforehand, our prior information}. We integrate the additional knowledge of the similar signal into CS via $\ell_1$-$\ell_1$ and $\ell_1$-$\ell_2$ minimization. We then establish bounds on the number of measurements required by these problems to successfully reconstruct the original signal. Our bounds and geometrical interpretations reveal that if the prior information has good enough quality, $\ell_1$-$\ell_1$ minimization improves the performance of CS dramatically. In contrast, $\ell_1$-$\ell_2$ minimization has a performance very similar to classical CS and brings no significant benefits. All our findings are illustrated with experimental results.
\end{abstract}


\isdraft{\pagebreak}{}

\section{Introduction}
\label{Sec:Intro}

	Nearly a decade ago, compressed sensing (CS) emerged as a new paradigm for signal acquisition~\cite{Donoho06-CompressedSensing,Candes06-RobustUncertaintyPrinciplesExactSignalReconstructionHighlyIncomplete}. By assuming that signals are compressible rather than bandlimited, CS enables signal acquisition using far less measurements than classical acquisition schemes~\cite{Nyquist28-CertainTopicsTelegraphTransmissionTheory,Shannon98-CommunicationPresenceNoise}. Since most signals of interest are indeed compressible, CS has found many applications, including medical imaging~\cite{Lustig07-SparseMRI}, radar~\cite{Baraniuk07-CompressiveRadarImaging}, camera design~\cite{Duarte08-SinglePixelImagingCS}, and sensor networks~\cite{Rabbat08-CompressedSensingNetworkedData}. 
								
	We show that, whenever a signal similar to the signal to reconstruct is available, the number of measurements can be reduced even further. Such additional knowledge is often called \textit{prior}~\cite{Borries07-CompressedSensingUsingPriorInformation,Chen08-PriorImageConstrainedCS,Vaswani10-ModifiedCS,Khajehnejad11-AnalyzingWeightedL1MinimizationSparseRecovery,Scarlett13-CompressedSensingPriorInformation,Mishra13-OffTheGridSpectralCSPriorInfo,Eldar14-ApplicationCSLongitudinalMRI} or \textit{side}~\cite{Stankovic09-CompressiveImageSamplingSideInformation,Rostami12-CompressedSensingPresenceSideInformation-thesis,Wang13-SideInformationAidedCS} \textit{information}.
	
	\mypar{Compressed Sensing (CS)}
	Let~$x^\star \in \mathbb{R}^n$ be an unknown $s$-sparse signal, i.e., it has at most~$s$ nonzero entries. Assume we have~$m$ linear measurements~$y = Ax^\star$, where the matrix~$A \in \mathbb{R}^{m\times n}$ is known. CS answers two fundamental questions: \textit{how to reconstruct the signal~$x^\star$ from the measurements~$y$? And how many measurements~$m$ are required for successful reconstruction?} A remarkable result states that if~$A$ satisfies a restricted isometry property (RIP)~\cite{Candes05-DecodingByLinearProgramming,Candes08-RIPAndItsImplicationsForCompressedSensing,Candes08-IntroductionToCompressiveSampling} or a nullspace property~\cite{Chandrasekaran12-ConvexGeometryLinearInverseProblems}, then~$x^\star$ can be reconstructed perfectly by solving \textit{Basis Pursuit} (BP)~\cite{Donoho98-AtomicDecompositionBasisPursuit}:
	\begin{equation}\label{Eq:BP}\tag{BP}
		\begin{array}[t]{ll}
			\underset{x}{\text{minimize}} & \|x\|_1 \\
			\text{subject to} & Ax = y\,,
		\end{array}		
	\end{equation}
	where~$\|x\|_1 := \sum_{i=1}^n |x_i|$ is the $\ell_1$-norm of~$x$; see~\cite{Candes05-DecodingByLinearProgramming,Candes08-RIPAndItsImplicationsForCompressedSensing,Candes08-IntroductionToCompressiveSampling,Chandrasekaran12-ConvexGeometryLinearInverseProblems}. 
	For example, if $m > 2 s \log(n/s) + (7/5)s$, and the entries of~$A \in \mathbb{R}^{m\times n}$ are drawn independently and identically distributed (i.i.d.) from the Gaussian distribution, then~$A$ satisfies a nullspace property (and thus BP recovers~$x^\star$) with high probability~\cite{Chandrasekaran12-ConvexGeometryLinearInverseProblems}.
	See~\cite{Donoho06-CompressedSensing,Candes06-RobustUncertaintyPrinciplesExactSignalReconstructionHighlyIncomplete,Candes06-NearOptimalSignalRecoveryRandomProjections,Candes07-SparsityAndIncoherenceCompressiveSampling,Baraniuk08-SimpleProofRIP,Rudelson08-SparseReconstructionFourierGaussianMeasurements,Donoho09-CountingFacesRandomlyProjectedPolytopes,Bah10-ImprovedBoundsRIPGaussian} for related results.
	
	\mypar{CS with prior information}
	Consider that, in addition to the set of measurements $y = Ax^\star$, we also have access to \textit{prior information}, that is, to a signal~$w \in \mathbb{R}^n$ similar to the original signal~$x^\star$. This occurs in many scenarios: for example, in video acquisition~\cite{Kang09-DistributedCompressiveVideoSensing,Stankovic09-CompressiveImageSamplingSideInformation} and estimation problems~\cite{Charles11-SparsityPenaltiesDynamicalSystemEstimation}, past signals are very similar to the signal to be acquired and, thus, they can be used as prior information; more concretely, if~$x^\star$ is a sparse representation of the signal we want to reconstruct, then~$w$ can be a sparse representation of an already reconstructed signal. Similarly, signals captured by nearby sensors in sensor networks~\cite{Baron06-DistributedCompressedSensing} and images captured by close-by cameras in multiview camera systems~\cite{Cevher08-CompressiveSensingBackgroundSubtraction,Trocan10-DisparityCompensatedCompressedSensingMultiviewImages} are also similar and hence can be used as prior information. The goal of this paper is to answer the following two key questions: 
	\begin{itemize}
		\item \textit{How to reconstruct the signal~$x^\star$ from the measurements $y = Ax^\star$ and the prior information~$w$?}
		
		\item \textit{And how many measurements~$m$ are required for successful reconstruction?}
	\end{itemize}

	\subsection{Overview of Our Approach and Main Results}
	
	We address CS with prior information by solving an appropriate modification of BP. Suppose~$g: \mathbb{R}^n \xrightarrow{} \mathbb{R}$ is a function that measures the similarity between~$x^\star$ and the prior information~$w$, in the sense that~$g(x^\star - w)$ is expected to be small. Then, given~$y = Ax^\star$ and~$w$, we solve
	\begin{equation}\label{Eq:BPSideInfoGeneric}
		\begin{array}[t]{ll}
			\underset{x}{\text{minimize}} & \|x\|_1 + \beta\, g(x-w) \\
			\text{subject to} &  Ax = y\,,
		\end{array}
	\end{equation}
	where~$\beta > 0$ establishes a tradeoff between signal sparsity and fidelity to prior information. 
	We consider two specific, convex models for~$g$: $g_1 := \|\cdot\|_1$ and $g_2 := \frac{1}{2}\|\cdot\|_2^2$, where~$\|z\|_2:=\sqrt{z^\top z}$ is the $\ell_2$-norm. Then, problem~\eqref{Eq:BPSideInfoGeneric} becomes
	\begin{align}
		&
		\begin{array}[t]{ll}
			\underset{x}{\text{minimize}} & \|x\|_1 + \beta \|x - w\|_1 \\
			\text{subject to} & Ax = y\vspace{0.2cm}
		\end{array}
		\label{Eq:L1L1}
		\\
		&
		\begin{array}[t]{ll}
			\underset{x}{\text{minimize}} & \|x\|_1 + \frac{\beta}{2}\|x - w\|_2^2 \\
			\text{subject to} & Ax = y\,,
		\end{array}
		\label{Eq:L1L2}
	\end{align}
	which we will refer to as $\ell_1$-$\ell_1$ and $\ell_1$-$\ell_2$ minimization, respectively. The use of the constraints~$Ax = y$ implicitly assumes that~$y$ was acquired without noise. However, our results also apply to the noisy scenario, i.e., when the constraints are~$\|Ax - y\|_2 \leq \sigma$ instead of $Ax = y$. 
		
	\mypar{Overview of results}
	Problems~\eqref{Eq:L1L1} and~\eqref{Eq:L1L2}, as well as their Lagrangian versions, have rarely appeared in the literature (see \sref{Sec:RelatedWork}). For instance, \cite{Chen08-PriorImageConstrainedCS,Eldar14-ApplicationCSLongitudinalMRI} (resp.\ \cite{Vaswani10-ModifiedCS}) considered problems very similar to~\eqref{Eq:L1L1} (resp.\ \eqref{Eq:L1L2}). Yet, to the best of our knowledge, no CS-type results have ever been provided for either~\eqref{Eq:L1L1} and~\eqref{Eq:L1L2}, their variations in~\cite{Chen08-PriorImageConstrainedCS,Eldar14-ApplicationCSLongitudinalMRI,Vaswani10-ModifiedCS}, or their Lagrangian versions. 
	
	Our goal is to establish bounds on the number of measurements that guarantee that~\eqref{Eq:L1L1} and~\eqref{Eq:L1L2} reconstruct~$x^\star$ with high probability, when~$A$ has i.i.d. Gaussian entries. Our bounds are a function of the prior information ``quality'' and the tradeoff parameter~$\beta$. Hence, they not only help us understand what ``good'' prior information is, but also to select a~$\beta$ that minimizes the number of measurements. The main elements of our contribution can be summarized as follows:
	\begin{itemize}
		\item 
			First, the bound on the number of measurements that~\eqref{Eq:L1L1} requires for perfect reconstruction can be much smaller than the bounds for both classical CS (i.e., BP) and~\eqref{Eq:L1L2}. For example, even when the prior information~$w$ has a relative error of~$50\%$ with respect to~$x^\star$, i.e., $\|w - x^\star\|_2/\|x^\star\|_2 \simeq 0.5$, \eqref{Eq:L1L1} can require much fewer measurements than both~BP and~\eqref{Eq:L1L2}. This superior performance is also observed experimentally, and we interpret it in terms of the underlying geometry of the problem.  
			
		\item
			Second, our bound on the number of measurements for~\eqref{Eq:L1L1} is minimized for~$\beta = 1$, a value that is independent of~$w$, $x^\star$, or any other problem parameter. We will see later that the best~$\beta$ in practice is indeed very close to~$1$. In contrast, the value of~$\beta$ that minimizes our bound on the number of measurements for~\eqref{Eq:L1L2} depends on several parameters, including the unknown entries of~$x^\star$. 
	\end{itemize}
	
	\mypar{A representative result}
	To give an example our results, we state a simplified version of our \thref{Thm:L1L1Reconstruction}, which establishes bounds on the number of measurements for successful $\ell_1$-$\ell_1$ reconstruction with high probability. Here, we rewrite it for~$\beta = 1$ which, incidentally, gives not only the simplest result, but also the best bound. Let us define
	\begin{align*}
		\overline{h} 
			&:= 
			\big|
				\{i\,:\, x_i^\star > 0, \,\,x_i^\star > w_i\} \cup \{i\,:\, x_i^\star < 0, \,\,x_i^\star < w_i\}
			\big|
		\\
		\xi &:= \big|\{i\,:\, w_i \neq x_i^\star = 0\}\big| - \big|\{i\,:\, w_i = x_i^\star \neq 0\}\big|\,,	
	\end{align*}
	where~$|\cdot|$ denotes the cardinality of a set. Note that~$\overline{h}$ is defined on the support $I := \{i\,:\, x_i^\star \neq 0\}$ of~$x^\star$. Recall that~$s = |I|$. Later, we will call~$\overline{h}$ the \textit{number of bad components} of~$w$. 
	For example, if $x^\star = (0, 3, -2, 0, 1, 0, 4)$ and $w = (0, 4, 3, 1, 1, 0, 0)$,
	then $\overline{h} = 2$ (due to $3$rd and last components) and $\xi = 1 - 1 = 0$ ($4$th and $5$th components).
				
	\begin{Theorem}[$\ell_1$-$\ell_1$ minimization: \textit{simplified}]
	\label{Thm:L1L1Simplified}
		Let~$x^\star \in \mathbb{R}^n$ be the vector to reconstruct and let~$w \in \mathbb{R}^n$ be the prior information. Assume~$\overline{h} > 0$ and that there exists at least one index~$i$ for which $x_i^\star = w_i = 0$. Let the entries of~$A \in \mathbb{R}^{m \times n}$ be i.i.d.\ Gaussian with zero mean and variance~$1/m$. If
		\begin{equation}\label{Eq:IntroL1L1Bound}
			m \geq 2\overline{h}\log\Big(\frac{n}{s + \xi/2}\Big) + \frac{7}{5}\Big(s + \frac{\xi}{2}\Big) + 1\,,
		\end{equation}
		then, with probability greater than $1 - \exp\big(-\frac{1}{2}(m - \sqrt{m})^2\big)$, $x^\star$ is the unique solution of~\eqref{Eq:L1L1} with~$\beta = 1$.
	\end{Theorem}
	
	Recall that classical CS requires 
	\begin{equation}\label{Eq:IntroClassicalCSBound}
		m \geq 2s\log\Big(\frac{n}{s}\Big) + \frac{7}{5}s + 1
	\end{equation} 
	measurements to reconstruct~$x^\star$ with a similar probability~\cite{Chandrasekaran12-ConvexGeometryLinearInverseProblems}; see also \thref{Thm:Chandrasekaran} and \pref{Prop:ChandrasekaranBound} in \sref{Sec:GeoInter} below. To compare~\eqref{Eq:IntroL1L1Bound} and~\eqref{Eq:IntroClassicalCSBound}, suppose~$\xi = 0$, i.e., the number of components in which~$x^\star$ and~$w$ differ outside~$I$ equals the number of components in which they coincide on~$I$. In this case, \eqref{Eq:IntroL1L1Bound} becomes equal to~\eqref{Eq:IntroClassicalCSBound}, except for the factor multiplying the $\text{log}$: it is~$2\overline{h}$ in~\eqref{Eq:IntroL1L1Bound} and~$2s$ in~\eqref{Eq:IntroClassicalCSBound}. Since, by definition, $\overline{h}$ is smaller than~$s$, \eqref{Eq:IntroL1L1Bound} is always smaller than~\eqref{Eq:IntroClassicalCSBound}. When~$\xi \neq 0$ and the dominant terms are the ones involving the $\text{log}$'s, \eqref{Eq:IntroL1L1Bound} is smaller than~\eqref{Eq:IntroClassicalCSBound} whenever~$\xi$ is larger than some small negative number. 
		
	\begin{figure}
	\centering
		
	\readdata{\data}{figures/ResultsLaTexSI.dat}
	\readdata{\dataCS}{figures/ResultsLaTexCS.dat}
	\readdata{\dataSILT}{figures/ResultsLaTexSIL2.dat}
		
	\psscalebox{1.0}{
	\begin{pspicture}(8.8,4.95)
					
		\def\xMax{600}                                
		\def\xMin{0}                                  
		\def\xNumTicks{6}                             
		\def\yMax{1.00}                               
		\def\yMin{0}                                  
		\def\yNumTicks{5}                             
		\def\xIncrement{100}                          
		\def\yIncrement{0.2}                          
					
		\def\xOrig{0.50}                              
		\def\yOrig{0.80}                              
		\def\SizeX{8.00}                              
		\def\SizeY{3.70}                              
		\def\xTickIncr{1.33}                          
		\def\yTickIncr{0.74}                          

		\input{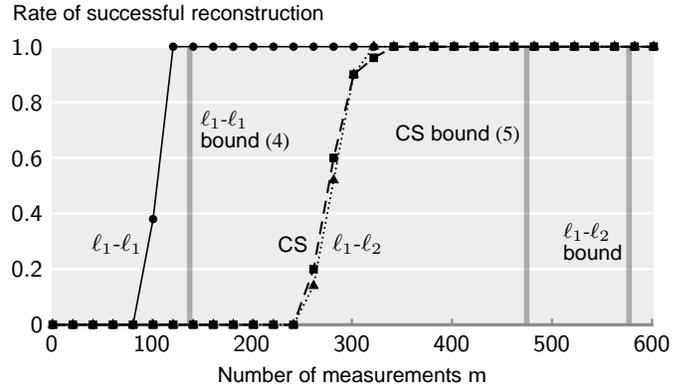}
		
		
		\psset{xunit=\xScale\psunit,yunit=\yScale\psunit,linewidth=0.8pt}
		\psline[linewidth=2.1pt,linecolor=black!90!white,strokeopacity=0.31]{-}(175,0.222)(175,1.212)
		\psline[linewidth=2.1pt,linecolor=black!90!white,strokeopacity=0.31]{-}(511,0.222)(511,1.212)
		\psline[linewidth=2.1pt,linecolor=black!90!white,strokeopacity=0.31]{-}(613,0.222)(613,1.212)
		\dataplot[origin={\xDataOrig,\yDataOrig},showpoints=true,linestyle=dotted,dotsep=1.1pt,dotstyle=triangle*]{\dataSILT}
		\dataplot[origin={\xDataOrig,\yDataOrig},showpoints=true,linestyle=dashed,fillcolor=black!40!white,dotstyle=square*]{\dataCS}			
		\dataplot[origin={\xDataOrig,\yDataOrig},showpoints=true,linewidth=0.6pt]{\data}
		\psset{xunit=\psunit,yunit=\psunit}
		
		\rput[lb](-0.02,4.85){\footnotesize \textbf{\sf Rate of successful reconstruction}}
		\rput[ct](4.5,0.14){\footnotesize \textbf{\sf Number of measurements {\small$\mathsf{m}$}}}
			
		\rput[rt](1.70,2.0){\small $\ell_1$-$\ell_1$}
		\rput[lt](4.23,2.00){\small $\ell_1$-$\ell_2$}
		\rput[rt](3.90,1.98){\footnotesize {\sf CS}}
		\rput[lt](2.48,3.64){\footnotesize $\ell_1$-$\ell_1$}
		\rput[lt](2.48,3.36){\sf\footnotesize bound \eqref{Eq:IntroL1L1Bound}}
		\rput[lb](5.07,3.2){\footnotesize {\sf CS bound \eqref{Eq:IntroClassicalCSBound}}}
		\rput[lt](7.30,2.15){\footnotesize $\ell_1$-$\ell_2$}
		\rput[lt](7.30,1.87){\sf\footnotesize bound}
	\end{pspicture}
	}
	\isdraft{\vspace{0.1cm}}{\vspace{-0.6cm}}    
	\caption{
		Experimental rate of reconstruction of classical CS~\eqref{Eq:BP}, $\ell_1$-$\ell_1$ minimization, and $\ell_1$-$\ell_2$ minimization, both with~$\beta=1$. The vertical lines are the bounds for classical CS, and $\ell_1$-$\ell_1$ and $\ell_1$-$\ell_2$ minimization. 
	}
	\label{Fig:IntroExperimentalResults}
	\end{figure}

	\mypar{A numerical example}	
	We now provide a numerical example to illustrate further our results. We generated~$x^\star$ with~$1000$ entries, $70$ of which were nonzero, i.e., $n = 1000$ and~$s=70$. The nonzero components of~$x^\star$ were drawn from a standard Gaussian distribution. The prior information~$w$ was generated as~$w=x^\star+z$, where~$z$ is a $28$-sparse vector whose nonzero entries were drawn from a zero-mean Gaussian distribution with standard deviation~$0.8$. The supports of~$x^\star$ and~$z$ coincided in~$22$ positions and differed in~$6$. Such a prior information differed significantly from~$x^\star$ in both the $\ell_2$- and $\ell_1$-norms: $\|w-x^\star\|_2/\|x^\star\|_2 \simeq 0.45$ and $\|w - x^\star\|_1/\|x^\star\|_1 \simeq 0.25$. This pair of~$x^\star$ and~$w$ yielded~$\overline{h} = 11$ and~$\xi = -42$. Plugging the previous values into~\eqref{Eq:IntroL1L1Bound} and~\eqref{Eq:IntroClassicalCSBound}, we see that $\ell_1$-$\ell_1$ minimization and classical CS require~$136$ and~$472$ measurements for perfect reconstruction with high probability, respectively. 
	
	\fref{Fig:IntroExperimentalResults} shows the experimental performance of classical CS and $\ell_1$-$\ell_1$ and $\ell_1$-$\ell_2$ minimization, i.e., problems~\eqref{Eq:BP}, \eqref{Eq:L1L1} and~\eqref{Eq:L1L2}, respectively. More specifically, it depicts the rate of success of each problem versus the number of measurements~$m$. For a fixed~$m$, the success rate is the number of times a given problem recovered~$x^\star$ with an error smaller than $1\%$ divided by the total number of~$50$ trials (each trial considered different pairs of~$A$ and~$b$). The plot shows that $\ell_1$-$\ell_1$ minimization required less measurements to reconstruct~$x^\star$ successfully than both CS and $\ell_1$-$\ell_2$ minimization. The curves of the last two, in fact, almost coincide, with $\ell_1$-$\ell_2$ minimization (line with triangles) having a slightly sharper phase transition. The vertical lines show the bounds~\eqref{Eq:IntroL1L1Bound}, \eqref{Eq:IntroClassicalCSBound}, and the bound for $\ell_1$-$\ell_2$ minimization, provided in \sref{Sec:MainResults}. We see that, for this particular example, the bound~\eqref{Eq:IntroL1L1Bound} is quite sharp, while the bound for $\ell_1$-$\ell_2$ minimization is quite loose (the sharpness of our bounds is discussed in Sections~\ref{Sec:MainResults} and~\ref{Sec:Proofs}). Most importantly, this example shows that using prior information properly can improve the performance of CS dramatically.

	\subsection{Outline}
	
	The rest of the paper provides a detailed treatment of CS with prior information, covering both an overview of related research and the statement and proof of the main results.	In \sref{Sec:RelatedWork} we discuss related work, including the use of other types of ``prior information'' in CS. \sref{Sec:GeoInter} introduces the fundamental tools in our analysis, which are also used to provide geometrical interpretations  of $\ell_1$-$\ell_1$ and $\ell_1$-$\ell_2$ minimization. The main results are stated and discussed in~\sref{Sec:MainResults}. In \sref{Sec:ExperimentalResults}, we provide some experimental results. The main theoretical results are proven in \sref{Sec:Proofs}, and some auxiliary results are proven in the appendices. 
	
\section{Related Work}
\label{Sec:RelatedWork}

	There is a clear analogy of CS with prior information and the distributed source coding problem. Namely, we can view the number of measurements and the reconstruction quality in CS as the information rate and the incurred distortion in coding theory, respectively. As such, our problem of CS with prior information at the reconstruction side is reminiscent of the problem of coding with side/prior information at the decoder, a field whose foundations were laid by Slepian and Wolf~\cite{Slepian73-NoiselessCodingCorrelatedInformationSources}, and Wyner and Ziv~\cite{WynerZiv}; see also~\cite{Cover91-ElementsInformationTheory}.
	
	The concept of prior information has appeared in CS under many guises~\cite{Chen08-PriorImageConstrainedCS,Eldar14-ApplicationCSLongitudinalMRI,Vaswani10-ModifiedCS,Charles11-SparsityPenaltiesDynamicalSystemEstimation,Wang13-SideInformationAidedCS}. The work in~\cite{Chen08-PriorImageConstrainedCS} was apparently the first to consider~\eqref{Eq:BPSideInfoGeneric}, in particular $\ell_1$-$\ell_1$ minimization. Specifically, \cite{Chen08-PriorImageConstrainedCS} considers dynamic computed tomography, where the image reconstructed in the previous time instant helps reconstructing the current one. That is accomplished by solving~\eqref{Eq:L1L1}. That work, however, neither provides any kind of analysis nor highlights the benefits of solving~\eqref{Eq:L1L1} with respect to classical CS, i.e., BP. Very recently, \cite{Eldar14-ApplicationCSLongitudinalMRI} considered a variation of~\eqref{Eq:L1L1} where the second term of the objective rather than penalizing differences between~$x$ and~$w$ in the sparse domain, penalizes differences in the signals' original domain. Specifically, \cite{Eldar14-ApplicationCSLongitudinalMRI} solves (a slightly more general version of)
	\begin{equation}\label{Eq:EldarsProb}
		\begin{array}[t]{ll}
			\underset{x}{\text{minimize}} & \|x\|_1 + \beta \|\Psi (x - w)\|_1 \\
			\text{subject to} & \Phi\Psi x = y\,,
		\end{array}
	\end{equation}
	where~$A$ was decomposed as the product of a sensing matrix~$\Phi$ and a transform matrix~$\Psi$ that sparsifies both~$x^\star$ and~$w$. Although~\cite{Eldar14-ApplicationCSLongitudinalMRI} shows experimentally that~\eqref{Eq:EldarsProb} requires less measurements than conventional CS to reconstruct MRI images, no analysis or reconstruction guarantees are given for~\eqref{Eq:EldarsProb}.
	
	In~\cite{Vaswani10-ModifiedCS}, prior information refers to an estimate~$T \subseteq \{1,\ldots,n\}$ of the support of~$x^\star$ (see~\cite{Borries07-CompressedSensingUsingPriorInformation,Mishra13-OffTheGridSpectralCSPriorInfo,Scarlett13-CompressedSensingPriorInformation} for related approaches). Using the restricted isometry constants of~$A$, \cite{Vaswani10-ModifiedCS} provides exact recovery conditions for BP when its objective is modified to~$\|x_{T^c}\|_1$, where~$x_T$ denotes the components of~$x$ indexed by the set~$T$, and~$T^c$ is the complement of~$T$ in $\{1,\ldots,n\}$. When~$T$ is a reasonable estimate of the support of~$x^\star$, those conditions are shown to be milder than the ones in~\cite{Candes05-DecodingByLinearProgramming,Candes08-RIPAndItsImplicationsForCompressedSensing} for standard BP. Then, \cite{Vaswani10-ModifiedCS} considers prior information as we do: there is an estimate of the support of~$x^\star$ as well as of the value of the respective nonzero components. However, it solves a problem slightly different from~\eqref{Eq:L1L2}. Namely, the objective of~\eqref{Eq:L1L2} is replaced with~$\|x_{T^c}\|_1 + \beta\|x_T - w_T\|_2^2$. Although some experimental results are presented, no analysis is given for that problem.
	
	A modification of~BP that has often appeared in the literature considers, instead of the $\ell_1$-norm, the weighted $\ell_1$-norm~$\|x\|_r := \sum_{i = 1}^n r_i x_i$, where~$r_i > 0$ is a known weight. This norm penalizes each component of~$x$ according to the magnitude of the corresponding weight and, thus, requires ``prior information'' about~$x$. The weight~$r_i$ associated to the component~$x_i$ can, for example, be proportional to the probability of~$x_i^\star = 0$. Several algorithms based on this idea have been proposed e.g., \cite{Candes08-EnhancingSparsityReweightedL1Minimization,Asig13-FastAccurateAlgorithmsReweightedL1NormMinimization}. Moreover, \cite{Khajehnejad11-AnalyzingWeightedL1MinimizationSparseRecovery} proved that the number of measurements for exact recovery is asymptotically smaller for the weighted $\ell_1$-norm than for the unweighted one. The work in~\cite{Scarlett13-CompressedSensingPriorInformation} also obtains asymptotic bounds and proposes setting~$r_i = -\log\,p_i$, where~$p_i$ is the probability that $x_i^\star \neq 0$.

	Alternative work has considered
	\begin{equation}\label{Eq:RelatedWorkLagrangianVersion}
		\underset{x}{\text{minimize}} \,\,\, \|x\|_1 + \beta\, g(x-w) + \lambda\|Ax - y\|_2^2\,,
	\end{equation}
	with~$\lambda > 0$,	which can be viewed as a Lagrangian version of
	\begin{equation}\label{Eq:RelatedWorkNoisy}
		\begin{array}[t]{ll}
			\underset{x}{\text{minimize}} & \|x\|_1 + \beta \, g(x-w) \\
			\text{subject to} & \|Ax - y\|_2 \leq \sigma\,.
		\end{array}
	\end{equation}
	Problem~\eqref{Eq:RelatedWorkNoisy} is a generalization of~\eqref{Eq:BPSideInfoGeneric} for noisy scenarios, and we will provide bounds on the number of measurements that it requires for successful reconstruction with $g = \|\cdot\|_1$ and $g = \frac{1}{2}\|\cdot\|_2^2$. Problem~\eqref{Eq:RelatedWorkLagrangianVersion} has appeared before in~\cite{Charles11-SparsityPenaltiesDynamicalSystemEstimation}, in the context of dynamical system estimation. Specifically, the state~$x^{(t)}$ of a system at time~$t$ evolves as~$x^{(t+1)} = f^{(t)}(x^{(t)}) + \epsilon^{(t)}$, where~$f^{(t)}$ models the system's dynamics at time~$t$ and~$\epsilon^{(t)}$ accounts for modeling errors. Observations of the state~$x^{(t)}$ are taken as~$y^{(t)} = A^{(t)} x^{(t)} + \eta^{(t)}$, where~$A^{(t)}$ is the observation matrix and~$\eta^{(t)}$ is noise. The goal is to estimate the state~$x^{(t)}$ given the observations~$y^{(t)}$. The state of the system in the previous instant, $x^{(t-1)}$, can be used as prior information by making~$w^{(t)} = f^{(t-1)}(x^{(t-1)})$. If the modeling error~$\epsilon^{(t)}$ is Gaussian and the state~$x^{(t)}$ is assumed sparse, then~$x^{(t)}$ can be estimated by solving~\eqref{Eq:RelatedWorkLagrangianVersion} with~$g = \|\cdot\|_2^2$; if the modeling noise is Laplacian, we set~$g = \|\cdot\|_1$ instead. Although~\cite{Charles11-SparsityPenaltiesDynamicalSystemEstimation} does not provide any analysis, their experimental results show that, among several strategies for state estimation including Kalman filtering, \eqref{Eq:RelatedWorkLagrangianVersion} with $g = \|\cdot\|_1$ yields the best results. If we take into account the relation between~\eqref{Eq:RelatedWorkLagrangianVersion} and~\eqref{Eq:RelatedWorkNoisy}, our theoretical analysis can be used to provide an explanation. We also mention that~\cite{Wang13-SideInformationAidedCS} proposed and analyzed an approximate message passing algorithm to solve problem~\eqref{Eq:RelatedWorkLagrangianVersion} with~$g = \|\cdot\|_2^2$. 
	
	The use of more complex signal models, rather than sparsity, can also be seen as an instance of prior information, and has attracted considerable attention~\cite{Duarte11-StructuredCompressedSensing}. Examples include the notion of block sparsity (see~\cite{Stojnic09-ReconstructionBlockSparseSignals} and~\cite{Eldar09-RobustRecoverySignalsStructuredUnionSubspaces} for nullspace-based and RIP-based reconstruction guarantees), model-based CS~\cite{Baraniuk10-ModelBasedCompressiveSensing}, multiple measurement vectors~\cite{Chen06-TheoreticalResultsSparseRepresentationsMMV,Friedlander09-JointSparseRecoveryMultipleMeasurements}, and Gaussian mixture models~\cite{Renna14-ReconstructionSignalsGMM}. The additional structure considered in these works can be used to reduce the number of measurements for successful reconstruction (see, e.g., \cite{Stojnic09-ReconstructionBlockSparseSignals} and~\cite{Eldar09-RobustRecoverySignalsStructuredUnionSubspaces} for block sparsity, \cite{Chen06-TheoreticalResultsSparseRepresentationsMMV} and~\cite{Friedlander09-JointSparseRecoveryMultipleMeasurements} for multiple measurement vectors, and~\cite{Baraniuk10-ModelBasedCompressiveSensing} for more general models) or even to design measurement matrices~\cite{Carson12-CommunicationsInspiredProjectionDesign,Wang13-DesignedMeasurementsVectorCountData,Wang14-NonlinearInformationTheoreticCompressiveMeasurementDesign}. 
	
	Finally, we mention that several authors have been using the same tools as we do, namely the concept of Gaussian width and Gordon's lemma~\cite{Gordon88-EscapeThroughTheMesh}, to derive CS results~\cite{Rudelson08-SparseReconstructionFourierGaussianMeasurements,Stojnic09-VariousThresholdsForL1OptimizationInCompressedSensing,Chandrasekaran12-ConvexGeometryLinearInverseProblems} and  analyze related problems~\cite{Tropp13-LivingOnTheEdge,Oymak13-SquaredErrorGeneralizedLASSO,Chandrasekaran13-ComputationalStatisticalTradeoffsViaConvexRelaxation,Oymak13-SharpMSEBoundsProximalDenoising,Oymak13-SimultaneousStructuredModelsSparseLowRankMatrices,Foygel14-CorruptedSensing,Bandeira14-CompressiveClassificationRareEclipseProblem,Tropp14-ConvexRecoveryStructuredSignalFromIndependentRandomLinearMeasurements,Kabanava14-RobustAnalysisL1RecoveryGaussianMeasurementsTV}. 

\section{The Geometry of $\ell_1$-$\ell_1$ and $\ell_1$-$\ell_2$ Minimization}
\label{Sec:GeoInter}

	This section introduces concepts and results in CS that will be used in our analysis. We follow the approach of~\cite{Chandrasekaran12-ConvexGeometryLinearInverseProblems}, since it leads to the current best CS bounds for Gaussian measurements,
	and provides the means to understand some of our definitions. 

	\subsection{Known Results and Tools}
	
	The concept of \textit{Gaussian width} plays a key role in~\cite{Chandrasekaran12-ConvexGeometryLinearInverseProblems}. Originally proposed in~\cite{Gordon88-EscapeThroughTheMesh} to quantify the probability of a randomly oriented subspace intersecting a cone, the Gaussian width has been used to prove CS results~\cite{Rudelson08-SparseReconstructionFourierGaussianMeasurements,Stojnic09-VariousThresholdsForL1OptimizationInCompressedSensing,Chandrasekaran12-ConvexGeometryLinearInverseProblems} and, more recently, to tackle problems in other areas~\cite{Tropp13-LivingOnTheEdge,Oymak13-SquaredErrorGeneralizedLASSO,Chandrasekaran13-ComputationalStatisticalTradeoffsViaConvexRelaxation,Oymak13-SharpMSEBoundsProximalDenoising,Oymak13-SimultaneousStructuredModelsSparseLowRankMatrices,Foygel14-CorruptedSensing,Bandeira14-CompressiveClassificationRareEclipseProblem,Tropp14-ConvexRecoveryStructuredSignalFromIndependentRandomLinearMeasurements,Kabanava14-RobustAnalysisL1RecoveryGaussianMeasurementsTV}. Before defining it, we analyze the optimality conditions of a linearly constrained convex optimization problem.

	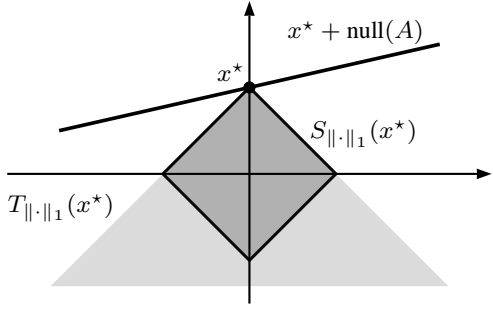
\begin{figure}[t]
		\centering			
		\psscalebox{1.15}{	
		\begin{pspicture}(4,3.5)
				
			\psset{blendmode=2}	
			
			\psset{linewidth=1.0pt}
			
			\def\radius{1.0}       
								
			\def\diamond{       
				\pspolygon*[linecolor=black!20!white](0,\radius)(\radius,0.0)(0.0,-\radius)(-\radius,0.0)
				\pspolygon(0,\radius)(\radius,0.0)(0.0,-\radius)(-\radius,0.0)
			}
					
			\rput(2.0,1.5){\diamond}        
	
			\pspolygon*[linecolor=black!80!white,opacity=0.18](2.0,2.5)(-0.3,0.2)(4.3,0.2)
				
			\psline[linewidth=1.2pt]{-}(-0.2,2.0)(4.2,3.0)
			\rput[br](4.02,3.0){\footnotesize $x^\star + \text{null}(A)$}
											
			\pscircle*(2.0,2.5){0.07}	
			\rput[br](1.95,2.58){\footnotesize $x^\star$}			
								                
			\psset{arrowsize=3.5pt,arrowinset=0.02}
			\psline[linewidth=0.7pt]{->}(-0.8,1.5)(4.8,1.5)
			\psline[linewidth=0.7pt]{->}(2.0,0.0)(2.0,3.5)
			
			\rput[lb](2.7,1.8){\footnotesize $S_{\|\cdot\|_1}(x^\star)$}
			\rput[rb](0.44,0.95){\footnotesize $T_{\|\cdot\|_1}(x^\star)$}
															                                  
		\end{pspicture}
		}		
		\vspace{0.1cm}		
		\caption{
			Visualization of the nullspace property in \pref{Prop:OptimalityCond} for BP.
		}
		\label{Fig:GeometricalConcepts}
	\end{figure}
			
	\mypar{The nullspace property}
	Consider a real-valued convex function~$f:\mathbb{R}^{n} \xrightarrow{} \mathbb{R}$ and the following optimization problem:
	\begin{equation}\label{Eq:OptimProbBeforeOptimalCond}
		\begin{array}[t]{ll}
			\underset{x}{\text{minimize}} & f(x) \\
			\text{subject to} & A x = y\,.
		\end{array}
	\end{equation}
	Assume~$Ax = y$ has at least one solution, say~$x^\star$. The set of all solutions of~$Ax = y$ is given by $\mathcal{A} := x^\star + \text{null}(A)$, where~$\text{null}(A) := \{x\,:\, Ax = 0\}$ is the \textit{nullspace} of~$A$. In other words, $\mathcal{A}$ is the feasible set of~\eqref{Eq:OptimProbBeforeOptimalCond}. To determine whether or not an arbitrary~$x^\star \in \mathcal{A}$ is a solution of~\eqref{Eq:OptimProbBeforeOptimalCond}, we can use the concept of \textit{tangent cone} of~$f$ at~$x^\star$~\cite[Prop.5.2.1,Thm.1.3.4]{Lemarechal04-FundamentalsConvexAnalysis}: 
	\begin{equation}\label{Eq:TangentConeSublevelSet}
		T_f(x^\star) := \text{cone}\big(S_f(x^\star) - x^\star\big)\,,
	\end{equation} 	
	where~$\text{cone}\,C := \{\alpha c\,:\, \alpha \geq 0,\, c \in C\}$ is the \textit{cone generated by the set}~$C$, and~$S_f(x^\star) := \{x\,:\, f(x) \leq f(x^\star)\}$ is the \textit{sublevel set of~$f$ at~$x^\star$}. In words, $d$ belongs to~$T_f(x^\star)$ if it can be written as~$d = \alpha (\overline{x} - x^\star)$ for some~$\alpha \geq 0$ and~$\overline{x} \in S_f(x^\star)$. In particular, if~$\alpha > 0$, then $\overline{x} = x^\star + \frac{1}{\alpha}d \in S_f(x^\star)$, that is, $f(x^\star + \frac{1}{\alpha}d) \leq f(x^\star)$. This means that~$T_f(x^\star)$ contains all the directions~$d$ such that $x^\star + \gamma d$, for some~$\gamma > 0$, leads to a possible decrease in the value of~$f(x^\star)$: $f(x^\star + \gamma d) \leq f(x^\star)$. If not such direction exists in~$\mathcal{A}$, then~$x^\star$ is the unique solution of~\eqref{Eq:OptimProbBeforeOptimalCond}, and vice-versa. That is,
	\begin{equation}\label{Eq:IntersectionTangentConeAffineSpace}
		T_{f}(x^\star)\, \cap\, \big(x^\star + \text{null}(A) \big) = \{x^\star\}
	\end{equation} 
	if and only if~$x^\star$ is the unique solution of~\eqref{Eq:OptimProbBeforeOptimalCond}. If we subtract~$x^\star$ from both sides of~\eqref{Eq:IntersectionTangentConeAffineSpace}, we obtain:
	\begin{Proposition}[Prop.\ 2.1 in \cite{Chandrasekaran12-ConvexGeometryLinearInverseProblems}]
	\label{Prop:OptimalityCond}
		$x^\star$ is the unique optimal solution of~\eqref{Eq:OptimProbBeforeOptimalCond} if and only if~$
			T_f(x^\star) \cap\text{\emph{null}}(A) = \{0\}
		$.
	\end{Proposition}
	Although this proposition was stated in \cite[Prop.2.1]{Chandrasekaran12-ConvexGeometryLinearInverseProblems} for~$f$ equal to an atomic norm, its proof holds for any real-valued convex function. \fref{Fig:GeometricalConcepts} illustrates~\eqref{Eq:IntersectionTangentConeAffineSpace} for BP, i.e., with~$f(x) = \|x\|_1$. It shows the respective sublevel set~$S_{\|\cdot\|_1}(x^\star)$ and tangent cone~$T_{\|\cdot\|_1}(x^\star)$ at a ``sparse'' point~$x^\star$. In the figure, $\mathcal{A} = x^\star + \text{null}(A)$ intersects~$T_{\|\cdot\|_1}(x^\star)$ at~$x^\star$ only, meaning that~\eqref{Eq:IntersectionTangentConeAffineSpace}, and hence \pref{Prop:OptimalityCond}, holds. 
	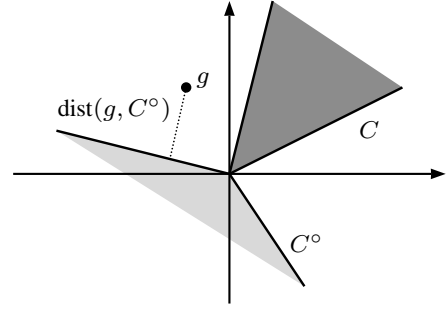
\begin{figure}[t]
		\centering			
		\psscalebox{1.15}{	
		\begin{pspicture}(4,3.5)
	
			\psset{blendmode=2}	
				
			\rput(1.5,2.5){\rnode{g}{\pscircle*(0,0){0.06}}}
	
			\rput(1.31,1.665){\rnode{p}{\pscircle*(0,0){0.0}}}
			
			\ncline[linewidth=0.6pt,linestyle=dotted,dotsep=0.6pt]{g}{p}
			\nbput[npos=0.41]{\footnotesize $\text{dist}(g,C^\circ)$\hspace{-0.12cm}}
	
			
	
	
			\psset{linewidth=0.8pt}
			\pspolygon*[linecolor=black!45!white](2.0,1.5)(2.5,3.5)(4.0,2.5)
			\psline(2.0,1.5)(2.5,3.5)
			\psline(2.0,1.5)(4.0,2.5)
			
			\pspolygon*[linecolor=black!15!white](2.0,1.5)(0.0,2.0)(2.867,0.2)				
			\psline(2.0,1.5)(0.0,2.0)
			\psline(2.0,1.5)(2.867,0.2)
																		                
			\psset{arrowsize=3.5pt,arrowinset=0.02,linestyle=solid}
			\psline[linewidth=0.7pt]{->}(-0.5,1.5)(4.5,1.5)
			\psline[linewidth=0.7pt]{->}(2.0,0.0)(2.0,3.5)
			
			\rput[tl](3.5,2.1){\footnotesize $C$}
			\rput[bl](2.7,0.6){\footnotesize $C^\circ$}
			\rput[bl](1.62,2.5){\footnotesize $g$}
	
		\end{pspicture}
		}
		\vspace{0.1cm}		
		\caption{
				Illustration of how the Gaussian width measures the width of a cone, according to \pref{Prop:GaussianWidthDistancePolarCone}.
		}
		\label{Fig:GaussianWidth}
	\end{figure}	

	\mypar{Gaussian width}
	When~$A$ is generated randomly, its nullspace~$\text{null}(A)$ has a random orientation, and~\eqref{Eq:IntersectionTangentConeAffineSpace} holds or not with a given probability. The smaller the width (or aperture) of~$T_f(x^\star)$, the more likely~\eqref{Eq:IntersectionTangentConeAffineSpace} will hold. Such a statement was formalized by Gordon in~\cite{Gordon88-EscapeThroughTheMesh} for Gaussian matrices~$A$. To measure the width of cone~$C \in \mathbb{R}^n$, Gordon defined the concept of \textit{Gaussian width}:
	\begin{equation}\label{Eq:GaussianWidth}
		w(C) := \mathbb{E}_g \Bigl[
		\sup_{z \in C \cap B_n(0,1)}\, g^\top z
		\Bigr]\,,
	\end{equation}
	where~$B_n(0,1) := \{x \in \mathbb{R}^n\,:\, \|x\|_2 \leq 1\}$ is the unit $\ell_2$-norm ball in~$\mathbb{R}^n$ and~$g \sim \mathcal{N}(0,I_n)$ is a vector of~$n$ independent, zero-mean, and unit-variance Gaussian random variables. The symbol~$\mathbb{E}_g[\cdot]$ denotes the expected value with respect to~$g$. The Gaussian width is usually defined for generic sets by taking their intersection with the spherical part of~$B_n(0,1)$, $\mathbb{S}_n(0,1):=\{x \in \mathbb{R}^n\,:\, \|x\|_2 = 1\}$, rather than with~$B_n(0,1)$. When the set is a cone, however, that is equivalent to intersecting it with~$B_n(0,1)$, as in~\eqref{Eq:GaussianWidth}.\footnote{That is because the maximizer of the problem in~\eqref{Eq:GaussianWidth} is always in~$\mathbb{S}_n(0,1)$. To see that, suppose it is not, i.e., for a fixed~$g$, $z_g := \sup\{g^\top z: z \in C \cap B_n(0,1)\}$ and $z_g \not\in \mathbb{S}_n(0,1)$. This means $\|z_g\|_2 < 1$. Since~$C$ is a cone, $\hat{z}_g:=z_g/\|z_g\|_2 \in C \cap \mathbb{S}_n(0,1)$. And $g^\top \hat{z}_g = (1/\|\hat{z}_g\|_2)g^\top z_g > g^\top z_g$, contradicting the fact that~$z_g$ is optimal.} 
	As a result, the Gaussian width of a cone~$C$ is the expected distance of a Gaussian vector~$g$ to the \textit{polar cone of}~$C$, defined as~$C^\circ := \{y\,:\, y^\top z \leq 0\,, \forall\, z \in C\}$:
	\begin{Proposition}[Example 2.3.1 in~\cite{Lemarechal04-FundamentalsConvexAnalysis}; Prop.\ 3.6 in~\cite{Chandrasekaran12-ConvexGeometryLinearInverseProblems}]\label{Prop:GaussianWidthDistancePolarCone}
	The Gaussian width of a cone~$C$ can be written as
	\begin{equation}\label{Eq:GaussianWidthPolar}
		w(C) = \mathbb{E}_g\Bigl[\text{\emph{dist}}(g,C^\circ)\Bigr]\,,
	\end{equation} 
	where $\text{\emph{dist}}(x,S) := \min\{\|z - x\|_2: z \in S\}$ denotes the distance of the point~$x$ to the set~$S$.
	\end{Proposition}

	This follows from the fact that the support function of a ``truncated'' cone is the distance to its polar cone~\cite[Ex.2.3.1]{Lemarechal04-FundamentalsConvexAnalysis}; and can be proved by computing the dual of the optimization problem in~\eqref{Eq:GaussianWidth} \cite[Prop.3.6]{Chandrasekaran12-ConvexGeometryLinearInverseProblems}. Besides providing a way easier than~\eqref{Eq:GaussianWidth} for computing Gaussian widths, \pref{Prop:GaussianWidthDistancePolarCone} also provides a geometrical explanation of why the Gaussian width measures the width of a cone. The wider the cone~$C$, the smaller its polar cone~$C^\circ$. Therefore, the expected distance of a Gaussian vector~$g$ to~$C^\circ$ increases as~$C^\circ$ gets smaller or, equivalently, as $C$ gets wider; see~\fref{Fig:GaussianWidth}. 
	
	\mypar{From geometry to CS bounds}
	In~\cite{Gordon88-EscapeThroughTheMesh}, Gordon used the concept of Gaussian width to compute bounds on the probability of a cone intersecting a subspace whose orientation is uniformly distributed, e.g., the nullspace of a Gaussian matrix. More recently, \cite{Tropp13-LivingOnTheEdge} showed that those bounds are sharp. 
	Based on Gordon's result, on \pref{Prop:OptimalityCond} (and its generalization for the case where the constraints of~\eqref{Eq:OptimProbBeforeOptimalCond} are~$\|Ax - y\|_2\leq \sigma$), and a concentration of measure result, \cite{Chandrasekaran12-ConvexGeometryLinearInverseProblems} establishes:
	\begin{Theorem}[Corollary 3.3 in \cite{Chandrasekaran12-ConvexGeometryLinearInverseProblems}]
	\label{Thm:Chandrasekaran}
		Let~$A \in \mathbb{R}^{m\times n}$ be a matrix whose entries are i.i.d., zero-mean Gaussian random variables with variance~$1/m$. Assume~$f:\mathbb{R}^n \xrightarrow{} \mathbb{R}$ is convex, and let $\lambda_m := \mathbb{E}_g[\|g\|_2]$ denote the expected length of a zero-mean, unit-variance Gaussian vector~$g \sim \mathcal{N}(0,I_m)$ in~$\mathbb{R}^m$.
		\begin{enumerate}
			\item 
				Suppose $y = Ax^\star$ and let
				\begin{equation}\label{Eq:ThmChandrasekaranNoiseless}
					\hat{x} = 
					\begin{array}[t]{cl}
						\underset{x}{\arg\min} & f(x) \\
						\text{\emph{s.t.}} &  A x = y\,,
					\end{array}
				\end{equation}
				and
				\begin{equation}\label{Eq:ThmChandrasekaranBoundNoiseless}
					m \geq w(T_f(x^\star))^2 + 1\,.
				\end{equation}				
				Then, $\hat{x} = x^\star$ is the unique solution of~\eqref{Eq:ThmChandrasekaranNoiseless} with probability greater than $1-\exp\bigl(-\frac{1}{2}\bigl[\lambda_m - w(T_f(x^\star))\bigr]^2\bigr)$.
				
			\item
				Suppose $y = Ax^\star + \eta$, where $\|\eta\|_2 \leq \sigma$ and let
				\begin{equation}\label{Eq:ThmChandrasekaranNoisy}
					\hat{x} \in 
					\begin{array}[t]{cl}
						\underset{x}{\arg\min} & f(x) \\
						\text{\emph{s.t.}} & \|A x - y\|_2 \leq \sigma\,.
					\end{array}
				\end{equation}
				Define~$0 < \epsilon < 1$ and let
				\begin{equation}\label{Eq:ThmChandrasekaranBoundNoisy}
					m \geq \frac{w(T_f(x^\star))^2 + 3/2}{(1-\epsilon)^2}\,.
				\end{equation}
				Then, $\|\hat{x} - x^\star\|_2 \leq 2\sigma/\epsilon$ with probability greater than $1 - \exp\bigl(-\frac{1}{2}\bigl[\lambda_m - w(T_f(x^\star)) - \epsilon\sqrt{m}\bigr]^2\bigr)$.
		\end{enumerate}
	\end{Theorem}
	\thref{Thm:Chandrasekaran} was stated in~\cite{Chandrasekaran12-ConvexGeometryLinearInverseProblems} for~$f$ equal to an atomic norm. Its proof, however, remains valid when~$f$ is any convex function. Note, in particular, that~\eqref{Eq:ThmChandrasekaranNoiseless} becomes~\eqref{Eq:BP}, \eqref{Eq:L1L1}, and~\eqref{Eq:L1L2} when~$f(x)$ is~$\|x\|_1$, $\|x\|_1 + \beta\|x - w\|_1$, and~$\|x\|_1 + \frac{\beta}{2}\|x - w\|_2^2$, respectively; and~\eqref{Eq:ThmChandrasekaranNoisy} becomes the noise-robust version of these problems.	The quantity~$\lambda_m$ can be bounded (sharply) with~\cite{Chandrasekaran12-ConvexGeometryLinearInverseProblems}: $m/\sqrt{m+1} \leq \lambda_m \leq \sqrt{m}$.	One of the steps of the proof shows that condition~\eqref{Eq:ThmChandrasekaranBoundNoiseless} implies $w(T_f(x^\star)) \leq \lambda_m$ and that condition~\eqref{Eq:ThmChandrasekaranBoundNoisy} implies $w(T_f(x^\star)) + \epsilon\sqrt{m} \leq \lambda_m$. Roughly, the theorem says that, given the noiseless (resp.\ noisy) measurements $y = Ax^\star$ (resp.\ $y = Ax^\star + \eta$), we can recover~$x^\star$ exactly (resp.\ with an error of~$2\sigma/\epsilon$), provided the number of measurements is larger than a function of the Gaussian width of~$T_f(x^\star)$. It is rare, however, to be able to compute Gaussian widths in closed-form; instead, one usually upper bounds it. As proposed in~\cite{Chandrasekaran12-ConvexGeometryLinearInverseProblems}, a useful tool to obtain such bounds is Jensen's inequality~\cite[Thm.B.1.1.8]{Lemarechal04-FundamentalsConvexAnalysis}, \pref{Prop:GaussianWidthDistancePolarCone}, and the following proposition. Before stating it, recall that the \textit{normal cone~$N_f(x)$ of a function~$f$ at a point~$x$} is the polar of its tangent cone: $N_f(x) := T_f(x)^\circ$. Also, $\partial f(x) := \{d\,:\, f(y) \geq f(x) + d^\top (y - x),\, \text{for all $y$}\}$ is the \textit{subgradient of~$f$ at a point~$x$}~\cite{Lemarechal04-FundamentalsConvexAnalysis}.	
	\begin{Proposition}[Theorem 1.3.5, Chapter~D, in~\cite{Lemarechal04-FundamentalsConvexAnalysis}]
	\label{Prop:Subdifferential}
		Let~$f:\mathbb{R}^n \xrightarrow{} \mathbb{R}$ be a convex function and suppose $0 \not\in \partial f(x)$ for a given~$x \in \mathbb{R}^n$. Then, $N_f(x) = \text{\emph{cone}}\, \partial f(x)$.
	\end{Proposition}
	Using Propositions \ref{Prop:GaussianWidthDistancePolarCone} and~\ref{Prop:Subdifferential}, \cite{Chandrasekaran12-ConvexGeometryLinearInverseProblems} proves:\footnote{We noticed an extra factor of~$\sqrt{\pi}$ in equation (73) of~\cite{Chandrasekaran12-ConvexGeometryLinearInverseProblems} (proof of Proposition 3.10). Namely, $\pi$ in (73) should be replaced by~$\sqrt{\pi}$. As a consequence, equation (74) in that paper can be replaced, for example, by our equation~\eqref{Eq:BoundChandrasekaran}. In that case, the number of measurements in Proposition 3.10 in~\cite{Chandrasekaran12-ConvexGeometryLinearInverseProblems} should be corrected from $2s\log(n/s) + (5/4)s$ to $2s\log(n/s) + (7/5)s$.}
	\begin{Proposition}[Proposition 3.10 in~\cite{Chandrasekaran12-ConvexGeometryLinearInverseProblems}]
	\label{Prop:ChandrasekaranBound}
		Let~$x^\star\neq 0$ be an $s$-sparse vector in~$\mathbb{R}^n$. Then, 
		\begin{equation}\label{Eq:PropChandrasekaranBound}
			w\bigl(T_{\|\cdot\|_1}(x^\star)\bigr)^2 \leq 2s\log\Bigl(\frac{n}{s}\Bigr) + \frac{7}{5}s\,.
		\end{equation} 
	\end{Proposition}
	Together with~\thref{Thm:Chandrasekaran}, this means that if $m \geq 2s\log(n/s) + (7/5)s + 1$, then BP recovers~$x^\star$ from~$m$ noiseless Gaussian measurements with high probability. A similar result holds for noisy measurements, i.e., for~\eqref{Eq:ThmChandrasekaranNoisy} with~$f(x) = \|x\|_1$.

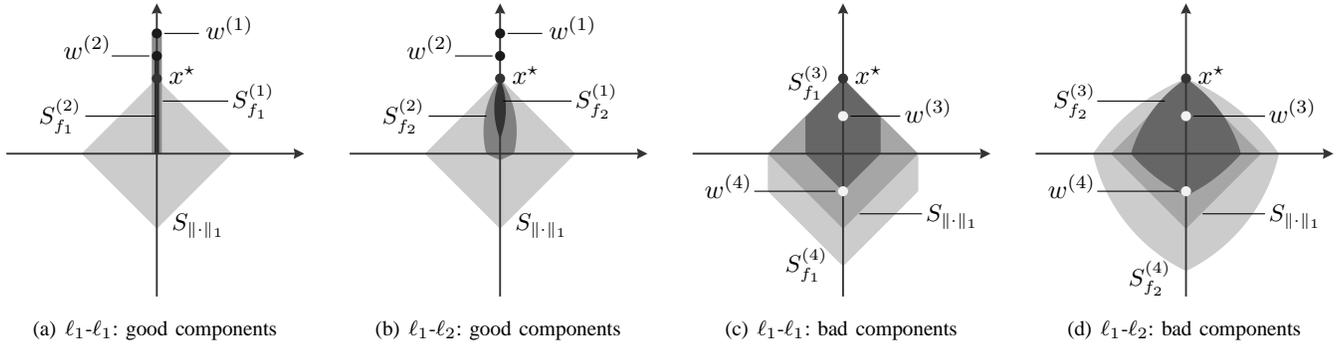
\begin{figure*}
	\centering	
		
	\subfigure[$\ell_1$-$\ell_1$: good components]{\label{SubFig:IllustrationL1L1Good}
	\psscalebox{1.0}{
	\begin{pspicture}(3.96,3.9)
		\psset{blendmode=2}	
	
	\def\radius{1}       
								
	\def\diamond{       
		\pspolygon*[linecolor=black!20!white](0,\radius)(\radius,0.0)(0.0,-\radius)(-\radius,0.0)	
	}

	\rput(2.0,2.0){\diamond}        
											
	\psset{linecolor=black!45!white}
	\psline[linewidth=3.8pt](2.0,2.0)(2.0,3.6)
	
	\psset{linecolor=black!85!white}
	\psline[linewidth=1.7pt](2.0,2.0)(2.0,3.3)

	\psset{linestyle=solid,linecolor=black!80!white,arrowsize=4pt,arrowinset=0.1,arrowlength=1.0}
	
  \psline[linewidth=0.7pt]{->}(0.0,2.0)(3.96,2.0)
  \psline[linewidth=0.7pt]{->}(2.0,0.1)(2.0,4.0)
		
	\rput(2.0,3.3){\pscircle*[linecolor=black!90!white](0,0){0.067}}
	\rput(2.0,3.6){\pscircle*[linecolor=black!90!white](0,0){0.067}}	
			
	\rput(2.0,3.0){\rnode{x}{\pscircle*(0,0){0.067}}}
	
	
	\psset{linestyle=solid,linecolor=black,linewidth=0.4pt}
	
	\rput[l](2.16,3.05){$x^\star$}
	
	\rput[l](2.65,3.68){$w^{(1)}$}
	\psline{-}(2.1,3.6)(2.56,3.6)
	
	\rput[r](1.4,3.42){$w^{(2)}$}
	\psline{-}(1.9,3.3)(1.4,3.3)
		
	\rput[l](3.0,2.7){\small $S_{f_1}^{(1)}$}
	\psline{-}(2.08,2.7)(2.95,2.7)
	\rput[r](1.0,2.5){\small $S_{f_1}^{(2)}$}	
	\psline{-}(1.05,2.5)(1.98,2.5)
	\rput[tl](2.20,1.20){\small $S_{\|\cdot\|_1}$}
	
	\end{pspicture}
	}
	}
	\subfigure[$\ell_1$-$\ell_2$: good components]{\label{SubFig:IllustrationL1L2Good}
	\psscalebox{1.0}{
	\begin{pspicture}(3.96,3.9)
		\psset{blendmode=2}	
	
	\def\radius{1}       
								
	\def\diamond{       
		\pspolygon*[linecolor=black!20!white](0,\radius)(\radius,0.0)(0.0,-\radius)(-\radius,0.0)	
	}

	\rput(2.0,2.0){\diamond}

	\psset{linecolor=black!50!white}
	\pspolygon*
	(1.923065,2.874740)
  (1.901287,2.831831)
  (1.880715,2.787145)
  (1.861901,2.741586)
  (1.845324,2.696098)
  (1.830520,2.649861)
  (1.817511,2.602927)
  (1.806320,2.555350)
  (1.796967,2.507185)
  (1.789470,2.458485)
  (1.784190,2.410246)
  (1.780763,2.361596)
  (1.779200,2.312591)
  (1.779803,2.264244)
  (1.782256,2.215664)
  (1.786569,2.166905)
  (1.792750,2.118023)
  (1.801014,2.070051)
  (1.811118,2.022074)
  (1.826634,1.994841)
  (1.845718,1.984655)
  (1.865142,1.974832)
  (1.884897,1.965383)
  (1.904976,1.956317)
  (1.925302,1.946645)
  (1.945976,1.937372)
  (1.966992,1.928508)
  (1.988329,1.919063)
  (2.009958,1.918046)
  (2.031331,1.927458)
  (2.052381,1.936289)
  (2.073021,1.946528)
  (2.093451,1.955171)
  (2.113455,1.965201)
  (2.133225,1.974619)
  (2.152664,1.984410)
  (2.171934,1.993581)
  (2.187889,2.018828)
  (2.198335,2.065822)
  (2.206527,2.114773)
  (2.213090,2.162690)
  (2.217558,2.211462)
  (2.220166,2.260060)
  (2.220922,2.308430)
  (2.219514,2.357462)
  (2.216239,2.406144)
  (2.210750,2.455353)
  (2.203766,2.503161)
  (2.194563,2.551373)
  (2.183521,2.599000)
  (2.170660,2.645988)
  (2.156001,2.692285)
  (2.139097,2.738719)
  (2.120896,2.783463)
  (2.100464,2.828221)
  (2.078823,2.871206)
  (2.054974,2.914082)
  (2.000000,3.000000)

	\psset{linecolor=black!80!white}
	\pspolygon*
	(1.964922,2.897855)
  (1.953822,2.856234)
  (1.944664,2.815103)
  (1.936870,2.772600)
  (1.930759,2.729728)
  (1.926342,2.686538)
  (1.923837,2.644057)
  (1.922816,2.600386)
  (1.923685,2.557533)
  (1.926239,2.514572)
  (1.930484,2.471553)
  (1.936533,2.429520)
  (1.944146,2.386537)
  (1.953526,2.344646)
  (1.964559,2.302900)
  (1.977245,2.261350)
  (1.991590,2.221045)
  (2.007197,2.218033)
  (2.021666,2.258316)
  (2.034478,2.299848)
  (2.045638,2.341580)
  (2.055144,2.383461)
  (2.062994,2.425443)
  (2.069061,2.468468)
  (2.073433,2.511488)
  (2.076116,2.554455)
  (2.077112,2.597317)
  (2.076427,2.640024)
  (2.073841,2.683500)
  (2.069797,2.725742)
  (2.063831,2.768644)
  (2.056182,2.811181)
  (2.046862,2.853303)
  (2.035885,2.894961)
  (2.000000,3.000000)
		
	\psset{linestyle=solid,linecolor=black!80!white,arrowsize=4pt,arrowinset=0.1,arrowlength=1.0}
	
  \psline[linewidth=0.7pt]{->}(0.0,2.0)(3.96,2.0)
  \psline[linewidth=0.7pt]{->}(2.0,0.1)(2.0,4.0)
		
	\rput(2.0,3.3){\pscircle*[linecolor=black!90!white](0,0){0.067}}
	\rput(2.0,3.6){\pscircle*[linecolor=black!90!white](0,0){0.067}}	
			
	\rput(2.0,3.0){\rnode{x}{\pscircle*(0,0){0.067}}}
	
	
	\psset{linestyle=solid,linecolor=black,linewidth=0.4pt}
	
	\rput[l](2.16,3.05){$x^\star$}
	
	\rput[l](2.65,3.68){$w^{(1)}$}
	\psline{-}(2.1,3.6)(2.56,3.6)
	
	\rput[r](1.4,3.42){$w^{(2)}$}
	\psline{-}(1.9,3.3)(1.4,3.3)
		
	\rput[l](3.0,2.7){\small $S_{f_2}^{(1)}$}
	\psline{-}(2.04,2.7)(2.95,2.7)
	\rput[r](1.0,2.5){\small $S_{f_2}^{(2)}$}	
	\psline{-}(1.05,2.5)(1.82,2.5)
	\rput[tl](2.20,1.20){\small $S_{\|\cdot\|_1}$}
	
	\end{pspicture}
	}
	}
	\subfigure[$\ell_1$-$\ell_1$: bad components]{\label{SubFig:IllustrationL1L1Bad}
	\psscalebox{1.0}{
	\begin{pspicture}(3.96,3.9)
		\psset{blendmode=2}	
	
	\def\radius{1}       
								
	\def\diamond{       
		\pspolygon*[linecolor=black!35!white](0,\radius)(\radius,0.0)(0.0,-\radius)(-\radius,0.0)	
	}

	\psset{linecolor=black!20!white}
	\pspolygon*	
	(0.999529,1.969877)
  (0.999864,1.928140)
  (0.999473,1.883780)
  (0.999813,1.838227)
  (0.999673,1.789940)
  (0.999830,1.739627)
  (0.999791,1.686462)
  (0.999720,1.630381)
  (0.999781,1.571330)
  (0.999588,1.508427)
  (1.022053,1.476938)
  (1.048521,1.450868)
  (1.075136,1.424387)
  (1.101942,1.397483)
  (1.128985,1.370142)
  (1.156311,1.342354)
  (1.184402,1.315012)
  (1.212417,1.286317)
  (1.241255,1.258074)
  (1.270529,1.229379)
  (1.299925,1.199300)
  (1.330234,1.169695)
  (1.360801,1.138697)
  (1.392036,1.107244)
  (1.423989,1.075346)
  (1.456710,1.043013)
  (1.489999,1.009290)
  (1.524194,0.975152)
  (1.559343,0.940617)
  (1.595117,0.903743)
  (1.632191,0.867487)
  (1.670105,0.828921)
  (1.709286,0.790039)
  (1.749670,0.749882)
  (1.791365,0.708478)
  (1.834404,0.664864)
  (1.879004,0.620074)
  (1.925269,0.574151)
  (1.973290,0.526144)
  (2.022806,0.522105)
  (2.070984,0.570037)
  (2.117353,0.616888)
  (2.162084,0.661611)
  (2.205253,0.705161)
  (2.247073,0.746504)
  (2.287581,0.786604)
  (2.326734,0.826420)
  (2.364917,0.863947)
  (2.401920,0.901136)
  (2.437999,0.936983)
  (2.473035,0.972449)
  (2.507323,1.006543)
  (2.540705,1.040223)
  (2.573515,1.072514)
  (2.605557,1.104373)
  (2.636880,1.135789)
  (2.667534,1.166752)
  (2.697567,1.197255)
  (2.727409,1.226370)
  (2.756368,1.255950)
  (2.785269,1.284156)
  (2.813347,1.312814)
  (2.841048,1.341013)
  (2.868888,1.367874)
  (2.895992,1.395182)
  (2.922859,1.422053)
  (2.949533,1.448501)
  (2.975520,1.475382)
  (3.000258,1.503510)
  (3.000231,1.566458)
  (3.000452,1.625556)
  (3.000534,1.681684)
  (3.000023,1.735683)
  (3.000307,1.786033)
  (3.000289,1.834359)
  (3.000080,1.880696)
  (3.000471,1.924352)
  (3.000220,1.966848)
  (2.000000,3.000000)
	
	\rput(2.0,2.0){\diamond}

	\psset{linecolor=black!60!white}
	\pspolygon*
	(1.499765,2.484938)
  (1.499932,2.464070)
  (1.499403,2.441518)
  (1.499580,2.418735)
  (1.499836,2.394970)
  (1.499604,2.369422)
  (1.499593,2.342833)
  (1.499860,2.315191)
  (1.499891,2.285665)
  (1.499794,2.254213)
  (1.499680,2.220798)
  (1.499662,2.185385)
  (1.499854,2.147945)
  (1.499884,2.107580)
  (1.499921,2.064243)
  (1.499683,2.016998)
  (1.510729,1.989187)
  (1.527283,1.971427)
  (1.544753,1.954844)
  (1.562317,1.937628)
  (1.580027,1.919767)
  (1.597934,1.901254)
  (1.616416,1.883029)
  (1.635160,1.864156)
  (1.654221,1.844633)
  (1.673919,1.825423)
  (1.694000,1.805574)
  (1.714516,1.785091)
  (1.735522,1.763979)
  (1.757070,1.742246)
  (1.779213,1.719901)
  (1.802003,1.696957)
  (1.825493,1.673429)
  (1.849736,1.649333)
  (1.874782,1.624689)
  (1.900614,1.598520)
  (1.927392,1.571844)
  (1.955168,1.544690)
  (1.983968,1.515087)
  (2.013686,1.513063)
  (2.042602,1.541622)
  (2.070431,1.569733)
  (2.097292,1.596368)
  (2.123205,1.622499)
  (2.148222,1.648101)
  (2.172523,1.672161)
  (2.196070,1.695654)
  (2.218917,1.718565)
  (2.241115,1.740878)
  (2.262716,1.762581)
  (2.283775,1.783664)
  (2.304344,1.804119)
  (2.324476,1.823941)
  (2.344223,1.843125)
  (2.363334,1.862624)
  (2.382128,1.881473)
  (2.400657,1.899675)
  (2.418613,1.918167)
  (2.436370,1.936007)
  (2.453980,1.953203)
  (2.471078,1.970675)
  (2.488095,1.987508)
  (2.500107,2.013527)
  (2.500468,2.059917)
  (2.500157,2.104166)
  (2.500302,2.144560)
  (2.500084,2.182883)
  (2.500156,2.218317)
  (2.500129,2.251755)
  (2.500116,2.283229)
  (2.500226,2.312778)
  (2.500569,2.340444)
  (2.500011,2.367841)
  (2.500471,2.392631)
  (2.500144,2.417180)
  (2.500373,2.439975)
  (2.500576,2.461810)
  (2.500110,2.483424)
  (2.000000,3.000000)

	\psset{linestyle=solid,linecolor=black!80!white,arrowsize=4pt,arrowinset=0.1,arrowlength=1.0}
	
  \psline[linewidth=0.7pt]{->}(0.0,2.0)(3.96,2.0)
  \psline[linewidth=0.7pt]{->}(2.0,0.1)(2.0,4.0)
		
	\rput(2.0,2.5){\pscircle*[linecolor=black!5!white](0,0){0.067}}
	\rput(2.0,1.5){\pscircle*[linecolor=black!5!white](0,0){0.067}}	
			
	\rput(2.0,3.0){\rnode{x}{\pscircle*(0,0){0.067}}}
	
	
	\psset{linestyle=solid,linecolor=black,linewidth=0.4pt}
	
	\rput[l](2.16,3.05){$x^\star$}
	
	\rput[l](2.82,2.57){$w^{(3)}$}
	\psline{-}(2.1,2.5)(2.76,2.5)
	
	\rput[r](0.8,1.57){$w^{(4)}$}
	\psline{-}(0.82,1.5)(1.9,1.5)
		
	\rput[rb](1.83,2.72){\small $S_{f_1}^{(3)}$}	
	\rput[rt](1.8,0.75){\small $S_{f_1}^{(4)}$}	
	\rput[l](3.1,1.16){\small $S_{\|\cdot\|_1}$}
	\psline{-}(2.22,1.2)(3.05,1.2)
	
	\end{pspicture}
	}
	}
	\subfigure[$\ell_1$-$\ell_2$: bad components]{\label{SubFig:IllustrationL1L2Bad}
	\psscalebox{1.0}{
	\begin{pspicture}(3.96,3.9)
		\psset{blendmode=2}	
	
	\def\radius{1}       
								
	\def\diamond{       
		\pspolygon*[linecolor=black!35!white](0,\radius)(\radius,0.0)(0.0,-\radius)(-\radius,0.0)	
	}

	\psset{linecolor=black!20!white}
	\pspolygon*
	(1.903289,2.959111)
  (1.805512,2.913147)
  (1.709575,2.863273)
  (1.616490,2.809990)
  (1.525458,2.752948)
  (1.436592,2.692209)
  (1.350003,2.627837)
  (1.265801,2.559900)
  (1.184941,2.488999)
  (1.106658,2.414710)
  (1.031881,2.337674)
  (0.959863,2.257437)
  (0.891504,2.174676)
  (0.826072,2.088908)
  (0.764437,2.000848)
  (0.771663,1.965386)
  (0.780571,1.930478)
  (0.789650,1.894842)
  (0.799672,1.859146)
  (0.810642,1.823414)
  (0.821869,1.786951)
  (0.834084,1.750472)
  (0.847292,1.714001)
  (0.860844,1.676807)
  (0.875428,1.639646)
  (0.891048,1.602545)
  (0.907706,1.565530)
  (0.924816,1.527822)
  (0.943003,1.490230)
  (0.961710,1.451953)
  (0.981530,1.413827)
  (1.002464,1.375882)
  (1.024514,1.338147)
  (1.047677,1.300652)
  (1.071483,1.262543)
  (1.095983,1.223824)
  (1.122100,1.186298)
  (1.148942,1.148205)
  (1.177361,1.111386)
  (1.206153,1.073121)
  (1.236512,1.036194)
  (1.268035,0.999718)
  (1.300389,0.962781)
  (1.333931,0.926347)
  (1.368653,0.890449)
  (1.404548,0.855119)
  (1.441357,0.819422)
  (1.479587,0.785323)
  (1.518751,0.750911)
  (1.559275,0.718172)
  (1.600747,0.685179)
  (1.643365,0.652941)
  (1.687114,0.621491)
  (1.731979,0.590863)
  (1.777945,0.561088)
  (1.824996,0.532197)
  (1.873115,0.504223)
  (1.922282,0.477197)
  (1.972481,0.451149)
  (2.023478,0.449108)
  (2.073758,0.475077)
  (2.123009,0.502027)
  (2.171212,0.529927)
  (2.218348,0.558745)
  (2.264402,0.588451)
  (2.309356,0.619013)
  (2.353194,0.650398)
  (2.395902,0.682574)
  (2.437467,0.715509)
  (2.478082,0.748191)
  (2.517339,0.782549)
  (2.555663,0.816597)
  (2.592568,0.852246)
  (2.628557,0.887529)
  (2.663374,0.923384)
  (2.697012,0.959777)
  (2.729463,0.996676)
  (2.761081,1.033115)
  (2.791157,1.070934)
  (2.820426,1.108243)
  (2.848523,1.145943)
  (2.875879,1.183100)
  (2.901639,1.221494)
  (2.926688,1.259301)
  (2.950590,1.297391)
  (2.973848,1.334869)
  (2.995470,1.373443)
  (3.016481,1.411365)
  (3.036379,1.449470)
  (3.055736,1.486908)
  (3.074015,1.524494)
  (3.091217,1.562199)
  (3.107345,1.599996)
  (3.123040,1.637085)
  (3.137698,1.674235)
  (3.151325,1.711421)
  (3.164606,1.747884)
  (3.176893,1.784358)
  (3.188903,1.820112)
  (3.199234,1.856546)
  (3.209326,1.892240)
  (3.219225,1.927216)
  (3.227451,1.962786)
  (3.236300,1.996991)
  (3.178788,2.082125)
  (3.114391,2.167525)
  (3.045456,2.251123)
  (2.973658,2.331606)
  (2.899094,2.408894)
  (2.821016,2.483441)
  (2.740352,2.554607)
  (2.657206,2.622318)
  (2.570808,2.686984)
  (2.482125,2.748024)
  (2.391265,2.805371)
  (2.297440,2.859392)
  (2.202561,2.909154)
  (2.104930,2.955439)
  (2.000000,3.000000)
	
	\rput(2.0,2.0){\diamond}

	\psset{linecolor=black!60!white}
	\pspolygon*
	(1.905590,2.932600)
  (1.849206,2.887727)
  (1.794602,2.840590)
  (1.741846,2.791241)
  (1.691769,2.740380)
  (1.642892,2.686792)
  (1.596796,2.631839)
  (1.552051,2.574247)
  (1.510181,2.515440)
  (1.471200,2.455527)
  (1.434439,2.393880)
  (1.399283,2.329821)
  (1.367788,2.265651)
  (1.338637,2.199973)
  (1.311878,2.132857)
  (1.287555,2.064372)
  (1.268663,1.998628)
  (1.285394,1.979287)
  (1.302608,1.960217)
  (1.320300,1.941430)
  (1.337942,1.922083)
  (1.356088,1.903022)
  (1.374733,1.884257)
  (1.393872,1.865802)
  (1.413047,1.846778)
  (1.432741,1.828069)
  (1.452951,1.809690)
  (1.473271,1.790737)
  (1.494132,1.772123)
  (1.515528,1.753864)
  (1.537453,1.735972)
  (1.559576,1.717517)
  (1.582254,1.699445)
  (1.605480,1.681770)
  (1.628979,1.663545)
  (1.653051,1.645736)
  (1.677687,1.628360)
  (1.702672,1.610454)
  (1.728245,1.593004)
  (1.754397,1.576025)
  (1.781121,1.559534)
  (1.808277,1.542556)
  (1.836025,1.526093)
  (1.864265,1.509166)
  (1.893116,1.492785)
  (1.922569,1.476967)
  (1.952612,1.461730)
  (1.983223,1.446091)
  (2.014312,1.445066)
  (2.044967,1.460657)
  (2.075055,1.475847)
  (2.104552,1.491619)
  (2.133450,1.507956)
  (2.161736,1.524840)
  (2.189530,1.541261)
  (2.216733,1.558199)
  (2.243503,1.574650)
  (2.269702,1.591591)
  (2.295322,1.609004)
  (2.320355,1.626875)
  (2.345039,1.644216)
  (2.369158,1.661992)
  (2.392705,1.680185)
  (2.415978,1.697828)
  (2.438704,1.715871)
  (2.460876,1.734298)
  (2.482848,1.752161)
  (2.503912,1.771319)
  (2.525200,1.788982)
  (2.545569,1.807911)
  (2.565826,1.826266)
  (2.585568,1.844952)
  (2.604791,1.863956)
  (2.623977,1.882390)
  (2.642164,1.901998)
  (2.660862,1.920177)
  (2.678552,1.939506)
  (2.696291,1.958277)
  (2.712979,1.978150)
  (2.730330,1.996656)
  (2.713976,2.059256)
  (2.690462,2.127056)
  (2.663903,2.194290)
  (2.634949,2.260091)
  (2.602980,2.325137)
  (2.568682,2.388587)
  (2.532106,2.450374)
  (2.492596,2.511139)
  (2.450889,2.570083)
  (2.407041,2.627144)
  (2.361113,2.682260)
  (2.312400,2.736017)
  (2.261704,2.787680)
  (2.209878,2.836576)
  (2.155422,2.883896)
  (2.099180,2.928956)
  (2.000000,3.000000)
		
	\psset{linestyle=solid,linecolor=black!80!white,arrowsize=4pt,arrowinset=0.1,arrowlength=1.0}
	
  \psline[linewidth=0.7pt]{->}(0.0,2.0)(3.96,2.0)
  \psline[linewidth=0.7pt]{->}(2.0,0.1)(2.0,4.0)
		
	\rput(2.0,2.5){\pscircle*[linecolor=black!5!white](0,0){0.067}}
	\rput(2.0,1.5){\pscircle*[linecolor=black!5!white](0,0){0.067}}	
			
	\rput(2.0,3.0){\rnode{x}{\pscircle*(0,0){0.067}}}
	
	
	\psset{linestyle=solid,linecolor=black,linewidth=0.4pt}
	
	\rput[l](2.16,3.05){$x^\star$}
	
	\rput[l](3.1,2.57){$w^{(3)}$}
	\psline{-}(2.1,2.5)(3.05,2.5)
	
	\rput[r](0.8,1.57){$w^{(4)}$}
	\psline{-}(0.82,1.5)(1.9,1.5)
		
	\rput[r](0.8,2.7){\small $S_{f_2}^{(3)}$}	
	\psline{-}(0.82,2.7)(1.72,2.7)
	\rput[rt](1.8,0.57){\small $S_{f_2}^{(4)}$}	
	\rput[l](3.1,1.16){\small $S_{\|\cdot\|_1}$}
	\psline{-}(2.22,1.2)(3.05,1.2)
	
	\end{pspicture}
	}
	}	
	\caption{
		Sublevel sets of functions~$f_1$ and~$f_2$ with~$\beta = 1$ for~$x^\star = (0,1)$. In both \text{(a)} and \text{(b)}, the prior information is $w^{(1)} = (0,1.6)$ and $w^{(2)} = (0,1.3)$, while in \text{(c)} and \text{(d)} it is $w^{(3)} = (0,0.5)$, and $w^{(4)} = (0,-0.5)$. For reference, the sublevel set~$S_{\|\cdot\|_1}$ of the $\ell_1$-norm at~$x^\star$ is also shown in all figures.
	}
	\label{Fig:GoodAndBadComponents}
	
	\end{figure*}

	\subsection{The Geometry of $\ell_1$-$\ell_1$ and $\ell_1$-$\ell_2$ Minimization}
	
	\thref{Thm:Chandrasekaran} applies to CS by making~$f(x) = \|x\|_1$. Since it is applicable to any convex function~$f$, we will use it to characterize problems~\eqref{Eq:L1L1} and~\eqref{Eq:L1L2}, that is, when~$f$ is $f_1(x) := \|x\|_1 + \beta \|x - w\|_1$ and $f_2(x):=\|x\|_1 + \frac{\beta}{2}\|x - w\|_2^2$, respectively. In particular, we want to understand the relation between the Gaussian widths of the tangent cones associated with these functions and the one associated with the $\ell_1$-norm. If the former is smaller, we might obtain reconstruction bounds for~\eqref{Eq:L1L1} and~\eqref{Eq:L1L2} smaller than the one in~\eqref{Eq:PropChandrasekaranBound}. In the same way that \pref{Prop:ChandrasekaranBound} bounded the squared Gaussian width of the $\ell_1$-norm in terms of the key parameters~$n$ and~$s$, we seek to do the same for~$f_1$ and~$f_2$. To find out the key parameters in this case and also to gain some intuition about the problem, \fref{Fig:GoodAndBadComponents} shows the sublevel sets of~$f_1$ and~$f_2$ with~$\beta = 1$ and in two dimensions, i.e., for~$n=2$. Recall that, according to~\eqref{Eq:TangentConeSublevelSet}, one can estimate tangent cones by observing the sublevel sets that generate them. We set $x^\star = (0,1)$ in all plots of \fref{Fig:GoodAndBadComponents} and consider four different vectors as prior information~$w$: $w^{(1)} = (0,1.6)$ and~$w^{(2)} = (0,1.3)$ in Figs.~\ref{SubFig:IllustrationL1L1Good} and~\ref{SubFig:IllustrationL1L2Good}; and $w^{(3)} = (0,0.5)$ and $w^{(4)} = (0,-0.5)$ in Figs.~\ref{SubFig:IllustrationL1L1Bad} and~\ref{SubFig:IllustrationL1L2Bad}. The sublevel sets are denoted with
	$$
		S_{f_i}^{(j)} := \{x\,:\, \|x\|_1 + g_i(x-w^{(j)}) \leq \|x^\star\|_1 + g_i(x^\star-w^{(j)})\}\,,
	$$
	where~$i=1,2$, $j = 1,2,3,4$, and~$g_1 = \|\cdot\|_1$ and~$g_2 = \frac{1}{2}\|\cdot\|_2^2$. For reference, we also show the sublevel set~$S_{\|\cdot\|_1}$ associated with~BP. The sublevel sets of~$f_1$ are shown in Figs.~\ref{SubFig:IllustrationL1L1Good} and~\ref{SubFig:IllustrationL1L1Bad}, whereas the sublevel sets of~$f_2$ are shown in Figs.~\ref{SubFig:IllustrationL1L2Good} and~\ref{SubFig:IllustrationL1L2Bad}. 
	For example, the sublevel set~$S_{f_1}^{(1)}$ in \fref{SubFig:IllustrationL1L1Good} can be computed in closed-form as	$S_{f_1}^{(1)} = \{(0,x_2):\, 0 \leq x_2 \leq 1.6\}$. The cone generated by this set is the axis~$x_1 = 0$. In the same figure, $S_{f_1}^{(2)} = \{(0,x_2):\, 0 \leq x_2 \leq 1.3\}$ and it generates the same cone as~$S_{f_1}^{(1)}$: the axis~$x_1 = 0$. Hence, both~$S_{f_1}^{(1)}$ and~$S_{f_1}^{(2)}$ generate the same (tangent) cone $\{(0,x_2)\,:\, x_2 \in \mathbb{R}\}$, which has zero Gaussian width in~$\mathbb{R}^2$. When we consider~$f_2$ and the same prior information vectors, as in \fref{SubFig:IllustrationL1L2Good}, the tangent cones no longer have zero width, but still have a width smaller than~$T_{\|\cdot\|_1}(x^\star)$. Since small widths are desirable, we say that the nonzero components of the $w$'s in Figs.~\ref{SubFig:IllustrationL1L1Good} and~\ref{SubFig:IllustrationL1L2Good} are \textit{good components}. On the other hand, the cones generated by the sublevel sets of \fref{SubFig:IllustrationL1L1Bad} coincide with~$T_{\|\cdot\|_1}(x^\star)$, and the cones generated by the sublevel sets of \fref{SubFig:IllustrationL1L2Bad} have widths larger than~$T_{\|\cdot\|_1}(x^\star)$. Therefore, we say that the nonzero components of the~$w$'s in Figs.~\ref{SubFig:IllustrationL1L1Bad} and~\ref{SubFig:IllustrationL1L2Bad} are \textit{bad components}. \fref{Fig:GoodAndBadComponents} illustrates the concepts of good and bad components only for~$x_i^\star > 0$. For~$x_i^\star < 0$, there is geometric symmetry. This motivates the following definition.
	\begin{Definition}[Good and bad components]
	\label{def:GoodBadComponents}
		Let~$x^\star \in \mathbb{R}^n$ be the vector to reconstruct and let~$w \in \mathbb{R}^n$ be the prior information. For $i=1,\ldots,n$, a component~$w_i$ is considered good if
		$$
			\text{$x_i^\star > 0$\, \text{and} \, $x_i^\star < w_i$}
			\qquad
			\text{or}
			\qquad
			\text{$x_i^\star < 0$\, \text{and} \, $x_i^\star > w_i$}\,,
		$$
		and~$w_i$ is considered bad if
		$$
			\text{$x_i^\star > 0$\, \text{and} \, $x_i^\star > w_i$}
			\qquad
			\text{or}
			\qquad
			\text{$x_i^\star < 0$\, \text{and} \, $x_i^\star < w_i$}\,.
		$$
	\end{Definition}
	Note that good and bad components are defined only on the support of~$x^\star$ and that the inequalities in the definition are strict. Although good and bad components were motivated geometrically, we will see that they arise naturally in our proofs. Notice that~\dref{def:GoodBadComponents} (and~\fref{Fig:GoodAndBadComponents}) consider only components~$w_i$ such that~$w_i \neq x_i^\star$ and for which~$x_i^\star \neq 0$. The other components will, of course, also influence the Gaussian width of~$T_{f_1}(x^\star)$ and~$T_{f_2}(x^\star)$ (see, e.g., the role of~$\xi$ in \thref{Thm:L1L1Simplified}). This will be clear when we present our main results in the next section.

\section{Main Results}
\label{Sec:MainResults}

	In this section we present our main results, namely reconstruction guarantees for $\ell_1$-$\ell_1$ and $\ell_1$-$\ell_2$ minimization. After some definitions and preliminary results, we present the results for $\ell_1$-$\ell_1$ minimization first, and the results for $\ell_1$-$\ell_2$ minimization next. All proofs are relegated to \sref{Sec:Proofs}.
		
	\subsection{Definitions and Preliminary Results}
	
	We start by defining the following sets.
	\begin{Definition}[Support sets]
	\label{def:SupportSets}
		Let~$x^\star \in \mathbb{R}^n$ be the vector to reconstruct and let~$w \in \mathbb{R}^n$ be the prior information. We define
		\begin{align*}
			  I &:= \bigl\{i\,:\, x_i^\star \neq 0\bigr\}
			&
			  J &:= \bigl\{j\,:\, x_j^\star \neq w_j\bigr\}			
			\\
			  I^c &:= \bigl\{i\,:\, x_i^\star = 0\bigr\} 
			&
				J^c &:= \bigl\{j\,:\, x_j^\star = w_j\bigr\}
			\\
				I_+ &:= \bigl\{i\,:\, x_i^\star > 0\bigr\}
			&
				J_+ &:= \bigl\{j\,:\, x_j^\star > w_j\bigr\}
			\\
				I_- &:= \bigl\{i\,:\, x_i^\star < 0\bigr\}
			&
				J_- &:= \bigl\{j\,:\, x_j^\star < w_j\bigr\}\,.			
		\end{align*}		
	\end{Definition}
	To simplify notation, we denote the intersection of two sets~$A$ and~$B$ with the product~$AB := A \cap B$. Then, the set of good components can be written as~$I_+J_- \cup I_- J_+$, and the set of bad components can be written as $I_+ J_+ \cup I_- J_-$.	
		
	\begin{Definition}[Cardinality of sets]\label{def:CardinalityOfsomeSets}
		The number of good components, the number of bad components, the sparsity of~$x^\star$, the sparsity of~$w$, and the cardinality of the union of the supports of~$x^\star$ and~$x^\star - w$ are represented, respectively, by
		\begin{align*}
			h &:= \bigl| I_+J_-\bigr| + \bigl| I_- J_+ \bigr|
			\\
			\overline{h} &:= \bigl| I_+J_+\bigr| + \bigl| I_- J_- \bigr|
			\\
			s &:= |I|
			\\
			l &:= |J|
			\\
			q &:= \bigl|I \cup J \bigr|\,.
		\end{align*}		
	\end{Definition}
	All these quantities are nonnegative. Before moving to our main results, we present the following useful lemma:
	\begin{Lemma}
	\label{lem:SetIdentities}
		For~$x^\star$ and~$w$ as in \dref{def:SupportSets},
		\begin{align}
			\big|IJ\big|  &= h + \overline{h}
			\label{Eq:LemSetIdentity1}			
			\\			
			|IJ^c|  &= s - (h + \overline{h})\,.
			\label{Eq:LemSetIdentity7}
			\\
			|I^c J| &= q - s
			\label{Eq:LemSetIdentity6}
			\\			
			\big|I^c J^c\big| &= n - q
			\label{Eq:LemSetIdentity3}
		\end{align}
	\end{Lemma}		
	\begin{proof}	
		Identity~\eqref{Eq:LemSetIdentity1} is proven by noticing that~$I_+$ and~$I_-$ partition~$I$, and $J_+$ and~$J_-$ partition~$J$. Then,
		\begin{align*}
				\big|IJ\big|
			&=
			  \big|I_+ J\big| + \big|I_- J\big|			
			\\
			&=
				\big|I_+ J_+\big| + \big|I_+ J_-\big| + \big|I_- J_+\big| + \big|I_- J_-\big|			
			\\
			\Shorter{}{
			&=
				\Bigl(\big|I_+ J_-\big| + \big|I_- J_+\big|\Bigr)
				+
				\Bigl(\big|I_+ J_+\big| + \big|I_- J_-\big|\Bigr)
			\\
			}
			&=
				h + \overline{h}\,.
		\end{align*}
		To prove~\eqref{Eq:LemSetIdentity7}, we use~\eqref{Eq:LemSetIdentity1} and the fact that~$J$ and~$J^c$ are a partition of $\{1,\ldots,n\}$:
		$$
			s = \big|I\big| = \big|IJ\big| + \big|IJ^c\big| = (h + \overline{h}) + \big|IJ^c\big|\,,
		$$
		from which~\eqref{Eq:LemSetIdentity7} follows. To prove~\eqref{Eq:LemSetIdentity6}, we use
		the identity $I \cup J = \big(I^cJ\big) \cup \big(IJ\big) \cup \big(IJ^c\big)$, where $I^cJ$, $IJ$ and~$IJ^c$ are pairwise disjoint. Then, using~\eqref{Eq:LemSetIdentity1} and~\eqref{Eq:LemSetIdentity7},
		\begin{align*}
				q 
			&=
				\big|I \cup J\big|
			=
			 \big|I^c J\big| + \big|IJ\big| + \big|I J^c\big|						 
			=
				\big|I^c J\big| + s\,.
		\end{align*}
		Finally, \eqref{Eq:LemSetIdentity3} holds because
		$$
			n = \big|I\big| + \big|I^c\big| = \big|IJ\big| + \big|IJ^c\big| + \big|I^c J\big| + \big|I^c J^c\big|
			=
			q + \big|I^c J^c\big|\,,
		$$			
		where we used~\eqref{Eq:LemSetIdentity1}, \eqref{Eq:LemSetIdentity7}, and~\eqref{Eq:LemSetIdentity6}.
	\end{proof}	
	From~\lref{lem:SetIdentities}, we can easily obtain the following identities, which will be used frequently:
	\begin{align}
		  \big|I^c J\big| + \big|IJ^c\big| &= q - (h + \overline{h})
		\label{Eq:LemSetIdentity2}
		\\		  
			\big|I^c J\big| + \big|IJ^c\big| + 2\big|I^c J^c\big| &= 2n - (q + h + \overline{h})
		\label{Eq:LemSetIdentity4}
		\\
			\big|I^cJ\big| - \big|IJ^c\big| &= q + h + \overline{h} - 2s\,.
		\label{Eq:LemSetIdentity5}  
	\end{align}
	Finally, note that~\eqref{Eq:LemSetIdentity3} allows interpreting~$q$ as the size of the union of the supports of~$x^\star$ and~$w$: since both~$x^\star$ and~$w$ are zero in~$I^cJ^c$, $q$ is the number of components in which at least one of them is not zero.
	
\subsection{$\ell_1$-$\ell_1$ Minimization}
\label{SubSec:MainResultsL1L1}
		
	We now state our result for $\ell_1$-$\ell_1$ minimization. Its proof can be found in \ssref{SubSec:ProofsL1L1}. 
	\begin{Theorem}[$\ell_1$-$\ell_1$ minimization]	
	\label{Thm:L1L1Reconstruction}
		Let~$x^\star \in \mathbb{R}^n$ be the vector to reconstruct and let~$w \in \mathbb{R}^n$ be the prior information. Let~$f_1(x) = \|x\|_1 + \beta \|x - w\|_1$ with~$\beta >0$, and assume~$x^\star \neq 0$, $w \neq x^\star$, and $q < n$.	
		\begin{enumerate}
			\item 
				Let $\beta = 1$, and assume there is at least one bad component, i.e., $\overline{h}>0$. Then,
				\begin{equation}\label{Eq:ThmL1L1Case1Bound}
						w\bigl(T_{f_1}(x^\star)\bigr)^2
					\leq
						2\overline{h} \log\Big(\frac{2n}{q + h + \overline{h}}\Big) + \frac{7}{10}(q + h + \overline{h})\,.
				\end{equation}
		
			\item 
				Let $\beta < 1$.
				\begin{enumerate}
					\item If
						\begin{equation}\label{Eq:ThmL1L1Case2Condition1}
							\frac{q - s}{2n - (q + h + \overline{h})}
							\leq
							\frac{1 - \beta}{1 + \beta}\bigg(\frac{q + h + \overline{h}}{2n}\bigg)^{\frac{4\beta}{(\beta + 1)^2}}\,,
						\end{equation}
						then
						\isdraft{                                              
							\begin{equation}\label{Eq:ThmL1L1Case2Bound1}
								w\bigl(T_{f_1}(x^\star)\bigr)^2
								\leq
								2\bigg[\overline{h} + (s - \overline{h})\frac{(1 - \beta)^2}{(1 + \beta)^2}\bigg]\log\Big(\frac{2n}{q + h + \overline{h}}\Big)
								+ s
								+ \frac{2}{5}(q + h + \overline{h})\,.
							\end{equation}                                       
						}{						
						\begin{multline}\label{Eq:ThmL1L1Case2Bound1}
								w\bigl(T_{f_1}(x^\star)\bigr)^2
							\leq
								2\bigg[\overline{h} + (s - \overline{h})\frac{(1 - \beta)^2}{(1 + \beta)^2}\bigg]\times
								\\\times\log\Big(\frac{2n}{q + h + \overline{h}}\Big)
								+ s
								+ \frac{2}{5}(q + h + \overline{h})\,.
						\end{multline}
						}
						
					\item If $q>s$ and
						\begin{equation}\label{Eq:ThmL1L1Case2Condition2}
							\frac{q - s}{2n - (q + h + \overline{h})}
							\geq
							\frac{1 - \beta}{1 + \beta}\bigg(\frac{s}{q}\bigg)^{\frac{4\beta}{(1 - \beta)^2}}\,,
						\end{equation}
						then
						\isdraft{                                              
							\begin{equation}\label{Eq:ThmL1L1Case2Bound2}
								w\bigl(T_{f_1}(x^\star)\bigr)^2
								\leq
								2\bigg[\overline{h}\frac{(1 + \beta)^2}{(1 - \beta)^2} + s - \overline{h}\bigg]
								\log\Big(\frac{q}{s}\Big) + \frac{7}{5}s\,.
							\end{equation}                                       
						}{
						\begin{multline}\label{Eq:ThmL1L1Case2Bound2}
							w\bigl(T_{f_1}(x^\star)\bigr)^2
							\leq
								2\bigg[\overline{h}\frac{(1 + \beta)^2}{(1 - \beta)^2} + s - \overline{h}\bigg]
								\log\Big(\frac{q}{s}\Big) \\+ \frac{7}{5}s\,.
						\end{multline}
						}
				\end{enumerate}
				
			\item 
				Let $\beta > 1$.
				\begin{enumerate}
					\item If
						\begin{equation}\label{Eq:ThmL1L1Case3Condition1}
							\frac{s - (h + \overline{h})}{2n - (q + h + \overline{h})}
							\leq
							\frac{\beta - 1}{\beta + 1}\bigg(\frac{q + h + \overline{h}}{2n}\bigg)^{\frac{4\beta}{(\beta + 1)^2}}\!,
						\end{equation}
						then
						\isdraft{                                              
							 \begin{equation}\label{Eq:ThmL1L1Case3Bound1}
								w\bigl(T_{f_1}(x^\star)\bigr)^2
								\leq
								2\bigg[\overline{h} + (q + h - s)\frac{(\beta-1)^2}{(\beta+1)^2}\bigg]								
								\log\Big(\frac{2n}{q + h + \overline{h}}\Big)  + l + \frac{2}{5}(q + h + \overline{h})\,.
								\end{equation}                                     
						}{
						\begin{multline}\label{Eq:ThmL1L1Case3Bound1}
							w\bigl(T_{f_1}(x^\star)\bigr)^2
							\leq
								2\bigg[\overline{h} + (q + h - s)\frac{(\beta-1)^2}{(\beta+1)^2}\bigg]\times
								\\\times
								\log\Big(\frac{2n}{q + h + \overline{h}}\Big)  + l + \frac{2}{5}(q + h + \overline{h})\,.
						\end{multline}
						}
						
					\item If~$s > h + \overline{h} > 0$ and
						\begin{equation}\label{Eq:ThmL1L1Case3Condition2}	
							\frac{s - (h + \overline{h})}{2n - (q + h + \overline{h})}
							\geq
							\frac{\beta - 1}{\beta + 1}\bigg(\frac{h + \overline{h}}{s}\bigg)^{\frac{4\beta}{(\beta - 1)^2}}\,,
						\end{equation}
						then
						\isdraft{                                               
							\begin{equation}\label{Eq:ThmL1L1Case3Bound2}
							w\bigl(T_{f_1}(x^\star)\bigr)^2
							\leq
								2\bigg[\overline{h}\frac{(\beta + 1)^2}{(\beta - 1)^2} + q + h - s\bigg]								
								\log\Big(\frac{s}{h + \overline{h}}\Big) + l + \frac{2}{5}(h + \overline{h})\,.
						\end{equation}                                          
						}{
						\begin{multline}\label{Eq:ThmL1L1Case3Bound2}
							w\bigl(T_{f_1}(x^\star)\bigr)^2
							\leq
								2\bigg[\overline{h}\frac{(\beta + 1)^2}{(\beta - 1)^2} + q + h - s\bigg]\times
								\\\times
								\log\Big(\frac{s}{h + \overline{h}}\Big) + l + \frac{2}{5}(h + \overline{h})\,.
						\end{multline}
						}
				\end{enumerate}
			
			\end{enumerate}
	\end{Theorem}
	Similarly to~\pref{Prop:ChandrasekaranBound}, the previous theorem establishes upper bounds on~$w(T_{f_1}(x^\star))^2$ that depend only on key parameters~$n$, $s$, $\beta$, $q$, $h$, and~$\overline{h}$. Together with \thref{Thm:Chandrasekaran}, it then provides (useful) bounds on the number of measurements that guarantee that~\eqref{Eq:L1L1} reconstructs~$x^\star$ with high probability. The assumption~$q<n$ means that the union of the supports of~$x^\star$ and~$w$ is not equal to the full set~$\{1,\dots,n\}$ or, equivalently, that there is at least one index~$i$ such that~$x_i^\star = w_i = 0$. Assuming~$w\neq x^\star$ and~$x^\star \neq 0$ is equivalent to assuming that the sets~$J$ and~$I$ are nonempty, respectively. 
	
	The theorem is divided into three cases: \text{1)} $\beta=1$, \text{2)} $\beta < 1$, and~\text{3)} $\beta > 1$. We will see that, although rare in practice, the theorem may not cover all possible values of~$\beta$, due to the conditions imposed in cases~\text{2)} and~\text{3)}. Recall that \thref{Thm:L1L1Simplified} in \sref{Sec:Intro} instantiates case~\text{1)}, i.e., $\beta =1$, but in a slightly different format. Namely, to obtain~\eqref{Eq:IntroL1L1Bound} from~\eqref{Eq:ThmL1L1Case1Bound}, notice that $\xi = |I^cJ|-|IJ^c|$ and that~\eqref{Eq:LemSetIdentity5} implies $(q + h + \overline{h})/2 = s + \xi/2$. Therefore, the observations made for \thref{Thm:L1L1Simplified} apply to case~\text{1)} of the previous theorem. We add to those observations that the assumption that there is at least one bad component, i.e., $\overline{h}>0$, is necessary to guarantee $0 \not\in \partial f_1(x^\star)$ and, hence, that we can use~\pref{Prop:Subdifferential}. In fact, it will be shown in part \text{1)} of \lref{lem:ZeroNotInSubGradient} that $0 \not\in \partial f_1(x^\star)$ if and only if~$\overline{h} > 0$ or~$\beta \neq 1$. Thus, the assumption~$\overline{h}>0$ can be dropped in cases \text{2)} and \text{3)}, where~$\beta\neq 1$. Note that the quantities on the right-hand side of~\eqref{Eq:ThmL1L1Case1Bound} are well defined and positive: the assumption that~$x^\star \neq 0$ implies~$q = |I \cup J| > 0$; and the assumption that $q<n$, i.e., $|I^cJ^c|>0$, and~\eqref{Eq:LemSetIdentity4} imply $2n > q + h + \overline{h}$.
		
	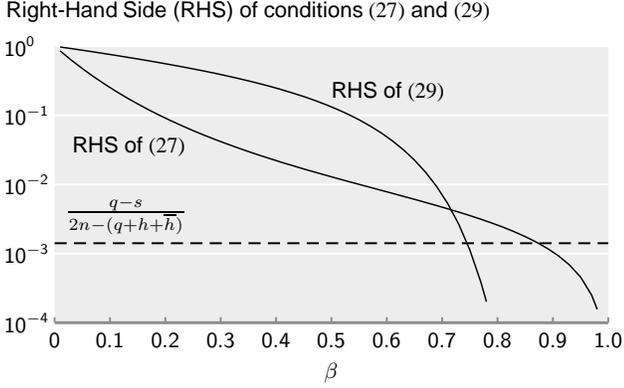
\begin{figure}[t]
	\centering
				
	\psscalebox{0.92}{
	\begin{pspicture}(8.8,4.95)

		\readdata{\CFirst}{figures/C1.dat}
		\readdata{\CSecon}{figures/C2.dat}
		\savedata{\quant}[{{0.01, -2.8795}, {1.0, -2.8795}}]
	
		\definecolor{colorXAxis}{gray}{0.55}           
		\definecolor{colorBackground}{gray}{0.93}      
		\def \distXLabels{0.15}		
		\def \distYLabels{0.12}		
		\def \xTickWidth{0.08}
		
		\def \xOrig{0.50}                            
		\def \yOrig{0.50}                            
		
		\fpSub{\yOrig}{\distXLabels}{\xPosLabels}
		\fpSub{\xOrig}{\distYLabels}{\yPosLabels}
		\fpAdd{\yOrig}{\xTickWidth}{\xTickTop}
		
		\psframe*[linecolor=colorBackground](0.5,0.5)(8.5,4.5)
									
 		\multido{\nx=\xOrig+0.80, \nB=0+0.1}{11}{	
 			\rput[t](\nx,\xPosLabels){\small $\mathsf{\nB}$}
 			\psline[linewidth=0.8pt,linecolor=colorXAxis](\nx,\yOrig)(\nx,\xTickTop)	
 		}						 
 		\multido{\ny=\yOrig+1.00, \iB=-4+1}{5}{	
 			\rput[l](-0.23,\ny){\small $\mathsf{10^{\iB}}$}
 			\psline[linecolor=white,linewidth=0.8pt](\xOrig,\ny)(8.5,\ny)
 		}						 
 		\psline[linewidth=1.2pt,linecolor=colorXAxis]{-}(\xOrig,\yOrig)(8.5,\yOrig)
											
 		\psset{xunit=8\psunit,yunit=\psunit,linewidth=0.8pt}
 		
 		\dataplot[origin={0.0625,4.5},showpoints=false,linewidth=0.6pt]{\CFirst}
 		\dataplot[origin={0.0625,4.5},showpoints=false,linewidth=0.6pt]{\CSecon} 		
 		\psset{xunit=\psunit,yunit=\psunit}
 		
 		\psline[linestyle=dashed](0.5,1.65)(8.5,1.65)
		
		\rput[lr](2.4,2.05){ $\frac{q-s}{2n - (q + h + \overline{h})}$}
		
		\rput[tr](2.39,3.20){\small {\sf RHS of \eqref{Eq:ThmL1L1Case2Condition1}}}
		\rput[lb](4.50,3.70){\small {\sf RHS of \eqref{Eq:ThmL1L1Case2Condition2}}}

 		\rput[lb](-0.2,4.85){\small \textbf{\sf Right-Hand Side (RHS) of conditions~\eqref{Eq:ThmL1L1Case2Condition1} and~\eqref{Eq:ThmL1L1Case2Condition2}}}
 		\rput[ct](4.5,-0.22){\small \textbf{\sf $\beta$}}
					
	\end{pspicture}
	}	
	\vspace{0.2cm}
	\caption{
		Values of the right-hand side (RHS) of conditions~\eqref{Eq:ThmL1L1Case2Condition1} and~\eqref{Eq:ThmL1L1Case2Condition2} from case~\text{2)} of \thref{Thm:L1L1Reconstruction}, for the example of~\fref{Fig:IntroExperimentalResults}.
	}
	\label{Fig:ConditionsThmL1L1}
	\end{figure}

	In case \text{2)}, $\beta < 1$ and we have two subcases: when condition~\eqref{Eq:ThmL1L1Case2Condition1} holds, $w(T_{f_1}(x^\star))^2$ is	bounded as in~\eqref{Eq:ThmL1L1Case2Bound1}; when condition~\eqref{Eq:ThmL1L1Case2Condition2} holds, it is bounded as in~\eqref{Eq:ThmL1L1Case2Bound2}. These subcases are not necessarily disjointed nor are they guaranteed to cover the entire interval~$0<\beta<1$.\footnote{If, for example, $n = 20$, $s = 15$, $q = 16$, $h = 10$, and~$\overline{h}=5$, neither~\eqref{Eq:ThmL1L1Case2Condition1} nor~\eqref{Eq:ThmL1L1Case2Condition2} hold for~$\beta = 0.9$. Note that $x^\star$ in this case is not ``sparse,'' i.e., $75\%$ of its entries are nonzero. In fact, increasing~$n$ to, e.g., $40$, makes~\eqref{Eq:ThmL1L1Case2Condition1} hold.} \fref{Fig:ConditionsThmL1L1} shows how conditions~\eqref{Eq:ThmL1L1Case2Condition1} and~\eqref{Eq:ThmL1L1Case2Condition2} vary with~$\beta$ for the example that was used to generate \fref{Fig:IntroExperimentalResults}. There, we had~$n = 1000$, $s=70$, $h = 11$, $\overline{h} = 11$, and~$q = 76$. The right-hand side of conditions~\eqref{Eq:ThmL1L1Case2Condition1} and~\eqref{Eq:ThmL1L1Case2Condition2} vary with~$\beta$ as shown in the figure, and the dashed line represents the left-hand side of~\eqref{Eq:ThmL1L1Case2Condition1} and~\eqref{Eq:ThmL1L1Case2Condition2}, which does not vary with~$\beta$. We can see that~\eqref{Eq:ThmL1L1Case2Condition1} holds in this case for $0<\beta \lessapprox 0.88$, and~\eqref{Eq:ThmL1L1Case2Condition2} holds for $ 0.75 \lessapprox \beta < 1$. Therefore, both conditions are valid in the interval~$0.75 < \beta < 0.88$. For instance, if~$\beta = 0.8$, the bounds in~\eqref{Eq:ThmL1L1Case2Bound1} and~\eqref{Eq:ThmL1L1Case2Bound2} give~$180$ and~$255$ (rounding up), respectively. Both values are larger than the one for~$\beta = 1$, which is given by~\eqref{Eq:ThmL1L1Case1Bound} and equal to~$135$. Indeed, the bound in~\eqref{Eq:ThmL1L1Case1Bound} is almost always smaller than the one in~\eqref{Eq:ThmL1L1Case2Bound1}: using~\eqref{Eq:LemSetIdentity5}, it can be shown that the linear, non-dominant terms in~\eqref{Eq:ThmL1L1Case1Bound} are smaller than the linear terms in~\eqref{Eq:ThmL1L1Case2Bound1} whenever
	\begin{equation}\label{Eq:ConditionForNonDominatingTermsL1L1}
		\xi < \frac{2}{5}(q + h +\overline{h})\,.
	\end{equation} 
	Furthermore, the dominant term in~\eqref{Eq:ThmL1L1Case1Bound}, namely the one involving the~$\log$, is always smaller than the dominant term in~\eqref{Eq:ThmL1L1Case2Bound1}. So, even if~\eqref{Eq:ConditionForNonDominatingTermsL1L1} does not hold, \eqref{Eq:ThmL1L1Case1Bound} is in general smaller than~\eqref{Eq:ThmL1L1Case2Bound1}. Curiously, the bound in~\eqref{Eq:ThmL1L1Case2Bound1} is minimized for~$\beta =1$ but, in that case, condition~\eqref{Eq:ThmL1L1Case2Bound1} will not hold unless~$q = s$ (according to~\eqref{Eq:LemSetIdentity6}, that would mean that~$w$ has exactly the same support as~$x^\star$). The bound in~\eqref{Eq:ThmL1L1Case2Bound2}, valid only if~$q>s$, can be much larger than both~\eqref{Eq:ThmL1L1Case1Bound} and~\eqref{Eq:ThmL1L1Case2Bound1} when~$\beta$ is close to~$1$: this is due to the term $(1+\beta)^2/(1-\beta)^2$ and to the fact that \eqref{Eq:ThmL1L1Case2Bound2} is valid only for values of~$\beta$ near~$1$ (cf.\ \eqref{Eq:ThmL1L1Case2Condition2} and \fref{Fig:ConditionsThmL1L1}). From this analysis, we conclude that the bounds given in case \text{2)} will not be sharp near~$1$. Yet, the bound for~$\beta = 1$, i.e., \eqref{Eq:ThmL1L1Case1Bound}, is the sharpest one in the theorem since, as we will see in its proof, it is the one whose derivation required the fewer number of approximations. Case~\text{3)} in the theorem is very similar to case~\text{2)}: the expression for both the conditions and the bounds are very similar. The observations made to case~\text{2)} then also apply to case \text{3)} similarly. Note, for example, that in case~\text{3b)} it is assumed~$s > h + \overline{h} > 0$. According to~\eqref{Eq:LemSetIdentity1} and~\eqref{Eq:LemSetIdentity7}, this is equivalent to saying that there is at least one index~$i$ for which~$x_i^\star \neq 0$ and~$w_i \neq x_i^\star$ and another index~$j$ for which~$x_j^\star \neq 0$ and~$w_j = x_j^\star$. The most striking fact about \thref{Thm:L1L1Reconstruction} is that its expressions depend only on the quantities given in \dref{def:CardinalityOfsomeSets}, which depend on the signs of~$x_i^\star$ and $x_i^\star - w_i$, but not on their magnitude. As we will see next, that is no longer the case for $\ell_1$-$\ell_2$ minimization.

\subsection{$\ell_1$-$\ell_2$ Minimization}
\label{SubSec:MainResultsL1L2}	
	
	Stating our results for $\ell_1$-$\ell_2$ minimization requires additional notation. Namely, we will use the following subsets of~$I^c J$:
	\begin{align}		
		\label{Eq:KEq}
		K^= &:= \Big\{i \in I^c J\,:\, |w_i| = \frac{1}{\beta}\Big\}
		\\
		K^{\neq} &:= \Big\{i \in I^c J\,:\, |w_i| > \frac{1}{\beta}\Big\}\,,
		\label{Eq:Kneq}
	\end{align}
	where we omit their dependency on~$\beta$ for notational simplicity. We will also use
	\isdraft{                                         
		\begin{equation}\label{Eq:DefvBeta}
			v_\beta := 
			\sum_{i \in I_+} (1 + \beta(x_i^\star - w_i))^2 + \sum_{i \in I_-} (1 - \beta(x_i^\star - w_i))^2 
			+ 
			\sum_{i \in K^{\neq}} (\beta |w_i| - 1)^2\,,
		\end{equation}                                  
	}{
	\begin{multline}\label{Eq:DefvBeta}
		v_\beta := 
		\sum_{i \in I_+} (1 + \beta(x_i^\star - w_i))^2 + \sum_{i \in I_-} (1 - \beta(x_i^\star - w_i))^2 
		\\
		+ 
			\sum_{i \in K^{\neq}} (\beta |w_i| - 1)^2\,,
	\end{multline}
	}
	and $\overline{w} := |w_k|$, where
	$$
		k:= \underset{i \in I^cJ}{\arg\min}\,\, \Big||w_i| - \frac{1}{\beta}\Big|\,.
	$$		
	In words, $\overline{w}$ is the absolute value of the component of~$w$ whose absolute value is closest to~$1/\beta$, in the set~$I^c J$. 
	
	\begin{Theorem}[$\ell_1$-$\ell_2$ minimization]	
	\label{Thm:L1L2Reconstruction}
		Let~$x^\star \in \mathbb{R}^n$ be the vector to reconstruct and let~$w \in \mathbb{R}^n$ be the prior information. Let~$f_2(x) = \|x\|_1 + \frac{\beta}{2}\|x - w\|_2^2$ with~$\beta > 0$, and assume~$x^\star \neq 0$ and $q < n$. Also, assume that there exists $i \in I^c$ such that $|w_i| > 1/\beta$ or that there exists~$i \in IJ$ such that $\beta \neq \text{\emph{sign}}(x_i^\star)/(w_i-x_i^\star)$.
		\begin{enumerate}
			\item If
				\begin{equation}\label{Eq:ThmL1L2Condition1}
					\frac{q-s}{n-q}
					\leq 
					|1-\beta \,\overline{w}|
					\exp\Big(\big((\beta \,\overline{w})^2 - 2\beta\,\overline{w}\big)\log\Big(\frac{n}{q}\Big)\Big)\,,
				\end{equation}
				then
				\begin{equation}\label{Eq:ThmL1L2Bound1}
					  w\bigl(T_{f_2}(x^\star)\bigr)^2
					\leq
					  2 v_\beta \log\Big(\frac{n}{q}\Big)
						+
						s
						+
						|K^{\neq}| + \frac{1}{2}|K^=|
						+
						\frac{4}{5}q\,.
				\end{equation}
				
			\item If $q > s$ and
				\begin{equation}\label{Eq:ThmL1L2Condition2}
					\frac{q-s}{n-q}
					\geq
					|1-\beta\, \overline{w}|
					\exp\Big(4\frac{(\beta\,\overline{w} -2)\beta\,\overline{w}}{|1-\beta \overline{w}|^2}
					\log\Big(\frac{q}{s}\Big)\Big)\,,
				\end{equation}
				then
				\isdraft{
					\begin{equation}\label{Eq:ThmL1L2Bound2}
					  w\bigl(T_{f_2}(x^\star)\bigr)^2
						\leq
						\frac{2v_\beta}{(1 - \beta \, \overline{w})^2} \log\Big(\frac{q}{s}\Big)
						+
						|K^{\neq}| + \frac{1}{2}|K^=|
						+
						\frac{9}{5}s\,.	
					\end{equation}
				}{
				\begin{multline}\label{Eq:ThmL1L2Bound2}
					  w\bigl(T_{f_2}(x^\star)\bigr)^2
					\leq
						\frac{2v_\beta}{(1 - \beta \, \overline{w})^2} \log\Big(\frac{q}{s}\Big)
						+
						|K^{\neq}| + \frac{1}{2}|K^=|
						\\+
						\frac{9}{5}s\,.	
				\end{multline}
				}
		\end{enumerate}
	\end{Theorem}
	
	Similarly to \thref{Thm:L1L1Reconstruction} and~\pref{Prop:ChandrasekaranBound}, this theorem upper bounds~$w(T_{f_2}(x^\star))^2$ with expressions that depend on key problem parameters, namely~$n$, $q$, $s$,  $\beta$, $v_\beta$, $\overline{w}$, $|K^{\neq}|$, and~$|K^=|$. Together with~\thref{Thm:Chandrasekaran}, it then provides a sufficient number of measurements that guarantee that~\eqref{Eq:L1L2} reconstructs~$x^\star$ with high probability. As in \thref{Thm:L1L1Reconstruction}, the previous theorem also assumes~$q<n$ or, equivalently, that~$I^cJ$ is nonempty. This makes~$\overline{w}$ well defined. It will be shown in~\lref{lem:ZeroNotInSubGradient} that the remaining assumptions are equivalent to $0 \not\in \partial f_2(x^\star)$ and, hence, that we can use \pref{Prop:Subdifferential}. It is relatively easy to satisfy one of these assumptions, namely that there exists~$i \in IJ$ such that~$\beta \neq \text{sign}(x_i^\star)/(w_i-x_i^\star)$; a sufficient condition is that there are at least two indices~$i, j$ in~$I$ such that $\text{sign}(x_i^\star)/(w_i - x_i^\star) \neq \text{sign}(x_j^\star)/(w_j - x_j^\star)$. The alternative is to set~$\beta > 1/|w_i|$ for all~$i \in I^c$. Setting large values for~$\beta$, however, will not only make the bounds in the theorem very large, but also degrade the performance of $\ell_1$-$\ell_2$ minimization significantly, as we will see in the next section. 
	
	The theorem is divided into two cases: \text{1)} if condition~\eqref{Eq:ThmL1L2Condition1} is satisfied, the bound in~\eqref{Eq:ThmL1L2Bound1} holds; \text{2)} if condition~\eqref{Eq:ThmL1L2Condition2} is satisfied, the bound in~\eqref{Eq:ThmL1L2Bound2} holds. As in $\ell_1$-$\ell_1$ minimization, the conditions~\eqref{Eq:ThmL1L2Condition1} and~\eqref{Eq:ThmL1L2Condition2} are neither necessarily disjointed nor are they guaranteed to cover all the possible values of~$\beta$ (although such a case is rare in practice). But in contrast, it is not easy to compare the bounds in the previous theorem with the one for classical CS in~\pref{Prop:ChandrasekaranBound}, or even with the ones for $\ell_1$-$\ell_1$ minimization. More specifically, $v_\beta$, defined in~\eqref{Eq:DefvBeta}, is always larger than~$s$: to see that, assume first that~$x_i^\star = w_i$ for all~$i \in I$ and observe that the first two terms in~\eqref{Eq:DefvBeta} will sum to~$s$; since the third term is nonnegative and the previous condition does not hold in general, $v_\beta$ is indeed larger than~$s$. Yet, it is not clear whether or not the dominant term of~\eqref{Eq:ThmL1L2Bound1}, i.e., $2v_\beta \log(n/q)$ is smaller than the corresponding term in~\eqref{Eq:BoundChandrasekaran}, i.e., $2s\log(n/s)$: while~$v_\beta$ is larger than~$s$, $n/q$ is smaller than~$n/s$ (since $q = |I\cup J| \geq |I| = s$). However, since $s + (4/5)q \geq s + (4/5)s = (9/5)s > (7/5)s$, the linear (non-dominant) terms in~\eqref{Eq:ThmL1L2Bound1} are always larger than the linear (non-dominant) terms in~\eqref{Eq:BoundChandrasekaran}. Similarly, it is also not clear how the bound in~\eqref{Eq:ThmL1L2Bound2} in case~\text{2)} compares with the one for classical CS: while~$q/s$ is always smaller than~$n/s$, the terms~$2v_\beta/|1-\beta \overline{w}|^2$ and~$2s$ do not compare easily. 
	
	Condition~\eqref{Eq:ThmL1L2Condition2} is likely to hold for values of~$\beta$ close to~$1/\overline{w}$. But, when that happens, the bound in~\eqref{Eq:ThmL1L2Bound2} becomes large, due to the term $2v_\beta/(1 - \beta \, \overline{w})^2$. On the other hand, condition~\eqref{Eq:ThmL1L2Condition1} is likely to hold for large values of~$\beta$. The bound in~\eqref{Eq:ThmL1L2Bound1}, however, gives large values when~$\beta$ is large, since~$v_\beta$ also becomes large; namely, a large~$\beta$ amplifies the differences between~$x_i^\star$ and~$w_i$, for all~$i \in I$. Curiously, the definition of~$v_\beta$ uses the notions of good and bad components: the first term considers~$i \in I_+$, i.e., $x_i^\star > 0$; therefore, if~$w_i$ is a bad component, i.e., $w_i<x_i^\star$, then it gets more penalized than a good component, i.e., $w_i>x_i^\star$. The same happens in the second term. Finally, note that~$v_\beta$ is the only term in~\eqref{Eq:ThmL1L2Bound1} that depends on~$\beta$. Therefore, that bound is minimized when~$v_\beta$ is minimized, which occurs for
	\begin{equation}\label{Eq:BetaThatMinimizesBound1L1L2}
		\beta^\star = \frac{1^\top w_{K^{\neq}} + 1^\top (x^\star_{I_-} -w_{I_-}) - 1^\top (x^\star_{I_+} - w_{I_+})}{\|x^\star_I - w_I\|^2 + \|w_{K^{\neq}}\|^2}\,,
	\end{equation}
	where~$z_S$ denotes the vector whose components are the components of~$z$ that are indexed by the set~$S$, and~$1$ denotes the vector of ones with appropriate dimensions. The bound in~\eqref{Eq:ThmL1L2Bound2} depends on~$\beta$ through the term $v_\beta/(1-\beta\overline{w})^2$. Although it can be minimized in closed-form, its expression is significantly more complicated than~\eqref{Eq:BetaThatMinimizesBound1L1L2}. Note that, in contrast with $\ell_1$-$\ell_1$ minimization, the~$\beta$ that minimizes these bounds depends on several unknown parameters. Given the interpretation of \fref{Fig:GoodAndBadComponents}, it is not surprising that the bounds we obtain for $\ell_1$-$\ell_2$ minimization depend on the differences between~$x^\star$ and~$w$. And, as we will see in its proof, it is exactly this fact that makes the bounds in~\eqref{Eq:ThmL1L2Bound1} and~\eqref{Eq:ThmL1L2Bound2} loose when compared with the ones for $\ell_1$-$\ell_1$ minimization.

\section{Experimental Results}
\label{Sec:ExperimentalResults}

	In this section, we describe experiments designed to assess the sharpness of our bounds for a wide range of $\beta$'s.
	
	\mypar{Experimental setup}
	The data was generated as the one in \fref{Fig:IntroExperimentalResults}, but for smaller dimensions. Namely, $x^\star$ had~$n = 500$ entries, $s=50$ of which were nonzero. The values of these entries were drawn from a zero-mean Gaussian distribution with unit variance. According to \thref{Thm:Chandrasekaran} and~\pref{Prop:ChandrasekaranBound}, this implies that standard CS requires at least~$302$ measurements for successful reconstruction. We then generated the prior information as~$w = x^\star + z$, where~$z$ was $20$-sparse, and whose support coincided with the one of~$x^\star$ in~$16$ entries and differed in~$4$ of them. The nonzero entries were zero-mean Gaussian with standard deviation~$0.8$. This yielded $h = 6$, $\overline{h} = 11$, $q = 53$, and~$l = 20$.
		
	The experiments were conducted as follows. We created a square matrix~$\overline{A} \in \mathbb{R}^{500 \times 500}$ with entries drawn independently from the standard Gaussian distribution. We then set~$\overline{y} = \overline{A}x^\star$. Next, for a fixed~$\beta$, we solved problem~\eqref{Eq:L1L1}, first by using only the first row of~$\overline{A}$ and the first entry of~$\overline{y}$. If the solution of~\eqref{Eq:L1L1}, say~$\hat{x}_1(\beta)$, did not satisfy $\|\hat{x}_1(\beta) - x^\star\|_2/\|x^\star\|_2 \leq 10^{-2}$, we proceeded by solving~\eqref{Eq:L1L1} with the first two rows of~$\overline{A}$ and the first two entries of~$\overline{y}$. This procedure was repeated until $\|\hat{x}_m(\beta) - x^\star\|_2/\|x^\star\|_2 \leq 10^{-2}$, where~$\hat{x}_m(\beta)$ denotes the solution of~\eqref{Eq:L1L1} when~$A$ (resp.\ $y$) consists of the first~$m$ rows (resp.\ entries) of~$\overline{A}$ (resp.\ $\overline{y}$). In other words, we stopped when we found the minimum number of measurements that $\ell_1$-$\ell_1$ minimization requires for successful reconstruction for a given~$\beta$, that is, $\min\,\{m\,:\, \|\hat{x}_m(\beta) - x^\star\|_2/\|x^\star\|_2 \leq 10^{-2}\}$. The values of~$\beta$ were $0.01$, $0.05$, $0.1$, $0.5$, $0.75$, $0.9$, $1$, $2.5$, $5$, $10$, $50$, and~$100$. We then repeated the entire procedure for~$4$ other randomly generated pairs $(\overline{A},\overline{y})$.
	
	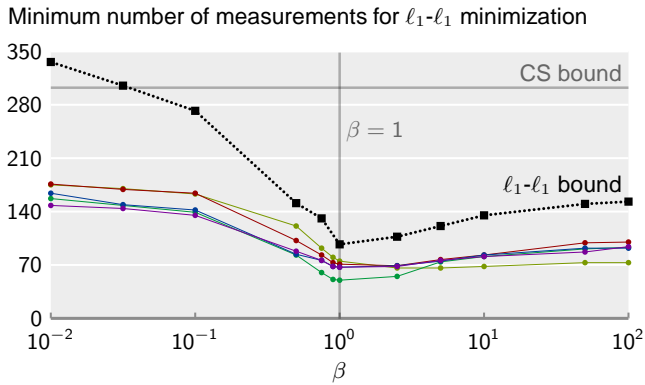
\begin{figure}[t]
	\centering
			
	\readdata{\betaA}{figures/BetaResults1.dat}
	\readdata{\betaB}{figures/BetaResults2.dat}
	\readdata{\betaC}{figures/BetaResults3.dat}
	\readdata{\betaD}{figures/BetaResults4.dat}
	\readdata{\betaE}{figures/BetaResults5.dat}
	\readdata{\bound}{figures/boundsL1L1.dat}
	
	\psscalebox{0.96}{
	\begin{pspicture}(8.8,5.2)
					
		\def\xMax{2}                                
		\def\xMin{-2}                               
		\def\xNumTicks{4}                           
		\def\yMax{350}                              
		\def\yMin{0}                                
		\def\yNumTicks{5}                           
		\def\xIncrement{1}                          
		\def\yIncrement{70}                         
					
		\def\xOrig{0.50}                              
		\def\yOrig{0.80}                              
		\def\SizeX{8.00}                              
		\def\SizeY{3.70}                              
		\def\xTickIncr{2.00}                          
		\def\yTickIncr{0.74}                          

		\definecolor{colorXAxis}{gray}{0.55}           
		\definecolor{colorBackground}{gray}{0.93}      

		\def \distXLabels{0.15}

		\def \distYLabels{0.12}

		\def \xTickWidth{0.08}

		\fpAdd{\xNumTicks}{1}{\xNumTicksPOne}         
		\fpAdd{\yNumTicks}{1}{\yNumTicksPOne}

		\FPadd \xEndPoint \xOrig \SizeX                
		\FPadd \yEndPoint \yOrig \SizeY                

		\FPset \unit 1

		\FPsub \xRange \xMax \xMin			
		\FPdiv \xScale \SizeX \xRange 
		\FPdiv \xMultByOrigin \unit \xScale
		\FPmul \xDataOrig  \xMultByOrigin \xOrig

		\FPsub \yRange \yMax \yMin
		\FPdiv \yScale \SizeY \yRange 
		\FPdiv \yMultByOrigin \unit \yScale
		\FPmul \yDataOrig  \yMultByOrigin \yOrig

		\fpSub{\yOrig}{\distXLabels}{\xPosLabels}
		\fpSub{\xOrig}{\distYLabels}{\yPosLabels}

		\fpAdd{\yOrig}{\xTickWidth}{\xTickTop}

		\psframe*[linecolor=colorBackground](\xOrig,\yOrig)(\xEndPoint,\yEndPoint)
									
		\multido{\nx=\xOrig+\xTickIncr, \iB=\xMin+\xIncrement}{\xNumTicksPOne}{	
			\rput[t](\nx,\xPosLabels){\small $\mathsf{10^{\iB}}$}
			\psline[linewidth=0.8pt,linecolor=colorXAxis](\nx,\yOrig)(\nx,\xTickTop)	
		}						

		\multido{\ny=\yOrig+\yTickIncr, \nB=\yMin+\yIncrement}{\yNumTicksPOne}{	
			\rput[r](\yPosLabels,\ny){\small $\mathsf{\nB}$}
			\psline[linecolor=white,linewidth=0.8pt](\xOrig,\ny)(\xEndPoint,\ny)
		}						

		\psline[linewidth=1.2pt,linecolor=colorXAxis]{-}(\xOrig,\yOrig)(\xEndPoint,\yOrig)
																	
		\definecolor{c1}{RGB}{153,0,0}
		\definecolor{c2}{RGB}{122,153,0}
		\definecolor{c3}{RGB}{0,153,61}
		\definecolor{c4}{RGB}{0,61,153}
		\definecolor{c5}{RGB}{122,0,153}
				
		\psset{xunit=\xScale\psunit,yunit=\yScale\psunit,linewidth=0.4pt,dotsize=2.2pt}		
		\psline[linewidth=1.0pt,linecolor=black!90!white,strokeopacity=0.31]{-}(2.25,75)(2.25,424)
		\psline[linewidth=1.0pt,linecolor=black!90!white,strokeopacity=0.31]{-}(0.25,378)(4.25,378)
		\dataplot[xlogBase=10,origin={\xDataOrig,\yDataOrig},showpoints=true,linecolor=c2]{\betaB}
		\dataplot[xlogBase=10,origin={\xDataOrig,\yDataOrig},showpoints=true,linecolor=c1]{\betaA}
		\dataplot[xlogBase=10,origin={\xDataOrig,\yDataOrig},showpoints=true,linecolor=c3]{\betaC}
		\dataplot[xlogBase=10,origin={\xDataOrig,\yDataOrig},showpoints=true,linecolor=c4]{\betaD}
		\dataplot[xlogBase=10,origin={\xDataOrig,\yDataOrig},showpoints=true,linecolor=c5]{\betaE}
		\dataplot[xlogBase=10,origin={\xDataOrig,\yDataOrig},showpoints=true,linecolor=black,linestyle=dotted,dotsep=0.7pt,dotstyle=square*,linewidth=1.2pt,dotsize=3.6pt]{\bound}
		\psset{xunit=\psunit,yunit=\psunit}
		
		\rput[lb](-0.09,4.85){\small \textbf{\sf Minimum number of measurements for $\ell_1$-$\ell_1$ minimization}}
		\rput[ct](4.5,0.05){\small \textbf{\sf $\mathsf{\beta}$}}
		
		\rput[lb](4.59,3.25){\small \textbf{\sf \color{black!60!white}{$\mathsf{\beta = 1}$}}}
		\rput[rb](8.40,4.08){\small \textbf{\sf \color{black!60!white}{CS bound}}}
		\rput[rb](8.4,2.52){\small \textbf{\sf $\ell_1$-$\ell_1$ bound}}
		
	\end{pspicture}
	}
	\vspace{-0.1cm}
	\caption{
		Experimental performance of $\ell_1$-$\ell_1$ minimization for~$5$ different Gaussian matrices as a function of~$\beta$ (solid lines). The dotted line depicts the bounds given in \thref{Thm:L1L1Reconstruction}, which are minimized for~$\beta = 1$ (vertical line). The horizontal line indicates the bound given by~\eqref{Eq:BoundChandrasekaran} for classical CS. 		
	}
	\label{Fig:BetasL1L1}
	\end{figure}

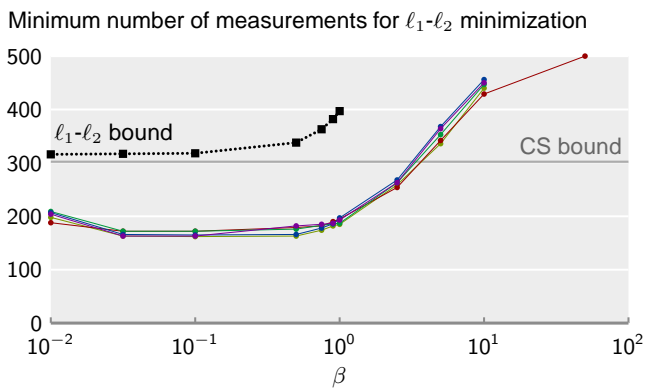
\begin{figure}[t]
	\centering
			
	\readdata{\betaA}{figures/BetaResultsL1L2-1.dat}
	\readdata{\betaB}{figures/BetaResultsL1L2-2.dat}
	\readdata{\betaC}{figures/BetaResultsL1L2-3.dat}
	\readdata{\betaD}{figures/BetaResultsL1L2-4.dat}
	\readdata{\betaE}{figures/BetaResultsL1L2-5.dat}
	\readdata{\bound}{figures/boundsL1L2.dat}
	
	\psscalebox{0.96}{
	\begin{pspicture}(8.8,5.2)
					
		\def\xMax{2}                                
		\def\xMin{-2}                               
		\def\xNumTicks{4}                           
		\def\yMax{500}                              
		\def\yMin{0}                                
		\def\yNumTicks{5}                           
		\def\xIncrement{1}                          
		\def\yIncrement{100}                         
					
		\def\xOrig{0.50}                              
		\def\yOrig{0.80}                              
		\def\SizeX{8.00}                              
		\def\SizeY{3.70}                              
		\def\xTickIncr{2.00}                          
		\def\yTickIncr{0.74}                          

		\definecolor{colorXAxis}{gray}{0.55}           
		\definecolor{colorBackground}{gray}{0.93}      

		\def \distXLabels{0.15}

		\def \distYLabels{0.12}

		\def \xTickWidth{0.08}

		\fpAdd{\xNumTicks}{1}{\xNumTicksPOne}         
		\fpAdd{\yNumTicks}{1}{\yNumTicksPOne}

		\FPadd \xEndPoint \xOrig \SizeX                
		\FPadd \yEndPoint \yOrig \SizeY                

		\FPset \unit 1

		\FPsub \xRange \xMax \xMin			
		\FPdiv \xScale \SizeX \xRange 
		\FPdiv \xMultByOrigin \unit \xScale
		\FPmul \xDataOrig  \xMultByOrigin \xOrig

		\FPsub \yRange \yMax \yMin
		\FPdiv \yScale \SizeY \yRange 
		\FPdiv \yMultByOrigin \unit \yScale
		\FPmul \yDataOrig  \yMultByOrigin \yOrig

		\fpSub{\yOrig}{\distXLabels}{\xPosLabels}
		\fpSub{\xOrig}{\distYLabels}{\yPosLabels}

		\fpAdd{\yOrig}{\xTickWidth}{\xTickTop}

		\psframe*[linecolor=colorBackground](\xOrig,\yOrig)(\xEndPoint,\yEndPoint)
									
		\multido{\nx=\xOrig+\xTickIncr, \iB=\xMin+\xIncrement}{\xNumTicksPOne}{	
			\rput[t](\nx,\xPosLabels){\small $\mathsf{10^{\iB}}$}
			\psline[linewidth=0.8pt,linecolor=colorXAxis](\nx,\yOrig)(\nx,\xTickTop)	
		}						

		\multido{\ny=\yOrig+\yTickIncr, \nB=\yMin+\yIncrement}{\yNumTicksPOne}{	
			\rput[r](\yPosLabels,\ny){\small $\mathsf{\nB}$}
			\psline[linecolor=white,linewidth=0.8pt](\xOrig,\ny)(\xEndPoint,\ny)
		}						

		\psline[linewidth=1.2pt,linecolor=colorXAxis]{-}(\xOrig,\yOrig)(\xEndPoint,\yOrig)

		\definecolor{c1}{RGB}{153,0,0}
		\definecolor{c2}{RGB}{122,153,0}
		\definecolor{c3}{RGB}{0,153,61}
		\definecolor{c4}{RGB}{0,61,153}
		\definecolor{c5}{RGB}{122,0,153}
				
		\psset{xunit=\xScale\psunit,yunit=\yScale\psunit,linewidth=0.4pt,dotsize=2.2pt}		
		\psline[linewidth=1.0pt,linecolor=black!90!white,strokeopacity=0.31]{-}(0.25,410)(4.25,410)
		\dataplot[xlogBase=10,origin={\xDataOrig,\yDataOrig},showpoints=true,linecolor=c2]{\betaB}
		\dataplot[xlogBase=10,origin={\xDataOrig,\yDataOrig},showpoints=true,linecolor=c1]{\betaA}
		\dataplot[xlogBase=10,origin={\xDataOrig,\yDataOrig},showpoints=true,linecolor=c3]{\betaC}
		\dataplot[xlogBase=10,origin={\xDataOrig,\yDataOrig},showpoints=true,linecolor=c4]{\betaD}
		\dataplot[xlogBase=10,origin={\xDataOrig,\yDataOrig},showpoints=true,linecolor=c5]{\betaE}
		\dataplot[xlogBase=10,origin={\xDataOrig,\yDataOrig},showpoints=true,linecolor=black,linestyle=dotted,dotsep=0.7pt,dotstyle=square*,linewidth=1.2pt,dotsize=3.6pt]{\bound}
		\psset{xunit=\psunit,yunit=\psunit}

		\rput[lb](-0.09,4.85){\small \textbf{\sf Minimum number of measurements for $\ell_1$-$\ell_2$ minimization}}
		\rput[ct](4.5,0.05){\small \textbf{\sf $\mathsf{\beta}$}}
		\rput[rb](8.40,3.13){\small \textbf{\sf \color{black!60!white}{CS bound}}}
		\rput[lb](0.55,3.3){\small \textbf{\sf $\ell_1$-$\ell_2$ bound}}
					
	\end{pspicture}
	}
	\vspace{-0.1cm}
	\caption{
		Same as \fref{Fig:BetasL1L1}, but for $\ell_1$-$\ell_2$ minimization. The data is the same as in \fref{Fig:BetasL1L1}, but the vertical scales are different. For~$\beta>1$, the bounds given by \thref{Thm:L1L2Reconstruction} are larger than $500$ and, hence, are not shown.
	}
	\label{Fig:BetasL1L2}
	\end{figure}

	\mypar{Results for $\ell_1$-$\ell_1$ minimization}
	\fref{Fig:BetasL1L1} shows the results of these experiments. It displays the minimum number of measurements for successful reconstruction, i.e., $\min\,\{m\,:\, \|\hat{x}_m(\beta) - x^\star\|_2/\|x^\star\|_2 \leq 10^{-2}\}$, versus~$\beta$. The~$5$ solid lines give the experimental performance of~\eqref{Eq:L1L1} for the~$5$ different pairs of~$(\overline{A},\overline{y})$. The dotted line displays the bounds given by \thref{Thm:L1L1Reconstruction}. When~$\beta \neq 1$, the subcases of cases~\text{2)} and~\text{3)} of that theorem may give two different bounds; in those cases, we always selected the smallest one. For reference, we use a vertical line to mark the value that minimizes the bounds in \thref{Thm:L1L1Reconstruction}: $\beta = 1$. The horizontal line marks the~$302$ measurements that classical CS requires. We point out that we removed the bound for $\beta = 0.9$, since it was~$576$, a value larger than signal dimensionality, $500$. As we had seen before, values of~$\beta$ close to~$1$ yield large bounds in \thref{Thm:L1L1Reconstruction}. We had also stated that the bound for~$\beta = 1$ is not only the sharpest one in that theorem, but also the smallest one. \fref{Fig:BetasL1L1} also shows that setting~$\beta$ to~$1$ leads to a performance in practice close to the optimal one. Indeed, three out of the five solid curves in the figure achieved their minimum at~$\beta = 1$; the remaining ones achieved it at~$\beta = 2.5$. We can also observe that the bound for~$\beta = 1$ is quite sharp: its value is~$97$, and the maximum among all of the solid lines for $\beta = 1$ was~$75$ measurements. The figure also shows that the bounds are looser for~$\beta < 1$ and, eventually, become larger than the bound for standard CS. For~$\beta > 1$, the bound is relatively sharp. Regarding the experimental performance of $\ell_1$-$\ell_1$ minimization, it degrades when~$\beta$ is small, towards standard CS, and achieves its minimum at around~$\beta =1$. Then, it degrades as~$\beta$ grows, but never performing worse than for~$\beta$ close to~$0.01$.

	\mypar{Results for $\ell_1$-$\ell_2$ minimization}
	\fref{Fig:BetasL1L2} shows the same experiments, with the same data, but for $\ell_1$-$\ell_2$ minimization. Notice the scale of the vertical axis is different from the one in \fref{Fig:BetasL1L1}. We do not show the bounds for~$\beta > 1$, because they were larger than~$500$ (e.g., the bound for~$\beta = 2.5$ was~$820$). The minimum value of the bound was~$315$ ($\beta = 0.01$), which is slightly larger than the bound for standard CS. In fact, for this example, the bounds given by \thref{Thm:L1L2Reconstruction} were always larger than the one for standard CS.\footnote{That is not always the case in practice: in general, the bounds in \thref{Thm:L1L2Reconstruction} can be smaller than the one for standard CS when~$x^\star$ is very sparse.} The experimental performance curves behaved differently from the ones for $\ell_1$-$\ell_1$ minimization: from~$\beta = 0.01$ to~$\beta = 0.05$, they decreased slightly and remained approximately constant until~$\beta = 1$. After that point, their performance degraded sharply. For instance, for~$\beta = 50$, \eqref{Eq:L1L2} was able to reconstruct~$x^\star$ for one pair~$(\overline{A},\overline{y})$ only; and this required using the full matrix~$\overline{A}$. In conclusion, although prior information helped (slightly) for~$\beta$ between~$0.01$ and~$1$, the bounds of \thref{Thm:L1L2Reconstruction} were not sharp. An interesting fact can be observed by comparing Figs.~\ref{Fig:BetasL1L1} and~\ref{Fig:BetasL1L2}: for all~$\beta > 0.01$, the maximum of the minimum number of measurements that $\ell_1$-$\ell_1$ minimization required, namely~$170$, was smaller than the minimum of the minimum number of measurements that $\ell_1$-$\ell_2$ minimization required, namely~$172$.

\section{Proof of Main Results}	
\label{Sec:Proofs}
	
	In this section we present the proofs of Theorems~\ref{Thm:L1L1Reconstruction} and~\ref{Thm:L1L2Reconstruction}. This is the content of Subsections~\ref{SubSec:ProofsL1L1} and~\ref{SubSec:ProofsL1L2}, respectively. Before presenting those proofs, we state some auxiliary results. 
	
	\subsection{Auxiliary Results}
	The following lemma will play an important role in our proofs. It upper bounds the expected squared distance of a scalar Gaussian random variable to an interval in~$\mathbb{R}$. Recall that the probability density function of a scalar Gaussian random variable with zero-mean and unit variance is given by
	\begin{equation}\label{Eq:AuxLemVarPhi}
		\varphi(x) := \frac{1}{\sqrt{2\pi}}\exp\Big(-\frac{x^2}{2}\Big)\,.
	\end{equation}
	We denote an interval in~$\mathbb{R}$ with
	\begin{equation}\label{Eq:AuxDefinitionIntervalR}
		\mathcal{I}(a,b) := \big\{x \in \mathbb{R}\,:\, |x - a| \leq b\big\} = \big[a-b,\, a+b\big]\,.
	\end{equation}
	
	\begin{Lemma}
	\label{lem:AuxLemmaIntervals}	
		Let~$g \sim \mathcal{N}(0,1)$ be a scalar, zero-mean Gaussian random variable with unit variance. Let~$a,b \in \mathbb{R}$ and~$b\geq 0$.
		\begin{enumerate}
			\item If $b = 0$, then $\mathcal{I}(a,b) = \{a\}$ and
				\begin{equation}\label{Eq:LemAuxDistanceToAPoint}
					\mathbb{E}_g\bigl[\text{\emph{dist}}(g,a)^2\bigr] = a^2 + 1\,.
				\end{equation}
		
			\item If~$b > 0$ and $|a| < b$, i.e., $0 \in \mathcal{I}(a,b)$, then
				\begin{equation}\label{Eq:LemAuxDistanceToIntervalZeroInside}
					\mathbb{E}_g\Bigl[\text{\emph{dist}}\bigl(g,\mathcal{I}(a,b)\bigr)^2\Bigr] 
					\leq 
					\frac{\varphi(b-a)}{b-a}				
					+
					\frac{\varphi(a+b)}{a+b}
					\,.
				\end{equation}							
				
			\item If~$b > 0$ and $a + b < 0$, then
				\begin{equation}\label{Eq:LemAuxDistanceToIntervalLeftOfZero}
					\mathbb{E}_g\Bigl[\text{\emph{dist}}\bigl(g,\mathcal{I}(a,b)\bigr)^2\Bigr]
					\leq 									
					1 + (a+b)^2	
					+ \frac{\varphi(b-a)}{b-a}
					\,.									
				\end{equation}
				
			\item If~$b > 0$ and $a - b > 0$, then
				\begin{equation}\label{Eq:LemAuxDistanceToIntervalRightOfZero}
					\mathbb{E}_g\Bigl[\text{\emph{dist}}\bigl(g,\mathcal{I}(a,b)\bigr)^2\Bigr]
					\leq 									
					1 + (a-b)^2
					+ \frac{\varphi(a+b)}{a+b}
					\,.									
				\end{equation}
			
			\item If $b > 0$ and $a + b = 0$, then
				\begin{equation}\label{Eq:LemAuxDistanceToIntervalTouchOnTheRight}
					\mathbb{E}_g\Bigl[\text{\emph{dist}}\bigl(g,\mathcal{I}(a,b)\bigr)^2\Bigr] 
					\leq										
					\frac{\varphi(b-a)}{b-a}					
					+\frac{1}{2}\,.
				\end{equation}
				
			\item If $b > 0$ and $a - b = 0$, then
				\begin{equation}\label{Eq:LemAuxDistanceToIntervalTouchOnTheLeft}
					\mathbb{E}_g\Bigl[\text{\emph{dist}}\bigl(g,\mathcal{I}(a,b)\bigr)^2\Bigr] 
					\leq					
					\frac{\varphi(a+b)}{a+b}					
					+\frac{1}{2}\,.
				\end{equation}				
		\end{enumerate}
	\end{Lemma}
	The proof can be found in~\aref{App:ProofLemmaIntervals}.	Each case in the lemma considers a different relative position between the interval~$\mathcal{I}(a,b)$ and zero, which is the mean of the random variable~$g$. In case \text{1)}, the interval is simply a point. In case~\text{2)}, $\mathcal{I}(a,b)$ contains zero. In cases \text{3)} and \text{4)}, $\mathcal{I}(a,b)$ does not contain zero. And, finally, in cases \text{5)} and \text{6)}, zero is one of the endpoints of~$\mathcal{I}(a,b)$. Notice that addressing cases \text{5)} and \text{6)} separately from cases \text{4)} and \text{5)} leads to sharper bounds on the former: for example, making~$a+b \xrightarrow{} 0$ in the right-hand side of~\eqref{Eq:LemAuxDistanceToIntervalLeftOfZero} gives $1 + \varphi(b-a)/(b-a)$, which is larger than the right-hand side of~\eqref{Eq:LemAuxDistanceToIntervalTouchOnTheRight}. We note that the proof of~\pref{Thm:Chandrasekaran} in~\cite{Chandrasekaran12-ConvexGeometryLinearInverseProblems} for standard CS uses the bound~\eqref{Eq:LemAuxDistanceToIntervalZeroInside} with~$a=0$. The following result will be used frequently. 
	\begin{Lemma}\label{lem:AuxLemmaBoundsFunctions}
		There holds
		\begin{equation}\label{Eq:BoundChandrasekaran}
			\frac{1 - \frac{1}{x}}{\sqrt{\pi \log\,x}} \leq \frac{1}{\sqrt{2\pi}}\leq \frac{2}{5}\,,
		\end{equation}
		for all~$x>1$.
	\end{Lemma}
	The proof can be found in~\aref{App:ProofLemmaBounds}. Recall the definitions of functions~$f_1$ and~$f_2$:
	\begin{align}
		f_1(x) &:= \|x\|_1 + \beta \|x - w\|_1
		\label{Eq:AuxDefF1}
		\\
		f_2(x) &:= \|x\|_1 + \frac{\beta}{2} \|x - w\|_2^2\,.
		\label{Eq:AuxDefF2}
	\end{align}	
	To apply \pref{Prop:Subdifferential} to these functions, i.e., to say that their normal cones at a given~$x^\star$ is equal to the cone generated by their subdifferentials at~$x^\star$, we need to guarantee that their subdifferentials do not contain the zero vector: $0 \not\in \partial f_j(x^\star)$, $j=1,2$. The next two lemmas give a characterization of this condition in terms of the problem parameters in \dref{def:CardinalityOfsomeSets}. Before that, let us compute~$\partial f_1 (x^\star)$ and~$\partial f_2 (x^\star)$. A key property of functions~$f_1$ and~$f_2$, and on which our results deeply rely, is that they admit a component-wise decomposition:
	$$
		f_1(x) = \sum_{i=1}^n f_1^{(i)}(x_i)
		\qquad
		\quad
		\,\,
		f_2(x) = \sum_{i=1}^n f_2^{(i)}(x_i)\,,
	$$
	where $f_1^{(i)} = |x_i| + \beta |x_i - w_i|$ and $f_2^{(i)} = |x_i| + \frac{\beta}{2} (x_i - w_i)^2$. Therefore, 
	\begin{align*}
		\partial f_1(x^\star) &= \Big( \partial f_1^{(1)}(x_1^\star), \partial f_1^{(2)}(x_2^\star), \ldots, \partial f_1^{(n)}(x_n^\star) \Big)
		\\
		\partial f_2(x^\star) &= \Big( \partial f_2^{(1)}(x_1^\star), \partial f_2^{(2)}(x_2^\star), \ldots, \partial f_2^{(n)}(x_n^\star) \Big)\,.
	\end{align*}
	Recall that $\partial|s| = \text{sign}(s)$ for~$s\neq 0$, and $\partial|s| = [-1,1]$ for~$s = 0$. The function~$\text{sign}(\cdot)$ returns the sign of a number, i.e., $\text{sign}(a) = 1$ if~$a> 0$, and~$\text{sign}(a) = -1$ if~$a< 0$. We then have
	\begin{equation}\label{Eq:SubgradientL1L1PerComp}
		\partial f_1^{(i)}(x_i^\star)
		=
		\left\{
			\begin{array}{ll}
				\text{sign}(x_i^\star) +\beta\, \text{sign}(x_i^\star - w_i) &,\,\, i \in IJ \vspace{0.1cm}\\
				\text{sign}(x_i^\star) + [-\beta,\beta] &,\,\, i \in IJ^c \vspace{0.1cm}\\
				\beta\,\text{sign}(x_i^\star - w_i) + [-1,1] &,\,\, i \in I^c J \vspace{0.1cm}\\
				\bigl[-\beta-1,\beta+1\bigr] &,\,\, i \in I^c J^c
			\end{array}
		\right.		
	\end{equation}
	and
	\begin{equation}\label{Eq:SubgradientL1L2PerComp}
		\partial f_2^{(i)}(x_i^\star)
		=
		\left\{
			\begin{array}{ll}
				\text{sign}(x_i^\star) +\beta(x_i^\star - w_i) &,\,\, i \in I \vspace{0.1cm}\\				
				\bigl[-1,1\bigr] - \beta w_i &,\,\, i \in I^c\,,
			\end{array}
		\right.
	\end{equation}
	for~$i = 1,\ldots,n$.
	
	\begin{Lemma}\label{lem:ZeroNotInSubGradient}
		Assume~$x^\star \neq 0$ or, equivalently, that~$I \neq \emptyset$. Assume also~$w \neq x^\star$ or, equivalently, that $J \neq \emptyset$. Consider~$f_1$ and~$f_2$ in~\eqref{Eq:AuxDefF1} and~\eqref{Eq:AuxDefF2}, respectively.
		\begin{enumerate}
			\item 
				$0 \not\in \partial f_1(x^\star)$ if and only if $\overline{h} > 0$ or $\beta \neq 1$.				
				
			\item 
				$0 \not\in \partial f_2(x^\star)$ if and only if there is $i \in IJ$ such that $\beta \neq \text{\emph{sign}}(x^\star_i)/(w_i - x_i^\star)$ or there is $i \in I^c$ such that $\beta > 1/|w_i|$.			
		\end{enumerate}
	\end{Lemma}
	The proof is in \aref{App:ProofLemmaZeroNotInSubGradient}.

	\subsection{Proof of Theorem~\ref{Thm:L1L1Reconstruction}}
	\label{SubSec:ProofsL1L1}
	
	\pref{Prop:GaussianWidthDistancePolarCone} establishes that~$w(C) = \mathbb{E}_g\bigl[\text{dist}(g,C^o)\bigr]$, for a cone~$C$ and its polar cone~$C^o$, where~$g \sim \mathcal{N}(0,I)$. Using Jensen's inequality~\cite[Thm.\ B.1.1.8]{Lemarechal04-FundamentalsConvexAnalysis}, $w(C)^2 \leq \mathbb{E}_g\bigl[\text{dist}(g,C^o)^2\bigr]$. The polar cone of the tangent cone~$T_{f_1}(x^\star)$ is the normal cone~$N_{f_1}(x^\star)$ which, according to \pref{Prop:Subdifferential}, coincides with the cone generated by the subdifferential~$\partial f_1(x^\star)$ whenever~$0 \not\in \partial f_1(x^\star)$. In other words, if~$0 \not\in \partial f_1(x^\star)$, then
	\begin{equation}\label{Eq:ProofL1L1FirstStep}
		w\bigl(T_{f_1}(x^\star)\bigr)^2 \leq \mathbb{E}_g\Bigl[\text{dist}\big(g,\text{cone}\,\partial f_1(x^\star)\big)^2\Bigr]\,.
	\end{equation} 
	Part \text{1)} of \lref{lem:ZeroNotInSubGradient} establishes that~$0 \not\in \partial f_1(x^\star)$ is equivalent to $\beta \neq 1$ or~$\overline{h} > 0$. So, provided we assume that~$\overline{h} > 0$ for part \text{1)} of the theorem, we can always use~\eqref{Eq:ProofL1L1FirstStep}. The proof is organized as follows. First, we compute a generic upper bound on~\eqref{Eq:ProofL1L1FirstStep}, using the several cases of \lref{lem:AuxLemmaIntervals}. This will give us three bounds, each one for a specific case of the theorem, i.e., $\beta = 1$, $\beta <1$, and~$\beta > 1$. These bounds, however, will be uninformative since they depend on unknown quantities and on a free variable. We then address each case separately, selecting a specific value for the free variable and ``getting rid'' of the unknown quantities. In this last step, we will use the bound in \lref{lem:AuxLemmaBoundsFunctions} frequently.
	
	\subsubsection{Generic Bound}
	
	A vector~$d \in \mathbb{R}^n$ belongs to the cone generated by~$\partial f_1(x^\star)$ if~$d = t y$ for some~$t \geq 0$ and some $y \in \partial f_1(x^\star)$. According to~\eqref{Eq:SubgradientL1L1PerComp}, each component~$d_i$ satisfies
	$$
		\left\{
			\begin{array}{ll}
				d_i = t\,\text{sign}(x_i^\star) + t\beta\,\text{sign}(x_i^\star - w_i) &,\,\, \text{if $i \in IJ$} 
				\vspace{0.1cm}\\
				|d_i - t\,\text{sign}(x_i^\star)| \leq t\beta &,\,\,\text{if $i \in I J^c$} 
				\vspace{0.1cm}\\
				|d_i - t\beta\,\text{sign}(x_i^\star - w_i)| \leq t  &,\,\, \text{if $i \in I^c J$}
				\vspace{0.1cm}\\
				|d_i| \leq t(\beta+1) &,\,\, \text{if $i \in I^c J^c$,}
			\end{array}
		\right.
	$$
	for some~$t\geq 0$.	Thus, the right-hand side of~\eqref{Eq:ProofL1L1FirstStep} is written as
	\begin{align*}
		\isdraft{}{&}
		  \mathbb{E}_g\Bigl[\text{dist}\big(g,\text{cone}\,\partial f_1(x^\star)\big)^2\Bigr]		  		  
		\isdraft{}{\\}
		&=
			\mathbb{E}_g\left[\,	
			\underset{t\geq 0}{\min}
			\Biggl\{
				\sum_{i \in I J}
					\text{dist}\Bigl(g_i\,,\, t\,\text{sign}(x_i^\star) + t\beta\,\text{sign}(x_i^\star - w_i)\Bigr)^2
			\right.
		\isdraft{}{\\
		&\quad\qquad}		
			+					
			\sum_{i \in I J^c}
				\text{dist}\Big(g_i\,,\, \mathcal{I}\big(t\,\text{sign}(x_i^\star), t\beta\big)\Big)^2
		\\
		&\isdraft{}{\quad\qquad}
			+					
			\sum_{i \in I^c J}
				\text{dist}\Big(g_i\,,\, \mathcal{I}\big(t\beta\,\text{sign}(x_i^\star - w_i), t\big)\Big)^2
		\isdraft{}{
		\\
		&\quad\qquad			
		}
			+				
			\sum_{i \in I^c J^c}
				\text{dist}\Big(g_i\,,\, \mathcal{I}\big(0, t(\beta+1)\big)\Big)^2
		\Biggr\}
		\Biggr]\,.
	\end{align*}	
	As in the proof of \pref{Prop:ChandrasekaranBound} (in~\cite{Chandrasekaran12-ConvexGeometryLinearInverseProblems}), we fix~$t$ now and select a particular value for it later. Our choice for~$t$ will not necessarily be optimal, but it will give bounds that can be expressed as a function of the parameters in \dref{def:CardinalityOfsomeSets}. In other words, if~$h$ is a function of~$t$ and~$g$, we have
	\begin{equation}\label{Eq:ProofL1L1RelationBetweenExpressionsFixingt}
		\mathbb{E}_g\Bigl[\underset{t}{\min} \,\,h(g,t)\Bigr]
		\leq
		\underset{t}{\min} \,\,\mathbb{E}_g\bigl[h(g,t)\bigr]
		\leq
		\mathbb{E}_g\bigl[h(g,t)\bigr]\,, \forall_t\,.
	\end{equation}
	The value we will select for~$t$ does not necessarily minimize the second term in~\eqref{Eq:ProofL1L1RelationBetweenExpressionsFixingt}, but allows deriving useful bounds.
	For a fixed~$t$, we then have:
	\begin{subequations}\label{Eq:ProofL1L1NormalConeDecompositionABCD}
	\begin{align}		
		&
		  \mathbb{E}_g\Bigl[\text{dist}\big(g,\text{cone}\,\partial f_1(x^\star)\big)^2\Bigr] 
		\notag
		\\	
		&\leq
		  \sum_{i \in I J}
 					\mathbb{E}_{g_i}\biggl[
 					\text{dist}\Bigl(g_i\,,\, t\,\text{sign}(x_i^\star) + t\beta\,\text{sign}(x_i^\star - w_i)\Bigr)^2
 					\biggr]
 		\label{NormalConeA}
 		\\
 		&\quad
			+					
 			\sum_{i \in I J^c}
 				\mathbb{E}_{g_i}\biggl[
 				\text{dist}\Big(g_i\,,\, \mathcal{I}\big(t\,\text{sign}(x_i^\star), t\beta\big)\Big)^2
 				\biggr]
    \label{NormalConeB}
 		\\
 		&\quad
			+
				\sum_{i \in I^c J}
					\mathbb{E}_{g_i}\biggl[
					\text{dist}\Big(g_i\,,\, \mathcal{I}\big(t \beta\,\text{sign}(x_i^\star - w_i), t \big)\Big)^2
					\biggr]
		\label{NormalConeC}
 		\\
 		&\quad
			+
			\sum_{i \in I^c J^c }
				\mathbb{E}_{g_i}\biggl[
				\text{dist}\Big(g_i\,,\, \mathcal{I}\big(0, t(\beta + 1)\big)\Big)^2
				\biggr].
		\label{NormalConeD}		
	\end{align}
	\end{subequations}
	Next, we use~\lref{lem:AuxLemmaIntervals} to compute~\eqref{NormalConeA} in closed-form and to upper bound~\eqref{NormalConeB}, \eqref{NormalConeC}, and~\eqref{NormalConeD}.

	\mypar{Expression for~\eqref{NormalConeA}} By partitioning the set~$IJ$ into~$I_+J_+ \cup I_-J_- \cup I_-J_+ \cup I_+J_-$, we obtain
	\begin{align*}
		  \eqref{NormalConeA}
		&=
			\sum_{i \in I_+J_+} 
			\mathbb{E}_{g_i}\biggl[
					\text{dist}\Bigl(g_i\,,\, t(\beta + 1)\Bigr)^2
 					\biggr]
 		\isdraft{}{
 		\\
 		&\quad}
			+
 			\sum_{i \in I_-J_-} 
			\mathbb{E}_{g_i}\biggl[
 					\text{dist}\Bigl(g_i\,,\,-t(\beta+1)\Bigr)^2
 					\biggr]
 		\isdraft{}{
 		\\
 		&\quad}
			+
			\sum_{i \in I_-J_+} 
			\mathbb{E}_{g_i}\biggl[
 					\text{dist}\Bigl(g_i\,,\, t(\beta-1)\Bigr)^2
 					\biggr] 		
 		\\
 		&\quad
			+
			\sum_{i \in I_+J_-} 
			\mathbb{E}_{g_i}\biggl[
 					\text{dist}\Bigl(g_i\,,\, t(1-\beta)\Bigr)^2
 					\biggr].
 	\end{align*}
 	And using~\eqref{Eq:LemAuxDistanceToAPoint} in~\lref{lem:AuxLemmaIntervals} and~$h$ and~$\overline{h}$ in~\dref{def:CardinalityOfsomeSets},
 	\begin{align}
			\eqref{NormalConeA}
		&=		 		
			\sum_{i \in I_+J_+}\Bigl[ t^2(\beta+1)^2 + 1 \Bigr]
			+
			\sum_{i \in I_-J_-}\Bigl[ t^2(\beta+1)^2 + 1 \Bigr]
		\isdraft{}{
		\notag
		\\
		&\quad}
			+
			\sum_{i \in I_-J_+}\Bigl[ t^2(\beta-1)^2 + 1 \Bigr]
			+
			\sum_{i \in I_+J_-}\Bigl[ t^2(\beta-1)^2 + 1 \Bigr]
		\notag
		\\
		&= 		
		 \big|IJ\big| + \Bigl(\bigl|I_+ J_+\bigr| + \bigl|I_- J_-\bigr|\Bigr)t^2(\beta + 1)^2
		\isdraft{}{
		\notag
		\\
		&\quad}
		 +
		 \Bigl(\bigl|I_- J_+\bigr| + \bigl|I_+ J_-\bigr|\Bigr)t^2(\beta - 1)^2
		\notag
		\\
		&=
			t^2\Big(\overline{h}(\beta + 1)^2 + h (\beta - 1)^2\Big) + \big|IJ\big|
		\,.
		\label{Eq:ProofL1L1ExpressionForA}
	\end{align}
	Note that~$h$ and~$\overline{h}$ appear here naturally, before selecting any~$t$.

\mypar{Bounding~\eqref{NormalConeB}}
	If we decompose $IJ^c = I_+J^c \cup I_- J^c$, we see that
	\isdraft{                                                     
		\begin{equation}\label{Eq:ProofL1L1ExpressionForBAux}       
		\eqref{NormalConeB}                                         
		=
		\sum_{i \in I_+ J^c}                                        
		   \mathbb{E}_{g_i} 
				\biggl[
					\text{dist}\Bigl(g_i\,,\, \mathcal{I}(t,t\beta)\Bigr)^2
				\biggr]				
		+
		\sum_{i \in I_-  J^c}                                       
				\mathbb{E}_{g_i}                                        
				\biggl[                                                 
					\text{dist}\Bigl(g_i\,,\, \mathcal{I}(-t,t\beta)\Bigr)^2
				\biggr]\,.
		\end{equation}                                              
	}{
	\begin{multline}\label{Eq:ProofL1L1ExpressionForBAux}
		\eqref{NormalConeB} 
		=
		\sum_{i \in I_+ J^c}
		   \mathbb{E}_{g_i} 
				\biggl[
					\text{dist}\Bigl(g_i\,,\, \mathcal{I}(t,t\beta)\Bigr)^2
				\biggr]		
		\\
		+
		\sum_{i \in I_-  J^c}
				\mathbb{E}_{g_i}
				\biggl[
					\text{dist}\Bigl(g_i\,,\, \mathcal{I}(-t,t\beta)\Bigr)^2
				\biggr]\,.
	\end{multline}
	}
	There are three cases: $\beta = 1$, $\beta < 1$, and $\beta > 1$. 
	\begin{itemize}
		\item 
			If $\beta = 1$, then $\mathcal{I}(t,t\beta) = [0,2t]$ and $\mathcal{I}(-t,t\beta) = [-2t,0]$. Applying~\eqref{Eq:LemAuxDistanceToIntervalTouchOnTheLeft} (resp.\ \eqref{Eq:LemAuxDistanceToIntervalTouchOnTheRight}) to each summand in the first (resp.\ second) term of~\eqref{Eq:ProofL1L1ExpressionForBAux} we conclude that \Shorter{}{all the summands are bounded by
			$
			\varphi(2t)/2t + 1/2.
			$
			Consequently,}
			\begin{equation}\label{Eq:ProofL1L1ExpressionForBBetaEqualTo1}
				\eqref{NormalConeB} 
				\leq
				\big|I J^c\big|
				\bigg[
					\frac{\varphi(2t)}{2t} + \frac{1}{2}
		  	\bigg]\,.
			\end{equation}			
			
		\item
			If $\beta < 1$, then $0 \not\in \mathcal{I}(t,t\beta)$ and $0 \not\in \mathcal{I}(-t,t\beta)$. If we apply~\eqref{Eq:LemAuxDistanceToIntervalRightOfZero} to the summands in the first term of~\eqref{Eq:ProofL1L1ExpressionForBAux} and~\eqref{Eq:LemAuxDistanceToIntervalLeftOfZero} to the summands in the second term, \Shorter{and take into account that~$|I_+J^c| + |I_-J^c| = |IJ^c|$,}{we find that all the summands are bounded by
			$
				1 + t^2(1-\beta)^2 + \varphi(t(\beta + 1))/(t(\beta+1))
			$.
			Taking into account that $|I_+J^c| + |I_-J^c| = |IJ^c|$, this implies}
			\begin{equation}\label{Eq:ProofL1L1ExpressionForBBetaLessThan1}
				\eqref{NormalConeB} 
				\leq
				\big|I J^c\big|
				\bigg[ 
					1 + t^2(1-\beta)^2 + \frac{\varphi(t(\beta + 1))}{t(\beta+1)}
				\bigg]\,.
			\end{equation}

		\item Finally, if $\beta > 1$, then $0 \in \mathcal{I}(t,t\beta)$ and $0 \in \mathcal{I}(-t,t\beta)$. Applying~\eqref{Eq:LemAuxDistanceToIntervalZeroInside} to each summand in both terms of~\eqref{Eq:ProofL1L1ExpressionForBAux} we conclude \Shorter{}{they are all bounded by 
		$
			\varphi(t(\beta - 1))/(t(\beta-1))
			+
			\varphi(t(\beta + 1))/(t(\beta+1))
		$.
		And thus}
		\begin{equation}\label{Eq:ProofL1L1ExpressionForBBetaLargerThan1}
				\eqref{NormalConeB} 
				\leq
				\big|I J^c\big|
				\bigg[
					\frac{\varphi(t(\beta - 1))}{t(\beta-1)}
					+
					\frac{\varphi(t(\beta + 1))}{t(\beta+1)}
		  	\bigg]\,.
		\end{equation}		
	\end{itemize}

\mypar{Bounding~\eqref{NormalConeC}}
	Decompose $I^c J = I^cJ_+ \cup I^c J_-$ and write
	\isdraft{                                                  
		\begin{equation}\label{Eq:ProofL1L1ExpressionForCAux}    
		\eqref{NormalConeC}                                      
		=
		\sum_{i \in I^c J_+}                                     
			\mathbb{E}_{g_i}\biggl[                                
			\text{dist}\Big(g_i\,,\, \mathcal{I}\big(t\beta,t\big)\Big)^2
			\biggr]			
		+                                                        
		\sum_{i \in I^c J_-}                                     
			\mathbb{E}_{g_i}\biggl[
			\text{dist}\Big(g_i\,,\, \mathcal{I}\big(-t\beta\,, t\big)\Big)^2
			\biggr]\,.
		\end{equation}                                           
	}{
	\begin{multline}\label{Eq:ProofL1L1ExpressionForCAux}
		\eqref{NormalConeC} 
		=
		\sum_{i \in I^c J_+}
			\mathbb{E}_{g_i}\biggl[
			\text{dist}\Big(g_i\,,\, \mathcal{I}\big(t\beta,t\big)\Big)^2
			\biggr]	
		\\
		+
		\sum_{i \in I^c J_-}
			\mathbb{E}_{g_i}\biggl[
			\text{dist}\Big(g_i\,,\, \mathcal{I}\big(-t\beta\,, t\big)\Big)^2
			\biggr]\,.
	\end{multline}
	}
	As before, we have three cases: $\beta = 1$, $\beta < 1$, and $\beta > 1$. 
	\begin{itemize}
		\item
			If $\beta = 1$, then $\mathcal{I}(t\beta,t) = [0,2t]$ and $\mathcal{I}(-t\beta,t) = [-2t,0]$. If we apply~\eqref{Eq:LemAuxDistanceToIntervalTouchOnTheLeft} (resp.\ \eqref{Eq:LemAuxDistanceToIntervalTouchOnTheRight}) to each summand in the first (resp.\ second) term of~\eqref{Eq:ProofL1L1ExpressionForCAux}, we conclude \Shorter{}{that all terms are bounded by
			$
			\varphi(2t)/2t + 1/2.
			$
			Therefore,}			
			\begin{equation}\label{Eq:ProofL1L1ExpressionForCBetaEqualTo1}
				\eqref{NormalConeC} 
				\leq
				\big|I^c J\big|				
				\bigg[
					\frac{\varphi(2t)}{2t} + \frac{1}{2}
		  	\bigg]\,.				
			\end{equation}			
	
		\item 
			If $\beta < 1$, then $0 \in \mathcal{I}(t\beta,t)$ and $0 \in \mathcal{I}(-t\beta,t)$. Therefore, according to~\eqref{Eq:LemAuxDistanceToIntervalZeroInside}, \Shorter{}{all terms in~\eqref{Eq:ProofL1L1ExpressionForCAux} are bounded by 
			$
				\varphi(t(1+\beta))/(t(1+\beta))
				+
				\varphi(t(1-\beta))/(t(1-\beta))
			$. Hence,}
			\begin{equation}\label{Eq:ProofL1L1ExpressionForCBetaLessThan1}
				\eqref{NormalConeC} 
				\leq
				\big|I^c J\big|
				\bigg[
					\frac{\varphi(t(1 + \beta))}{t(1+\beta)}
					+
					\frac{\varphi(t(1 - \beta))}{t(1-\beta)}
		  	\bigg]\,.
			\end{equation}			
		
		\item
			If $\beta > 1$, then $0 \not\in \mathcal{I}(t\beta,t)$ and $0 \not\in \mathcal{I}(-t\beta,t)$. If we apply~\eqref{Eq:LemAuxDistanceToIntervalRightOfZero} to each summand in the first term of~\eqref{Eq:ProofL1L1ExpressionForCAux} and~\eqref{Eq:LemAuxDistanceToIntervalLeftOfZero} to each summand in the second term, we find \Shorter{}{that all the summands are bounded by
			$
				1 + t^2(\beta-1)^2 
				+
				\varphi(t(\beta+1))/(t(\beta+1))
			$.
			Therefore,}
			\begin{equation}\label{Eq:ProofL1L1ExpressionForCBetaLargerThan1}
				\eqref{NormalConeC} 
				\leq
				\big|I^c J\big|
				\bigg[
					1 + t^2(\beta-1)^2
					+
					\frac{\varphi{(t(\beta + 1))}}{t(\beta+1)}
		  	\bigg]\,.
			\end{equation}
			
	\end{itemize}

\mypar{Bounding~\eqref{NormalConeD}}
	The interval $\mathcal{I}(0,t(\beta+1))$ contains the origin, so we can apply~\eqref{Eq:LemAuxDistanceToIntervalZeroInside} directly to each summand in~\eqref{NormalConeD}:
	\begin{equation}\label{Eq:ProofL1L1ExpressionForD}
		\eqref{NormalConeD} 
		\leq
		2\big|I^cJ^c\big|
		\frac{\varphi(t(\beta+1))}{t(\beta + 1)}
		\,.
	\end{equation} 
	
\mypar{Bounding \eqref{NormalConeA} + \eqref{NormalConeB} + \eqref{NormalConeC} + \eqref{NormalConeD}}
	Given all the previous bounds, we can now obtain a generic bound for~\eqref{Eq:ProofL1L1FirstStep}. Naturally, there are three cases: $\beta = 1$, $\beta < 1$, and~$\beta>1$. 
	\begin{itemize}
		\item For~$\beta = 1$, we sum~\eqref{Eq:ProofL1L1ExpressionForA} (with~$\beta = 1$), \eqref{Eq:ProofL1L1ExpressionForBBetaEqualTo1}, \eqref{Eq:ProofL1L1ExpressionForCBetaEqualTo1}, and~\eqref{Eq:ProofL1L1ExpressionForD} (with~$\beta = 1$):	
		\Shorter{
			\isdraft{
				\begin{equation}                                
				\mathbb{E}_g\Bigl[\text{dist}\big(g,\text{cone}\,\partial f_1(x^\star)\big)^2\Bigr]
				\leq							                              
				4\overline{h}t^2 + \big|IJ\big| 				        
				+                                               
				\frac{1}{2}\big[|I J^c|+|I^c J|\big]				    
				+                                               
				\big[                                           
					|I J^c|				
					+
					|I^c J|
					+
					2|I^cJ^c|
				\big]
				\frac{\varphi(2t)}{2t}			
				\,.
				\label{Eq:ProofL1L1GenericBoundBetaEqualTo1}    
				\end{equation}                                  
			}{
			\begin{multline}
				\mathbb{E}_g\Bigl[\text{dist}\big(g,\text{cone}\,\partial f_1(x^\star)\big)^2\Bigr]
				\leq							
				4\overline{h}t^2 + \big|IJ\big| 
				\\
				+
				\frac{1}{2}\big[|I J^c|+|I^c J|\big]				
				+
				\big[
					|I J^c|				
					+
					|I^c J|
					+
					2|I^cJ^c|
				\big]
				\frac{\varphi(2t)}{2t}			
				\,.
				\label{Eq:ProofL1L1GenericBoundBetaEqualTo1}
			\end{multline}
			}
		}{			
		\begin{align}
		  &\mathbb{E}_g\Bigl[\text{dist}\big(g,\text{cone}\,\partial f_1(x^\star)\big)^2\Bigr]
		\notag
		\\		
		&\leq		
			4\overline{h}t^2 + \big|IJ\big| 
			+
			\big[
				|I J^c|
				+
				|I^c J|
			\big]
			\Big[
				\frac{\varphi(2t)}{2t} + \frac{1}{2}
			\Big]
		\notag
		\\
		&\qquad
			+
			2\big|I^cJ^c\big|
			\frac{\varphi(2t)}{2t}
		\notag
		\\
		&=		
			4\overline{h}t^2 + \big|IJ\big| 
			+
			\frac{1}{2}\big[|I J^c|+|I^c J|\big]
		\notag
		\\
		&\qquad
			+
			\big[
				|I J^c|
				+
				|I^c J|
				+
				2|I^cJ^c|
			\big]
			\frac{\varphi(2t)}{2t}			
			\,.
		\label{Eq:ProofL1L1GenericBoundBetaEqualTo1}
		\end{align}
		}
		
	\item 
		For $\beta < 1$, we sum~\eqref{Eq:ProofL1L1ExpressionForA}, \eqref{Eq:ProofL1L1ExpressionForBBetaLessThan1}, \eqref{Eq:ProofL1L1ExpressionForCBetaLessThan1}, and~\eqref{Eq:ProofL1L1ExpressionForD}:
		\Shorter{			
			\begin{multline}\label{Eq:ProofL1L1GenericBoundBetaLessThan1}
				\mathbb{E}_g\Bigl[\text{dist}\big(g,\text{cone}\,\partial f_1(x^\star)\big)^2\Bigr]		
				\leq				
				t^2\Big[\overline{h}(\beta + 1)^2 \isdraft{}{\\}+ (h + |IJ^c|)(\beta - 1)^2\Big]
				+
				|I|								
				+
				|I^c J|
				\frac{\varphi(t(1-\beta))}{t(1-\beta)}		
				\\
				+
				\Big[|I J^c| + |I^c J| + 2|I^cJ^c|\Big]
				\frac{\varphi(t(\beta+1))}{t(\beta+1)}\,.				
			\end{multline}
		}{
		\begin{align}
		  &\mathbb{E}_g\Bigl[\text{dist}\big(g,\text{cone}\,\partial f_1(x^\star)\big)^2\Bigr]
		\notag
		\\
		&\leq
			t^2\Big[\overline{h}(\beta + 1)^2 + h (\beta - 1)^2\Big] + \big|IJ\big|
		\notag
		\\
		&\qquad
			+
			|I J^c|
					\Big[
						1 + t^2(\beta-1)^2
					\Big]					
					+
				|I^c J|
				\frac{\varphi(t(1-\beta))}{t(1-\beta)}
		\notag
		\\
		&\qquad
				+				
				\Big[|I J^c| + |I^c J|\Big]
				\frac{\varphi(t(\beta+1))}{t(\beta+1)}
			+
			2\big|I^cJ^c\big|
			\frac{\varphi(t(\beta+1))}{t(\beta + 1)}
		\notag
		\\
		&=
			t^2\Big[\overline{h}(\beta + 1)^2 + (h + |IJ^c|)(\beta - 1)^2\Big]
			+
			|I|		
		\notag
		\\
		&\qquad
			+
			|I^c J|
			\frac{\varphi(t(1-\beta))}{t(1-\beta)}
		\notag
		\\
		&\qquad
			+
			\Big[|I J^c| + |I^c J| + 2|I^cJ^c|\Big]
				\frac{\varphi(t(\beta+1))}{t(\beta+1)}\,.
		\label{Eq:ProofL1L1GenericBoundBetaLessThan1}
	\end{align}
	}
		
	\item For~$\beta > 1$, we sum~\eqref{Eq:ProofL1L1ExpressionForA}, \eqref{Eq:ProofL1L1ExpressionForBBetaLargerThan1}, \eqref{Eq:ProofL1L1ExpressionForCBetaLargerThan1}, and~\eqref{Eq:ProofL1L1ExpressionForD}:
	\Shorter{
			\begin{multline}\label{Eq:ProofL1L1GenericBoundBetaLargerThan1}
				\mathbb{E}_g\Bigl[\text{dist}\big(g,\text{cone}\,\partial f_1(x^\star)\big)^2\Bigr]
				\leq
				t^2\Big[\overline{h}(\beta + 1)^2 \isdraft{}{\\}+ (h + |I^c J|) (\beta - 1)^2\Big] + |J|		  								
				+
				|I J^c|
				\frac{\varphi(t(\beta-1))}{t(\beta -1)}			
				\\
				+
				\big[|I J^c| + |I^c J| + 2|I^cJ^c|\big]				
					\frac{\varphi(t(\beta+1))}{t(\beta+1)}\,.
			\end{multline}
	}{
	\begin{align}
				&\mathbb{E}_g\Bigl[\text{dist}\big(g,\text{cone}\,\partial f_1(x^\star)\big)^2\Bigr]
			\notag
			\\
			&\leq
				t^2\Big[\overline{h}(\beta + 1)^2 + h (\beta - 1)^2\Big] + \big|IJ\big|
			\notag
			\\
			&\qquad
				+
				|I^c J|
						\big[
							1 + t^2(\beta-1)^2
						\big]
			\notag
			\\
			&\qquad
				+
				|I J^c|
				\frac{\varphi(t(\beta-1))}{t(\beta -1)}											
				+				
				\big[|I J^c| + |I^c J|\big]				
					\frac{\varphi(t(\beta+1))}{t(\beta+1)}
			\notag
			\\
			&\qquad
				+
				2\big|I^cJ^c\big|
				\frac{\varphi(t(\beta+1))}{t(\beta + 1)}	
			\notag
			\\
			&=
				t^2\Big[\overline{h}(\beta + 1)^2 + (h + |I^c J|) (\beta - 1)^2\Big] + |J|
		  \notag
			\\
			&\qquad
				+
				|I J^c|
				\frac{\varphi(t(\beta-1))}{t(\beta -1)}
			\notag
			\\
			&\qquad
				+
				\big[|I J^c| + |I^c J| + 2|I^cJ^c|\big]				
					\frac{\varphi(t(\beta+1))}{t(\beta+1)}\,.
			\label{Eq:ProofL1L1GenericBoundBetaLargerThan1}
	\end{align}
	}
	\end{itemize}

\subsubsection{Specification of the Bound for Each Case}

	We now address each one of the cases $\beta = 1$, $\beta < 1$, and~$\beta > 1$ individually. Before that, recall from~\eqref{Eq:LemSetIdentity4} that~$|I^c J| + |IJ^c| + 2|I^c J^c| = 2n - (q + h + \overline{h})$, a term that appears in~\eqref{Eq:ProofL1L1GenericBoundBetaEqualTo1}, \eqref{Eq:ProofL1L1GenericBoundBetaLessThan1}, and~\eqref{Eq:ProofL1L1GenericBoundBetaLargerThan1}. That term is always positive due to our assumption that $n-q  = |I^c J^c| > 0$. 
		
	\mypar{Case 1: \boldmath{$\beta = 1$}} 
	Notice that, according to~\eqref{Eq:LemSetIdentity1} and~\eqref{Eq:LemSetIdentity2},
	$$
		\big|IJ\big| 
		+
		\frac{1}{2}\big[|I J^c|+|I^c J|\big]
		=
		h + \overline{h} + \frac{1}{2}q  - \frac{1}{2}(h + \overline{h})
		=
		\frac{1}{2}(q + h + \overline{h})\,.
	$$
	This allows rewriting~\eqref{Eq:ProofL1L1GenericBoundBetaEqualTo1} as
	\isdraft{                                                      
		\begin{equation}\label{Eq:ProofL1L1Case1Aux}                 
		  \mathbb{E}_g\Bigl[\text{dist}\big(g,\text{cone}\,\partial f_1(x^\star)\big)^2\Bigr]				
		\leq                                                         
			4 \overline{h} t^2 + \frac{1}{2}(q + h + \overline{h})     
			+                                                          
			\frac{1}{2}\Big[2n - (q + h + \overline{h})\Big]           
			\frac{1}{t\sqrt{2\pi}}                                     
			\exp(-2t^2)\,,                                             
		\end{equation}                                               
	}{
	\begin{multline}\label{Eq:ProofL1L1Case1Aux}
		  \mathbb{E}_g\Bigl[\text{dist}\big(g,\text{cone}\,\partial f_1(x^\star)\big)^2\Bigr]				
		\leq
			4 \overline{h} t^2 + \frac{1}{2}(q + h + \overline{h})
		\\
			+
			\frac{1}{2}\Big[2n - (q + h + \overline{h})\Big]
			\frac{1}{t\sqrt{2\pi}}
			\exp(-2t^2)\,,
	\end{multline}
	}
	where we used the definition of~$\varphi$ in~\eqref{Eq:AuxLemVarPhi}. We now select~$t$ as
	$$
		t^\star := \sqrt{\frac{1}{2}\log\Big(\frac{2n}{q + h + \overline{h}}\Big)}
		=
		\sqrt{\frac{1}{2}\log\,r}\,,
	$$
	where $r := 2n/(q + h + \overline{h})$. Notice that~$t^\star$ is well defined because $2n > q + h + \overline{h}$, i.e., $r>0$. It is also finite, as our assumption that~$x^\star \neq 0$, or~$|I|>0$, implies $q = |I \cup J| > 0$. Replacing~$t^\star$ in~\eqref{Eq:ProofL1L1Case1Aux}, we obtain
	\begin{align*}
		  \isdraft{}{&}\mathbb{E}_g\Bigl[\text{dist}\big(g,\text{cone}\,\partial f_1(x^\star)\big)^2\Bigr]
		\isdraft{}{\\}
		&\leq
			2 \overline{h} \log\,r + \frac{1}{2}(q + h + \overline{h})
		\isdraft{}{\\
		&\qquad}
			+
			\frac{1}{2}\Big[2n - (q + h + \overline{h})\Big]
			\frac{1}{\sqrt{\pi\log\,r}}
			\frac{1}{\frac{2n}{q + h + \overline{h}}}
		\\
		&=
			2 \overline{h} \log\,r + \frac{1}{2}(q + h + \overline{h})
			+
			\frac{1}{2}(q + h + \overline{h})
			\frac{1 - \frac{1}{r}}{\sqrt{\pi \log \,r}}
		\\
		&\leq
			2 \overline{h} \log\,r + \frac{1}{2}(q + h + \overline{h})
			+
			\frac{1}{5}(q + h + \overline{h})
		\\
		&=
			2 \overline{h} \log\Big(\frac{2n}{q + h + \overline{h}}\Big) + \frac{7}{10}(q + h + \overline{h})\,,
	\end{align*}
	where we used~\eqref{Eq:BoundChandrasekaran} in the second inequality. This is~\eqref{Eq:ThmL1L1Case1Bound}.

\mypar{Case 2: \boldmath{$\beta < 1$}}
	We rewrite~\eqref{Eq:ProofL1L1GenericBoundBetaLessThan1} as	
	\isdraft{
		\begin{equation}\label{Eq:ProofL1L1Case2Aux}
			\mathbb{E}_g\Bigl[\text{dist}\big(g,\text{cone}\,\partial f_1(x^\star)\big)^2\Bigr]		
		\leq			
			s
			+
			F(\beta,t)			
			+
			G(\beta,t)		
			t^2\Big[\overline{h}(\beta + 1)^2 + (s - \overline{h})(\beta - 1)^2\Big]\,,
		\end{equation}
	}{
	\begin{multline}\label{Eq:ProofL1L1Case2Aux}
			\mathbb{E}_g\Bigl[\text{dist}\big(g,\text{cone}\,\partial f_1(x^\star)\big)^2\Bigr]		
		\leq			
			s
			+
			F(\beta,t)			
			+
			G(\beta,t)
		\\
			t^2\Big[\overline{h}(\beta + 1)^2 + (s - \overline{h})(\beta - 1)^2\Big]\,,
	\end{multline}
	}
	where we used~$s:=|I|$, $|IJ^c| = s - (h + \overline{h})$ (cf.\ \eqref{Eq:LemSetIdentity7}), and
	\begin{align}
		  F(\beta,t) 
		&:=
		  (q-s)
			\frac{\varphi(t(1-\beta))}{t(1-\beta)}
		\notag
		\\
		  G(\beta,t)
		&:=
			(2n - (q + h + \overline{h}))	
			\frac{\varphi(t(\beta+1))}{t(\beta+1)}\,.
		\label{Eq:ProofL1L1Case2DefG}
	\end{align}
	Note that we used~\eqref{Eq:LemSetIdentity6} and~\eqref{Eq:LemSetIdentity4} when defining~$F$ and~$G$. We will consider two cases: $F(\beta,t) \leq G(\beta,t)$ and~$F(\beta,t) \geq G(\beta,t)$. Note that	
	\begin{align}
		\isdraft{}{&}\frac{F(\beta,t)}{G(\beta,t)} \lesseqgtr 1
		\isdraft{\quad}{\notag
		\\}
		\Shorter{}{
		\Longleftrightarrow\quad&
			\frac{q-s}{2n - (q + h + \overline{h})} \frac{1 + \beta}{1 - \beta}
			\frac{\varphi(t(1 - \beta))}{\varphi(t(1 + \beta))}
			\lesseqgtr 1
		\notag
		\\
		\Longleftrightarrow\quad&
			\frac{q-s}{2n - (q + h + \overline{h})} 
			\lesseqgtr
			\frac{1 - \beta}{1 + \beta}\frac{\varphi(t(1 + \beta))}{\varphi(t(1 - \beta))}
		\notag
		\\
		}
		\Longleftrightarrow\quad\isdraft{}{&}
			\frac{q-s}{2n - (q + h + \overline{h})} 
			\lesseqgtr
				\frac{1-\beta}{1 + \beta}\exp\big(-2\beta t^2\big)\,.
		\label{Eq:ProofL1L1BetaSmallerThanOneEquationThatDistinguishesSubcases}
	\end{align}
	\begin{itemize}
		\item Suppose $F(\beta,t) \leq G(\beta,t)$, i.e., \eqref{Eq:ProofL1L1BetaSmallerThanOneEquationThatDistinguishesSubcases} is satisfied with~$\leq$. The bound in \eqref{Eq:ProofL1L1Case2Aux} implies
		\begin{align}
			\isdraft{}{&}\mathbb{E}_g\Bigl[\text{dist}\big(g,\text{cone}\,\partial f_1(x^\star)\big)^2\Bigr]
		\isdraft{}{\notag
		\\}
		&\leq			
			t^2\Big[\overline{h}(\beta + 1)^2 + (s - \overline{h})(\beta - 1)^2\Big]
			+
			s
			+
			2G(\beta,t)
		\notag
		\\
		\Shorter{}{
		&=
			t^2\Big[\overline{h}(\beta + 1)^2 + (s - \overline{h})(\beta - 1)^2\Big]
			+
			s
		\notag
		\\
		&\qquad
			+
			2
			\Big[2n-(q + h + \overline{h})\Big]
			\frac{\varphi(t(\beta+1))}{t(\beta+1)}
		\notag
		\\
		}
		&=
			t^2\Big[\overline{h}(\beta + 1)^2 + (s - \overline{h})(\beta - 1)^2\Big]
			+
			s
		\isdraft{}{
		\notag
		\\
		&\qquad}
			+
			2
			\Big[2n - (q + h + \overline{h})\Big]			
			\frac{\exp\big(-\frac{t^2}{2}(\beta + 1)^2\big)}{\sqrt{2\pi}t(\beta+1)}
			\,,
		\label{Eq:ProofL1L1Case2Aux3}
	\end{align}
	where we used the definition of~$\varphi$. We now select~$t$ as
	$$
		t^\star = \frac{1}{\beta + 1}\sqrt{2\log\Big(\frac{2n}{q + h + \overline{h}}\Big)}
		=
		\frac{1}{\beta + 1}\sqrt{2\log\,r}\,,
	$$
	where $r := 2n/(q + h + \overline{h})$ is as before. Replacing~$t^\star$ in~\eqref{Eq:ProofL1L1Case2Aux3} yields
	\begin{align*}
			\isdraft{}{&}\mathbb{E}_g\Bigl[\text{dist}\big(g,\text{cone}\,\partial f_1(x^\star)\big)^2\Bigr]
		\isdraft{}{\\}
		&\leq
			2\bigg[\overline{h} + (s - \overline{h})\frac{(\beta-1)^2}{(\beta + 1)^2}\bigg]\log\,r
			+
			s
		\isdraft{}{
		\\
		&\qquad}
			+
			2\Big[ 2n - (q + h + \overline{h}) \Big]\frac{1}{\sqrt{2\log\,r}}\frac{1}{\sqrt{2\pi}}\frac{1}{\frac{2n}{q + h + \overline{h}}}
		\\
		&=
			2\bigg[\overline{h} + (s - \overline{h})\frac{(\beta-1)^2}{(\beta + 1)^2}\bigg]\log\,r
			+
			s
		\isdraft{}{
		\\
		&\qquad}
			+
			(q + h + \overline{h})\frac{1 - \frac{1}{r}}{\sqrt{\pi \log \, r}}
		\\
		&\leq
			2\bigg[\overline{h} + (s - \overline{h})\frac{(\beta-1)^2}{(\beta + 1)^2}\bigg]\log\Big(\frac{2n}{q + h + \overline{h}}\Big)
			+
			s
		\isdraft{}{		
		\\
		&\qquad}
			+
			\frac{2}{5}(q + h + \overline{h})\,,
	\end{align*}
	which is~\eqref{Eq:ThmL1L1Case2Bound1}.	We used~\eqref{Eq:BoundChandrasekaran} in the last inequality. This bound is valid only when~\eqref{Eq:ProofL1L1BetaSmallerThanOneEquationThatDistinguishesSubcases} with~$\leq$ is satisfied with~$t = t^\star$, i.e.,
	\begin{align*}
		  \frac{q - s}{2n - (q + h + \overline{h})} 
		&\leq
			\Shorter{}{
		  \frac{1-\beta}{1 + \beta}\Big(\frac{1}{r}\Big)^{4\beta/(\beta+1)^2}
		\\
		&=
			}
			\frac{1-\beta}{1 + \beta}\Big(\frac{q + h + \overline{h}}{2n}\Big)^{\frac{4\beta}{(\beta + 1)^2}}\,,
	\end{align*}
	which is condition~\eqref{Eq:ThmL1L1Case2Condition1}.
	
	\item Suppose now that $F(\beta,t) \geq G(\beta,t)$, i.e., \eqref{Eq:ProofL1L1BetaSmallerThanOneEquationThatDistinguishesSubcases} is satisfied with $\geq$. Then, \eqref{Eq:ProofL1L1Case2Aux} becomes 
	\begin{align}
			\isdraft{}{&}\mathbb{E}_g\Bigl[\text{dist}\big(g,\text{cone}\,\partial f_1(x^\star)\big)^2\Bigr]
		\isdraft{}{
		\notag
		\\}
		&\leq			
			t^2\Big[\overline{h}(\beta + 1)^2 + (s - \overline{h})(\beta - 1)^2\Big]
			+
			s
			+
			2F(\beta,t)
		\notag
		\\
		\Shorter{}{
		&=
			t^2\Big[\overline{h}(\beta + 1)^2 + (s - \overline{h})(\beta - 1)^2\Big]
			+
			s
		\notag
		\\
		&\qquad
			+			
			2(q-s)\frac{\varphi(t(1-\beta))}{t(1-\beta)}
		\notag
		\\
		}
		&=
			t^2\Big[\overline{h}(\beta + 1)^2 + (s - \overline{h})(\beta - 1)^2\Big]
			+
			s
		\isdraft{}{
		\notag
		\\
		&\qquad}
			+
			2(q-s)\frac{\exp\big(-\frac{t^2}{2}(1-\beta)^2\big)}{\sqrt{2\pi}\,t(1-\beta)}\,.
		\label{Eq:ProofL1L1Case2Aux4}
	\end{align}
	We select~$t$ as
	$$
		t^\star = \frac{1}{1-\beta}\sqrt{2\log\Big(\frac{q}{s}\Big)} = \frac{1}{1 - \beta}\sqrt{2\log\,r}\,,
	$$
	where~$r$ is now~$r := q/s$. Since in case \text{2b)} of the theorem, we assume~$ 0 < |I^c J| = q - s$, we have $t^\star > 0$. Notice that~$t^\star$ is finite, because~$s>0$ (given that~$x^\star \neq 0$).  Replacing~$t^\star$ into~\eqref{Eq:ProofL1L1Case2Aux4} yields
	\begin{align*}
			\isdraft{}{&}\mathbb{E}_g\Bigl[\text{dist}\big(g,\text{cone}\,\partial f_1(x^\star)\big)^2\Bigr]
		\isdraft{}{
		\\}
		&\leq
			2\bigg[\overline{h}\frac{(1 + \beta)^2}{(1-\beta)^2} + s - \overline{h}\bigg]\log\,r	
			+
			s
		\isdraft{}{
		\\
		&\qquad}
			+
			2(q-s)\frac{1}{\sqrt{2\log\,r}}\frac{1}{\sqrt{2\pi}}\frac{1}{\frac{q}{s}}
		\\
		&=
			2\bigg[\overline{h}\frac{(1 + \beta)^2}{(1-\beta)^2} + s - \overline{h}\bigg]\log\,r	
			+
			s
			+
			s\frac{1 - \frac{1}{r}}{\sqrt{\pi\log\,r}}
		\\
		&\leq
			2\bigg[\overline{h}\frac{(1 + \beta)^2}{(1-\beta)^2} + s - \overline{h}\bigg]\log\,r	
			+
			s
			+
			\frac{2}{5}s
		\\
		&=
			2\bigg[\overline{h}\frac{(1 + \beta)^2}{(1-\beta)^2} + s - \overline{h}\bigg]\log\Big(\frac{q}{s}\Big)	
			+
			\frac{7}{5}s\,,
	\end{align*}
	which is~\eqref{Eq:ThmL1L1Case2Bound2}. Again, we used~\eqref{Eq:BoundChandrasekaran} in the last inequality. This bound is valid only if~\eqref{Eq:ProofL1L1BetaSmallerThanOneEquationThatDistinguishesSubcases} with~$\geq$ is satisfied for~$t = t^\star$, i.e.,
	\begin{align*}
		  \frac{q - s}{2n - (q + h + \overline{h})} 
		&\geq 
			\Shorter{}{
		  \frac{1-\beta}{1 + \beta}\Big(\frac{1}{r}\Big)^{\frac{4\beta}{(1-\beta)^2}}
		\\
		&=}
			\frac{1-\beta}{1 + \beta}\Big(\frac{s}{q}\Big)^{\frac{4\beta}{(1-\beta)^2}}\,,
	\end{align*}
	which is condition~\eqref{Eq:ThmL1L1Case2Condition2}.
	
	\end{itemize}

\mypar{Case 3: \boldmath{$\beta > 1$}}
	We rewrite~\eqref{Eq:ProofL1L1GenericBoundBetaLargerThan1} as
	\begin{align}
		  \isdraft{}{&}\mathbb{E}_g\Bigl[\text{dist}\big(g,\text{cone}\,\partial f_1(x^\star)\big)^2\Bigr]
		\isdraft{}{
		\notag
		\\}
		&\leq
			t^2\Big[\overline{h}(\beta + 1)^2 + (q + h - s) (\beta - 1)^2\Big] + l
			+
			H(\beta,t)
		\isdraft{}{
		\notag
		\\
		&\qquad}
			+
			G(\beta,t)\,,
		\label{Eq:ProofL1L1Case3Aux}
	\end{align}
	where we used~$l := |J|$, $|I^cJ| = q - s$ (cf.\ \eqref{Eq:LemSetIdentity6}), $G$ is defined in~\eqref{Eq:ProofL1L1Case2DefG}, and
	\begin{align*}
		H(\beta,t) &:= (s - (h + \overline{h}))\frac{\varphi(t(\beta-1))}{t(\beta -1)}\,.
	\end{align*}
	Note that we used~\eqref{Eq:LemSetIdentity7} when defining~$H$. We also consider two cases: $H(\beta,t) \leq G(\beta,t)$ and~$H(\beta,t) \geq H(\beta,t)$. Note that
	\begin{align}
		\isdraft{}{&}\frac{H(\beta,t)}{G(\beta,t)} \lesseqgtr 1
		\isdraft{\quad}{
		\notag
		\\}
		\Shorter{}{
		\Longleftrightarrow\quad&
			\frac{s-(h+\overline{h})}{2n - (q + h + \overline{h})} \frac{\beta + 1}{\beta - 1}
			\frac{\varphi(t(\beta - 1))}{\varphi(t(\beta+1))}
			\lesseqgtr 1
		\notag
		\\
		\Longleftrightarrow\quad&
			\frac{s-(h+\overline{h})}{2n - (q + h + \overline{h})} 
			\lesseqgtr
			\frac{\beta - 1}{\beta + 1}\frac{\varphi(t(\beta + 1))}{\varphi(t(\beta - 1))}
		\notag
		\\
		}
		\Longleftrightarrow\quad\isdraft{}{&}
			\frac{s-(h+\overline{h})}{2n - (q + h + \overline{h})} 
			\lesseqgtr
				\frac{\beta - 1}{\beta + 1}\exp\big(-2\beta t^2\big)\,.
		\label{Eq:ProofL1L1BetaSmallerThanOneEquationThatDistinguishesSubcasesCase3}
	\end{align}
	
	\begin{itemize}
	
		\item Suppose $H(\beta,t) \leq G(\beta,t)$, i.e., \eqref{Eq:ProofL1L1BetaSmallerThanOneEquationThatDistinguishesSubcasesCase3} is satisfied with~$\leq$. Then,
		\eqref{Eq:ProofL1L1Case3Aux} implies
		\begin{align}
			\isdraft{}{&}\mathbb{E}_g\Bigl[\text{dist}\big(g,\text{cone}\,\partial f_1(x^\star)\big)^2\Bigr]
		\isdraft{}{
		\notag
		\\}
		&\leq
			t^2\Big[\overline{h}(\beta + 1)^2 + (q + h - s) (\beta - 1)^2\Big] + l
			+
			2G(\beta,t)
		\notag
		\\
		&=
			t^2\Big[\overline{h}(\beta + 1)^2 + (q + h -s) (\beta - 1)^2\Big] + l
		\isdraft{}{
		\notag
		\\
		&\qquad}
			+
			2(2n - (q + h + \overline{h}))\frac{\exp\Big(-\frac{t^2}{2}(\beta+1)^2\Big)}{\sqrt{2\pi}t(\beta+1)}\,.
		\label{Eq:ProofL1L1Case3Aux3}
	\end{align}
	Now we select
	$$
		t^\star = \frac{1}{\beta + 1}\sqrt{2\log\Big(\frac{2n}{q + h + \overline{h}}\Big)}
		=
		\frac{1}{\beta + 1}\sqrt{2\log\,r}\,,
	$$
	where $r := 2n/(q + h + \overline{h})$. Again, note that our assumptions imply that~$t^\star$ is well defined and positive. Replacing~$t^\star$ into~\eqref{Eq:ProofL1L1Case3Aux3} yields
	\begin{align*}
		  \isdraft{}{&}\mathbb{E}_g\Bigl[\text{dist}\big(g,\text{cone}\,\partial f_1(x^\star)\big)^2\Bigr]
		\isdraft{}{\\}
		&\leq
			2\bigg[\overline{h} + (q + h - s)\Big(\frac{\beta-1}{\beta+1}\Big)^2\bigg]\log\,r
			+
			l
		\isdraft{}{
		\\
		&\qquad}
			+
			(2n - (q + h + \overline{h}))\frac{1}{\sqrt{\pi\log\,r}}\frac{1}{\frac{2n}{q + h + \overline{h}}}
		\\
		&=
			2\bigg[\overline{h} + (q + h - s)\Big(\frac{\beta-1}{\beta+1}\Big)^2\bigg]\log\,r
			+
			l
		\isdraft{}{
		\\
		&\qquad}
			+
			(q + h + \overline{h})\frac{1 - \frac{1}{r}}{\sqrt{\pi\log\,r}}
		\\
		&\leq
			2\bigg[\overline{h} + (q + h - s)\Big(\frac{\beta-1}{\beta+1}\Big)^2\bigg]\log\Big(\frac{2n}{q + h + \overline{h}}\Big)
			+
			l
		\isdraft{}{
		\\
		&\qquad}
			+
			\frac{2}{5}(q + h + \overline{h})\,.
	\end{align*}
	This is~\eqref{Eq:ThmL1L1Case3Bound1}. Again, \eqref{Eq:BoundChandrasekaran} was used in the last step. This bound is valid only when~\eqref{Eq:ProofL1L1BetaSmallerThanOneEquationThatDistinguishesSubcasesCase3} with~$\leq$ is satisfied for~$t = t^\star$, i.e.,
	\begin{align*}
		\frac{s-(h+\overline{h})}{2n - (q + h + \overline{h})} 
		&\leq
		\Shorter{}{
		\frac{\beta - 1}{\beta + 1} \Big(\frac{1}{r}\Big)^{\frac{4\beta}{(\beta + 1)^2}}
		\\
		&=
		}
		\frac{\beta - 1}{\beta + 1} \Big(\frac{q + h + \overline{h}}{2n}\Big)^{\frac{4\beta}{(\beta + 1)^2}}\,,		
	\end{align*}
	which is condition~\eqref{Eq:ThmL1L1Case3Condition1}.

	\item 
		Suppose now that~$H(\beta, t) \geq G(\beta,t)$. Then, \eqref{Eq:ProofL1L1Case3Aux} implies
		\begin{align}
		  \isdraft{}{&}\mathbb{E}_g\Bigl[\text{dist}\big(g,\text{cone}\,\partial f_1(x^\star)\big)^2\Bigr]		  
		\isdraft{}{
		\notag
		\\}
		&\leq
			t^2\Big[\overline{h}(\beta + 1)^2 + (q + h - s) (\beta - 1)^2\Big] + l
			+
			2H(\beta,t)
		\notag
		\\		
		&=
			t^2\Big[\overline{h}(\beta + 1)^2 + (q + h - s) (\beta - 1)^2\Big] + l
		\isdraft{}{
		\notag
		\\
		&\qquad}
			+
			2(s - (h + \overline{h}))\frac{\exp\Big(-\frac{t^2}{2}(\beta - 1)^2\Big)}{\sqrt{2\pi}t(\beta-1)}\,.
		\label{Eq:ProofL1L1Case3Aux4}
	\end{align}
	Given our assumption that~$|IJ| = h + \overline{h} > 0$ in case \text{3b)}, we can select~$t$ as
	$$
		t^\star = \frac{1}{\beta - 1}\sqrt{2\log\Big(\frac{s}{h + \overline{h}}\Big)}
		=
		\frac{1}{\beta - 1}\sqrt{2\log\,r}\,,
	$$
	where~$r := s/(h + \overline{h})$. We also assume that~$|IJ^c| = s - (h + \overline{h}) > 0$, making~$t^\star >0$. Replacing~$t^\star$ into~\eqref{Eq:ProofL1L1Case3Aux4} gives
	\begin{align*}
		  \isdraft{}{&}\mathbb{E}_g\Bigl[\text{dist}\big(g,\text{cone}\,\partial f_1(x^\star)\big)^2\Bigr]
		\isdraft{}{\\}
		&\leq
			2\bigg[\overline{h}\Big(\frac{\beta + 1}{\beta - 1}\Big)^2 + q + h -s\bigg]\log\,r
			+
			l
		\isdraft{}{
		\\
		&\qquad}
			+
			2(s - (h + \overline{h}))
			\frac{1}{\sqrt{2\log\,r}}
			\frac{1}{\sqrt{2\pi}}
			\frac{1}{\frac{s}{h + \overline{h}}}
		\\
		&=
			2\bigg[\overline{h}\Big(\frac{\beta + 1}{\beta - 1}\Big)^2 + q + h - s\bigg]\log\,r
			+
			l			
		\isdraft{}{
		\\
		&\qquad}
			+
			(h + \overline{h})\frac{1 - \frac{1}{r}}{\sqrt{\pi \log\,r}}
		\\
		&\leq
			2\bigg[\overline{h}\Big(\frac{\beta + 1}{\beta - 1}\Big)^2 + q + h - s\bigg]\log\,r
			+
			l
			+
			\frac{2}{5}(h + \overline{h})\,,
	\end{align*}
	which is~\eqref{Eq:ThmL1L1Case3Bound2}. Again, \eqref{Eq:BoundChandrasekaran} was employed in the last inequality. This bound is valid only when~\eqref{Eq:ProofL1L1BetaSmallerThanOneEquationThatDistinguishesSubcasesCase3} with~$\geq$ holds for~$t = t^\star$, that is,
	\begin{align*}
		  \frac{s-(h+\overline{h})}{2n - (q + h + \overline{h})} 
		&\geq
			\Shorter{}{
			\frac{\beta - 1}{\beta + 1}
			\Big(\frac{1}{r}\Big)^{\frac{4\beta}{(\beta - 1)^2}}
		\\
		&=
			}
			\frac{\beta - 1}{\beta + 1}
			\Big(\frac{h + \overline{h}}{s}\Big)^{\frac{4\beta}{(\beta - 1)^2}}
			\,,
	\end{align*}
	which is condition~\eqref{Eq:ThmL1L1Case3Condition2}. This concludes the proof. \hfill \qed
	\end{itemize}

	\mypar{Remarks}
	The bound for case~\text{1)}, i.e., $\beta = 1$, is clearly the sharpest one, since it does not use inequalities like~\eqref{Eq:ProofL1L1BetaSmallerThanOneEquationThatDistinguishesSubcases} or~\eqref{Eq:ProofL1L1BetaSmallerThanOneEquationThatDistinguishesSubcasesCase3}. Perhaps the ``loosest'' inequality it uses is~\eqref{Eq:BoundChandrasekaran} in \lref{lem:AuxLemmaBoundsFunctions}. According to its proof in \aref{App:ProofLemmaBounds}, that bound is exact when~$x = 2$, which means~$2n/(q + h + \overline{h}) = 2$, for~$\beta = 1$. The bounds for~$\beta \neq 1$ are not as sharp, due to~\eqref{Eq:ProofL1L1BetaSmallerThanOneEquationThatDistinguishesSubcases} and~\eqref{Eq:ProofL1L1BetaSmallerThanOneEquationThatDistinguishesSubcasesCase3}. Note also that some sharpness is lost by selecting specific values of~$t$ and not the optimal ones (cf.\ \eqref{Eq:ProofL1L1RelationBetweenExpressionsFixingt}).

	\subsection{Proof of Theorem~\ref{Thm:L1L2Reconstruction}}
	\label{SubSec:ProofsL1L2}
	
	The steps to prove the theorem are the same steps as the ones in the proof of \thref{Thm:L1L1Reconstruction}. So, we will omit some details. Whenever~$0 \not\in \partial f_2(x^\star)$, we can use the bound in~\eqref{Eq:ProofL1L1FirstStep} with~$f_1$ replaced by~$f_2$, i.e., 
	\begin{equation}\label{Eq:ProofL1L2FirstStep}
		w\bigl(T_{f_2}(x^\star)\bigr)^2 \leq \mathbb{E}_g\Bigl[\text{dist}\big(g,\text{cone}\,\partial f_2(x^\star)\big)^2\Bigr]\,.
	\end{equation} 
	Note that this bound results from the characterization of the normal cone provided in \pref{Prop:Subdifferential} and from Jensen's inequality. Part \text{2)} of \lref{lem:ZeroNotInSubGradient} establishes that our assumptions guarantee that~$0 \not\in \partial f_2(x^\star)$ and, thus, that we can use~\eqref{Eq:ProofL1L2FirstStep}. Next, we express the right-hand side of~\eqref{Eq:ProofL1L2FirstStep} component-wise, and then we establish bounds for each term.
	
	A vector~$d \in \mathbb{R}^n$ belongs to the cone generated by $\partial f_2(x^\star)$ if, for some~$t \geq 0$ and some~$y \in \partial f_2(x^\star)$, $d = ty$. According to~\eqref{Eq:SubgradientL1L2PerComp}, each component~$d_i$ satisfies
	$$
		\left\{
			\begin{array}{ll}
				d_i = t\,\text{sign}(x_i^\star) + t\beta (x_i^\star - w_i) &,\,\, i \in I
				\vspace{0.2cm}
				\\
				|d_i + t\beta w_i| \leq t &,\,\, i \in I^c\,,
   		\end{array}
		\right.
	$$
	for some~$t\geq 0$.
	This allows expanding the right-hand side of~\eqref{Eq:ProofL1L2FirstStep} as
	\begin{align*}
			\isdraft{}{&}\mathbb{E}_g\Bigl[\text{dist}\big(g,\text{cone}\,\partial f_2(x^\star)\big)^2\Bigr]
		\isdraft{}{\\}
		&=
		\mathbb{E}_g
		\left[\,	
			\underset{t\geq 0}{\min}
			\Biggl\{
				\sum_{i \in I}
					\text{dist}\Bigl(g_i\,,\, t\,\text{sign}(x_i^\star) + t\beta(x_i^\star - w_i)\Bigr)^2
		\isdraft{}{
		\right.
		\\
		&\qquad
		\left.}
				+
				\sum_{i \in I^c}
					\text{dist}\Bigl(g_i\,,\, \mathcal{I}(-t\beta w_i, t)\Bigr)^2
			\Biggl\}			
		\right]\,.	
	\end{align*}
	As in the proof of \thref{Thm:L1L1Reconstruction}, we fix~$t$ and select it later (cf.\ \eqref{Eq:ProofL1L1RelationBetweenExpressionsFixingt}). Doing so, gives
	\begin{subequations}\label{Eq:ProofL1L2NormalConeDecompositionAB}
	\begin{align}
		  &\mathbb{E}_g\Bigl[\text{dist}\big(g,\text{cone}\,\partial f_2(x^\star)\big)^2\Bigr]		
		\notag
		\\
		&\leq
			\sum_{i \in I}
			\mathbb{E}_{g_i}
			\Big[
				\text{dist}\Bigl(g_i\,,\, t\,\text{sign}(x_i^\star) + t\beta(x_i^\star - w_i)\Bigr)^2
			\Big]
		\label{NormalConeL2A}
		\\
		&\qquad+
			\sum_{i \in I^c}
			\mathbb{E}_{g_i}
			\Big[
				\text{dist}\Bigl(g_i\,,\, \mathcal{I}(-t\beta w_i, t)\Bigr)^2
			\Big]\,.
		\label{NormalConeL2B}
	\end{align}
	\end{subequations}
	Next, we use \lref{lem:AuxLemmaIntervals} to derive a closed-form expression for~\eqref{NormalConeL2A} and establish a bound on~\eqref{NormalConeL2B}.
	
	\mypar{Expression for~\eqref{NormalConeL2A}}
	Using~\eqref{Eq:LemAuxDistanceToAPoint},
	\begin{align}
		\eqref{NormalConeL2A} 
		&=
		\sum_{i \in I}
			\mathbb{E}_{g_i}
			\Big[
				\text{dist}\Bigl(g_i\,,\, t\,\text{sign}(x_i^\star) + t\beta(x_i^\star - w_i)\Bigr)^2
			\Big]
		\notag
		\\
		&=
			\sum_{i \in I}
			\Big[(t\,\text{sign}(x_i^\star) + t\beta(x_i^\star - w_i))^2 + 1\Big]
		\notag
		\\
		&=
			t^2 \bigg[\sum_{i \in I_+} (1 + \beta(x_i^\star - w_i))^2 + \sum_{i \in I_-} (1 - \beta(x_i^\star - w_i))^2\bigg]
		\isdraft{}{
		\notag
		\\
		&\qquad}
		+ |I|\,,
		\label{Eq:ProofL1L2ExpressionA}
	\end{align}
	where we decomposed~$I = I_+ \cup I_-$.

	\mypar{Bounding~\eqref{NormalConeL2B}}
	We have
	\isdraft{                                                  
		\begin{equation}\label{Eq:ProofL1L2BoundBAux}            
		\eqref{NormalConeL2B}                                    
		=                                                        
		\sum_{i \in I^cJ}                                        
			\mathbb{E}_{g_i}                                       
			\Big[
				\text{dist}\Bigl(g_i\,,\, \mathcal{I}(-t\beta w_i, t)\Bigr)^2
			\Big]		
		+
		\sum_{i \in I^cJ^c}
			\mathbb{E}_{g_i}
			\Big[
				\text{dist}\Bigl(g_i\,,\, \mathcal{I}(0, t)\Bigr)^2  
			\Big]\,.                                               
		\end{equation}                                           
	}{
	\begin{multline}\label{Eq:ProofL1L2BoundBAux}
		\eqref{NormalConeL2B}
		=
		\sum_{i \in I^cJ}
			\mathbb{E}_{g_i}
			\Big[
				\text{dist}\Bigl(g_i\,,\, \mathcal{I}(-t\beta w_i, t)\Bigr)^2
			\Big]
		\\
		+
		\sum_{i \in I^cJ^c}
			\mathbb{E}_{g_i}
			\Big[
				\text{dist}\Bigl(g_i\,,\, \mathcal{I}(0, t)\Bigr)^2
			\Big]\,.
	\end{multline}
	}
	The second term in the right-hand side of~\eqref{Eq:ProofL1L2BoundBAux} can be bounded according to~\eqref{Eq:LemAuxDistanceToIntervalZeroInside}:
	\begin{equation}\label{Eq:ProofL1L2BoundBAux3}
		\sum_{i \in I^c J^c} \mathbb{E}_{g_i}\Big[\text{dist}(g_i, \mathcal{I}(0,t))^2\Big]
		\leq
		2|I^cJ^c| \frac{\varphi(t)}{t}\,.
	\end{equation} 
	The first term, however, is more complicated. Recall that $I^cJ = \{i\,:\, w_i \neq x_i^\star = 0\}$. Let us analyze the several possible situations for the interval~$\mathcal{I}(-t\beta w_i, t) = [t(-\beta w_i - 1), t(-\beta w_i + 1)]$. It does not contain zero whenever
	\begin{align*}
		t(-\beta w_i - 1) > 0
		\quad
		&\Longleftrightarrow
		\quad
		t \neq 0
		\quad\text{and}\quad
		w_i < -\frac{1}{\beta}\,,	
	\intertext{or}
		t(-\beta w_i + 1) < 0
		\quad
		&\Longleftrightarrow
		\quad
		t \neq 0
		\quad\text{and}\quad
		w_i > \frac{1}{\beta}\,.	
	\end{align*}
	In addition to the subsets of~$I^c J$ defined in~\eqref{Eq:KEq}-\eqref{Eq:Kneq}, define
	\begin{align*}
		K_- &:= \Big\{i \in I^c J\,:\, w_i < -\frac{1}{\beta}\Big\}		
		\\
		K_+ &:= \Big\{i \in I^c J\,:\, w_i > \frac{1}{\beta}\Big\}		
		\\						
		K_-^= &:= \Big\{i \in I^c J\,:\, w_i = -\frac{1}{\beta}\Big\}
		\\
		K_+^= &:= \Big\{i \in I^c J\,:\, w_i = \frac{1}{\beta}\Big\}
		\\
		L &:= \Big\{i \in I^c J\,:\, |w_i| < \frac{1}{\beta}\Big\}\,,
	\end{align*}	
	where we omit the dependency of these sets on~$\beta$ for notational simplicity. Noticing that 
	$I^c J = K_- \cup K_+ \cup K_-^= \cup K_+^= \cup L$ 
	and using \lref{lem:AuxLemmaIntervals}, we obtain
	\begin{align}
		\isdraft{}{&}
			\sum_{i \in I^cJ}
				\mathbb{E}_{g_i}
				\Big[
					\text{dist}\Bigl(g_i\,,\, \mathcal{I}(-t\beta w_i, t)\Bigr)^2
				\Big]
		\isdraft{}{
		\notag
		\\}
		&\leq
			\sum_{i \in K_-}
			\bigg[
				1 + t^2(\beta w_i + 1)^2 + \frac{\varphi(t(1-\beta w_i))}{t(1 - \beta w_i)}
			\bigg]		
		\notag
		\\
		&\qquad
		+
			\sum_{i \in K_+}
			\bigg[
				1 + t^2(1-\beta w_i)^2 + \frac{\varphi(t(1+\beta w_i))}{t(1+\beta w_i)}
			\bigg]
		\notag
		\\
		&\qquad
		+
			\sum_{i \in K_-^=}
			\bigg[
				\frac{1}{2} + \frac{\varphi(t(1-\beta w_i))}{t(1 - \beta w_i)}
			\bigg]
		\isdraft{}{
		\notag
		\\
		&\qquad}
		+
			\sum_{i \in K_+^=}
			\bigg[
				\frac{1}{2} + \frac{\varphi(t(1+\beta w_i))}{t(1 + \beta w_i)}
			\bigg]
		\notag
		\\
		&\qquad
		+
			\sum_{i \in L}
			\bigg[
				\frac{\varphi(t(1 + \beta w_i))}{t(1 + \beta w_i)}
				+
				\frac{\varphi(t(1 - \beta w_i))}{t(1 - \beta w_i)}
			\bigg]		
		\notag
		\\
		&=
			|K^{\neq}(\beta)| + \frac{1}{2}|K^=(\beta)| 
			+
			t^2 
			\bigg[
				\sum_{i \in K_-} (\beta w_i + 1)^2
		\isdraft{}{
		\notag
		\\
		&\qquad}
				+
				\sum_{i \in K_+} (\beta w_i - 1)^2
			\bigg]
			\isdraft{\notag\\&\qquad}{}
			+
			\sum_{i \in K_- \cup K_-^=}\frac{\varphi(t(1 - \beta w_i))}{t(1 - \beta w_i)}
		\isdraft{}{
		\notag		
		\\
		&\qquad}
			+
			\sum_{i \in K_+ \cup K_+^=}\frac{\varphi(t(1 + \beta w_i))}{t(1 + \beta w_i)}
		\notag
		\\
		&\qquad
			+
			\sum_{i \in L}\bigg[\frac{\varphi(t(1 + \beta w_i))}{t(1 + \beta w_i)} + \frac{\varphi(t(1 - \beta w_i))}{t(1 - \beta w_i)}\bigg]
		\notag
		\\
		&\leq
			|K^{\neq}(\beta)| + \frac{1}{2}|K^=(\beta)| 
			+
			t^2 
			\bigg[
				\sum_{i \in K_-} (\beta w_i + 1)^2
		\isdraft{}{
		\notag
		\\
		&\qquad}
				+
				\sum_{i \in K_+} (\beta w_i - 1)^2
			\bigg]
		\notag
		\\
		&\qquad
			+
			\Big[|K_-| + |K_-^=| + |L|\Big]
			\frac{\varphi(t(1 - \beta\, \overline{w}_p))}{t(1 - \beta\, \overline{w}_p)}
		\notag
		\\
		&\qquad
			+
			\Big[|K_+| + |K_+^=| + |L|\Big]
			\frac{\varphi(t(1 + \beta\, \overline{w}_m))}{t(1 + \beta\, \overline{w}_m)}\,,		
		\label{Eq:ProofL1L2IntroductionOfwbar}
	\end{align}
	where, in the second step, we used the definitions of~$K^=$ and~$K^{\neq}$ in~\eqref{Eq:KEq} and~\eqref{Eq:Kneq}, respectively. In the last step, we used~$\overline{w}_p := |w_p|$ and~$\overline{w}_m := |w_m|$, for
	\begin{align*}
		p &:= \underset{i \in K_- \,\cup\, K_-^= \,\cup\, L}{\arg\min} 1 - \beta w_i
		\\
		m &:= \underset{i \in K_+ \,\cup\, K_+^= \,\cup\, L}{\arg\min} 1 + \beta w_i\,.
	\end{align*}
	Note that~$\overline{w} = \big|\arg\min_{w = \overline{w}_p, \overline{w}_m} \big\{|w| - 1/\beta\big\}\big|$, since the union of the sets $K_-$, $K_-^=$, $L$, $K_+$, and~$K_+^=$ gives~$I^cJ$. Therefore, $\varphi(t(1 - \beta\, \overline{w}_j))/(t(1 - \beta\, \overline{w}_j)) \leq \varphi(t(1 - \beta\, \overline{w}))/(t|1 - \beta\, \overline{w}|)$, for~$j = p,m$. Using this in the last two terms of~\eqref{Eq:ProofL1L2IntroductionOfwbar}, and noticing that
	$$
		|K_-| + |K_-^=| + |K_+| + |K_+^=| = \Big\{i \in I^c J\,:\, |w_i| \geq \frac{1}{\beta}\Big\} =: K(\beta)\,,
	$$
	we obtain		
	\begin{multline}\label{Eq:ProofL1L2BoundBAux2}
		\sum_{i \in I^cJ}
				\mathbb{E}_{g_i}
				\Big[
					\text{dist}\Bigl(g_i\,,\, \mathcal{I}(-t\beta w_i, t)\Bigr)^2
				\Big]
		\leq
		|K^{\neq}(\beta)| + \frac{1}{2}|K^=(\beta)| 
		\isdraft{}{\\}+
			t^2 
			\bigg[
				\sum_{i \in K_-} (\beta w_i + 1)^2
				+
				\sum_{i \in K_+} (\beta w_i - 1)^2
			\bigg]
		\\
		+
		\Big[|K(\beta)| + 2|L|\Big]
		\frac{\varphi(t(1 - \beta\, \overline{w}))}{t|1 - \beta\, \overline{w}|}\,.
	\end{multline}	
	
	\mypar{Bounding \eqref{NormalConeL2A} + \eqref{NormalConeL2B}}
	Adding up~\eqref{Eq:ProofL1L2ExpressionA}, \eqref{Eq:ProofL1L2BoundBAux3}, and~\eqref{Eq:ProofL1L2BoundBAux2}, we obtain
	\begin{align}
		  \isdraft{}{&}\mathbb{E}_g\Bigl[\text{dist}\big(g,\text{cone}\,\partial f_2(x^\star)\big)^2\Bigr]
		\isdraft{}{
		\notag
		\\}
		&\leq
			|I| + t^2 \bigg[\sum_{i \in I_+} (1 + \beta(x_i^\star - w_i))^2 + \sum_{i \in I_-} (1 - \beta(x_i^\star - w_i))^2\bigg]		
		\notag
		\\
		&\qquad
		+
			2|I^cJ^c| \frac{\varphi(t)}{t}		
			+
			|K^{\neq}(\beta)| + \frac{1}{2}|K^=(\beta)| 
		\notag
		\\
		&\qquad
		+
			t^2 
			\bigg[
				\sum_{i \in K_-} (\beta w_i + 1)^2
				+
				\sum_{i \in K_+} (\beta w_i - 1)^2
			\bigg]
		\isdraft{}{
		\notag
		\\
		&\qquad}
		+
		\Big[|K(\beta)| + 2|L|\Big]
		\frac{\varphi(t(1 - \beta\, \overline{w}))}{t|1 - \beta\, \overline{w}|}
		\notag
		\\		
		&=
			v_\beta t^2
			+
			|I|
			+
			|K^{\neq}(\beta)| + \frac{1}{2}|K^=(\beta)|
			+
			\Big[|K(\beta)| 		
			+ 
			2|L|\Big]
		\isdraft{}{
		\times
		\notag
		\\
		&\qquad
		\times}
			\frac{\varphi(t(1 - \beta\, \overline{w}))}{t|1 - \beta\, \overline{w}|}
			+
			2|I^cJ^c| \frac{\varphi(t)}{t}
		\notag
		\\
		&\leq
			v_\beta t^2
			+
			|I|
			+
			|K^{\neq}(\beta)| + \frac{1}{2}|K^=(\beta)|
			+
			2F(t,\beta,\overline{w})
			+
			2G(t)\,,
		\label{Eq:ProofL1L2BoundBAux4}
	\end{align}
	where we used $|K(\beta)| + 2|L| \leq 2|I^c J| = 2(q-s)$ (cf.\ \eqref{Eq:LemSetIdentity6}) in the last inequality. Note that~$v_\beta$ is defined in~\eqref{Eq:DefvBeta} and that we defined
	\begin{align*}
		F(t,\beta,\overline{w}) &:= (q-s)
			\frac{\varphi(t(1 - \beta\, \overline{w}))}{t|1 - \beta\, \overline{w}|}
		\\
		G(t) &:= (n-q) \frac{\varphi(t)}{t}\,.
	\end{align*}
	We consider two scenarios: $F(t,\beta, \overline{w}) \leq G(t)$ and $F(t,\beta, \overline{w}) \geq G(t)$. Note that 
	\begin{align}
		\isdraft{}{&}
		  \frac{F(t,\beta,\overline{w})}{G(t)} \lesseqgtr 1
		\isdraft{\quad}{
		\notag
		\\}
		\Shorter{}{
		&\Longleftrightarrow		
		\qquad
			\frac{q-s}{n-q}
			\frac{1}{|1-\beta \overline{w}|}
			\frac{\varphi(t(1-\beta \overline{w}))}{\varphi(t)}
			\lesseqgtr 1
		\notag
		\\
		&\Longleftrightarrow
		\qquad
			\frac{q-s}{n-q}
			\lesseqgtr 
			|1-\beta \overline{w}| \frac{\varphi(t)}{\varphi(t(1-\beta \overline{w}))}
		\notag
		\\
		}
		\isdraft{}{&}\Longleftrightarrow
		\quad
			\frac{q-s}{n-q}
			\lesseqgtr 
			|1-\beta \overline{w}|
			\exp\Big(t^2 \beta \overline{w}(\frac{\beta \overline{w}}{2} - 1)\Big)\,,
		\label{Eq:ProofL1L2Condition1}
	\end{align}
	
	\begin{itemize}
		\item Suppose $F(t,\beta, \overline{w}) \leq G(t)$, i.e., \eqref{Eq:ProofL1L2Condition1} is satisfied with~$\leq$. The bound in~\eqref{Eq:ProofL1L2BoundBAux4} implies
		\begin{align*}
		  \mathbb{E}_g\Bigl[\text{dist}\isdraft{}{&}\big(g,\text{cone}\,\partial f_2(x^\star)\big)^2\Bigr]
		\isdraft{}{\\}
		&\leq
			v_\beta t^2
			+
			s
			+
			|K^{\neq}(\beta)| + \frac{1}{2}|K^=(\beta)|
			+
			4G(t)
		\\
		&=
			v_\beta t^2
			+
			s
			+
			|K^{\neq}(\beta)| + \frac{1}{2}|K^=(\beta)|
		\isdraft{}{
		\\
		&\qquad}
		+
			4(n-q)\frac{1}{t}\frac{1}{\sqrt{2\pi}}\exp\Big(-\frac{t^2}{2}\Big)\,.
	\end{align*}
	We now select~$t$ as
	$$
		t^\star = \sqrt{2\log\Big(\frac{n}{q}\Big)} = \sqrt{2\log\,r}\,,
	$$
	where~$r := n/q$. Note that~$r$ is well defined, since~$x^\star \neq 0$ implies $q > 0$. Also, the assumption~$q < n$ implies~$t^\star > 0$. Setting~$t$ to~$t^\star$ and using~\eqref{Eq:BoundChandrasekaran}, we get
	\begin{align*}
		  \isdraft{}{&}\mathbb{E}_g\Bigl[\text{dist}\big(g,\text{cone}\,\partial f_2(x^\star)\big)^2\Bigr]
		\isdraft{}{\\}
		&\leq			
			2 v_\beta \log\Big(\frac{n}{q}\Big)
			+
			s
			+
			|K^{\neq}(\beta)| + \frac{1}{2}|K^=(\beta)|
		\isdraft{}{
		\\
		&\qquad}
		+
			4(n-q)
			\frac{1}{\sqrt{2\log\,r}}\frac{1}{\sqrt{2\pi}}\frac{1}{\frac{n}{q}}
		\\
		&=
			2 v_\beta \log\Big(\frac{n}{q}\Big)
			+
			s
			+
			|K^{\neq}(\beta)| + \frac{1}{2}|K^=(\beta)|
		\isdraft{}{
		\\
		&\qquad}
		+
			2q\frac{1-\frac{1}{r}}{\sqrt{\pi\log\,r}}
		\\
		&\leq
			2 v_\beta \log\Big(\frac{n}{q}\Big)
			+
			s
			+
			|K^{\neq}(\beta)| + \frac{1}{2}|K^=(\beta)|
			+
			\frac{4}{5}q\,,
	\end{align*}
	which is~\eqref{Eq:ThmL1L2Bound1}. This bound is valid only if~\eqref{Eq:ProofL1L2Condition1} with~$\leq$ is satisfied for~$t^\star$, i.e.,
	$$
		\frac{q-s}{n-q}
			\leq 
			|1-\beta \,\overline{w}|
			\exp\Big(2\beta \,\overline{w}\log\Big(\frac{n}{q}\Big)\Big(\frac{\beta	\, \overline{w}}{2} - 1\Big)\Big)\,,
	$$
	which is condition~\eqref{Eq:ThmL1L2Condition1}.
	
	\item
		Suppose now that $F(t,\beta, \overline{w}) \geq G(t)$, i.e., \eqref{Eq:ProofL1L2Condition1} is satisfied with~$\geq$. The bound in~\eqref{Eq:ProofL1L2BoundBAux4} implies
		\begin{align*}
		  \isdraft{}{&}\mathbb{E}_g\Bigl[\text{dist}\big(g,\text{cone}\,\partial f_2(x^\star)\big)^2\Bigr]
		\isdraft{}{\\}
		&\leq
			v_\beta t^2
			+
			s
			+
			|K^{\neq}(\beta)| + \frac{1}{2}|K^=(\beta)|
			+
			4F(t,\beta,\overline{w})
		\\
		&=
			v_\beta t^2
			+
			s
			+
			|K^{\neq}(\beta)| + \frac{1}{2}|K^=(\beta)|
		\isdraft{}{
		\\
		&\qquad}
			+
			4(q-s)
			\frac{1}{\sqrt{2\pi}}
			\frac{1}{t|1-\beta\,\overline{w}|}
			\exp\Big(-\frac{t^2}{2}(1 - \beta\,\overline{w})^2\Big)\,.
		\end{align*}
		And we select~$t$ as
		$$
		t^\star 
		= 
		\frac{1}{|1 - \beta \,\overline{w}|}\sqrt{2\log\Big(\frac{q}{s}\Big)}
		=
		\frac{1}{|1 - \beta \,\overline{w}|}\sqrt{2\log\,r}\,,
	$$
	where~$r := q/s$. Again, $r$ is well defined because~$s>0$. Since we assume~$q>s$, $t^\star > 0$. Setting~$t$ to~$t^\star$ and using~\eqref{Eq:BoundChandrasekaran} again, we obtain
	\begin{align*}
		  \isdraft{}{&}\mathbb{E}_g\Bigl[\text{dist}\big(g,\text{cone}\,\partial f_2(x^\star)\big)^2\Bigr]
		\isdraft{}{\\}
		&\leq			
			\frac{2v_\beta}{|1 - \beta \, \overline{w}|^2} \log\Big(\frac{q}{s}\Big)
			+
			s
			+
			|K^{\neq}(\beta)| + \frac{1}{2}|K^=(\beta)|
		\isdraft{}{
		\\
		&\qquad}
		+
			4(q-s)
			\frac{1}{\sqrt{2\pi}}
			\frac{1}{\sqrt{2\log\,r}}
			\frac{1}{\frac{q}{s}}
		\\
		&=
			\frac{2v_\beta}{|1 - \beta \, \overline{w}|^2} \log\Big(\frac{q}{s}\Big)
			+
			s
			+
			|K^{\neq}(\beta)| + \frac{1}{2}|K^=(\beta)|
		\isdraft{}{
		\\
		&\qquad}
			+
			2s			
			\frac{1-\frac{1}{r}}{\sqrt{\pi\log\,r}}			
		\\
		&\leq			
			\frac{2v_\beta}{|1 - \beta \, \overline{w}|^2} \log\Big(\frac{q}{s}\Big)
			+
			|K^{\neq}(\beta)| + \frac{1}{2}|K^=(\beta)|
			+
			\frac{9}{5}s\,,			
	\end{align*}
	which is~\eqref{Eq:ThmL1L2Bound2}. This bound is valid only when~\eqref{Eq:ProofL1L2Condition1} with~$\geq$ is satisfied for~$t^\star$, i.e.,
	$$
		\frac{q-s}{n-q}
			\geq
			|1-\beta \overline{w}|
			\exp\Big(4\frac{(\beta\,\overline{w} -2)\beta\,\overline{w}}{|1-\beta \overline{w}|^2}
			\log\Big(\frac{q}{s}\Big)\Big)\,,
	$$
	which is condition~\eqref{Eq:ThmL1L2Condition2}.\qed
	\end{itemize}

\mypar{Remarks}
	Although these bounds were derived using the same techniques as the ones for $\ell_1$-$\ell_1$ minimization, they are much looser. The main reason is their dependency on the magnitudes of~$x^\star$, $w$, and~$x^\star - w$. This forced us to consider a worst-case scenario in the last step of~\eqref{Eq:ProofL1L2IntroductionOfwbar}. 
	
%
	
\section{Acknowledgments} 
The authors would like to thank Jo\~ao Xavier for insightful discussions regarding \pref{Prop:GaussianWidthDistancePolarCone} and the interpretation of \fref{Fig:GaussianWidth}, and to Volkan Cevher for fruitful discussions about relevant related literature.
	
\appendices
\section{Proof of Lemma~\ref{lem:AuxLemmaIntervals}}
\label{App:ProofLemmaIntervals}

	\subsection{Proof of \text{1)}:}	
	To show~\eqref{Eq:LemAuxDistanceToAPoint}, we simply use the linearity of the expected value and the fact that~$g$ has zero mean and unit variance:
		\begin{align*}
			\mathbb{E}_g\bigl[\text{dist}(g,a)^2\bigr] 
			&=
			\mathbb{E}_g\bigl[(a-g)^2\bigr]
			=
			\mathbb{E}_g\bigl[a^2 - 2ag + g^2\bigr]
			\isdraft{}{
			\\
			&}= a^2 + 1\,.
		\end{align*}

	\subsection{Proofs of \text{2)}, \text{3)}, and \text{4)}:}
	In cases \text{2)}, \text{3)}, and \text{4)}, we have~$b>0$ and $|a| \neq b$.	Hence,
	\begin{align}
		\mathbb{E}_g \isdraft{}{&} \Bigl[\text{dist}\bigl(g,\mathcal{I}(a,b)\bigr)^2\Bigr] 
		\isdraft{&}{}=
		\mathbb{E}_{g} 
		\Biggl[
			\begin{array}[t]{cl}
				\underset{u}{\min} & (u - g)^2 \\
				\text{s.t.} & |u - a| \leq b
			\end{array}
		\Biggr]
		\notag
		\\
		&=
			\frac{1}{\sqrt{2\pi}}\int_{a+b}^{+\infty}\bigl(g - (a + b)\bigr)^2 \exp\Bigl(-\frac{g^2}{2}\Bigr)\,dg
		\isdraft{}{
		\notag
		\\
		&\quad}
			+
			\frac{1}{\sqrt{2\pi}}\int_{-\infty}^{a-b}\bigl(g - (a - b)\bigr)^2 \exp\Bigl(-\frac{g^2}{2}\Bigr)\,dg
		\notag
		\\
		&=
			A(a+b) + B(a-b)\,,
		\label{Eq:ProofLemAuxDistance1}
	\end{align}
	where
			\begin{align}
					A(x) 
				&:=
					\frac{1}{\sqrt{2\pi}}\int_{x}^{+\infty}(g - x)^2 \,\exp\Bigl(-\frac{g^2}{2}\Bigr)\,dg							
					\notag
				\\
				&=
					\frac{1}{\sqrt{2\pi}}\int_{x}^{+\infty} g^2 \,\exp\Bigl(-\frac{g^2}{2}\Bigr)\,dg
					\tag{$A_1(x)$}
 				\\
 				&\quad
 					-\frac{2x}{\sqrt{2\pi}}\int_{x}^{+\infty} g \,\exp\Bigl(-\frac{g^2}{2}\Bigr)\,dg						
 					\tag{$-A_2(x)$}
				\\
				&\quad
					+\frac{x^2}{\sqrt{2\pi}}\int_{x}^{+\infty} \exp\Bigl(-\frac{g^2}{2}\Bigr)\,dg
					\tag{$A_3(x)$}
				\\
				&=:
					A_1(x) - A_2(x) + A_3(x)\,,
				\notag
			\end{align}
			and
			\begin{align}
					B(x)
				&:=
					\frac{1}{\sqrt{2\pi}}\int_{-\infty}^{x}(g - x)^2 \exp\Bigl(-\frac{g^2}{2}\Bigr)\,dg
				\notag
				\\
				&=
					\frac{1}{\sqrt{2\pi}}\int_{-\infty}^{x}g^2\, \exp\Bigl(-\frac{g^2}{2}\Bigr)\,dg
				\tag{$B_1(x)$}
				\\
				&\quad
					-
					\frac{2x}{\sqrt{2\pi}}\int_{-\infty}^{x}g\, \exp\Bigl(-\frac{g^2}{2}\Bigr)\,dg
				\tag{$-B_2(x)$}
				\\
				&\quad
					+
					\frac{x^2}{\sqrt{2\pi}}\int_{-\infty}^{x} \exp\Bigl(-\frac{g^2}{2}\Bigr)\,dg
				\tag{$B_3(x)$}
				\\
				&=:
					B_1(x) - B_2(x) + B_3(x)\,.
				\notag
			\end{align}
			Using symmetry arguments for even and odd functions, it can be shown that $B_1(x) = A_1(-x)$, $B_2(x) = -A_2(x)$, and $B_3(x) = A_3(-x)$. Therefore, 
			\begin{align}
					A(x) &= \bigl(A_1(x) + A_3(x)\bigr) - A_2(x) 
				\label{Eq:ProofLemAuxDistanceAx}
				\\
					B(x) &= \bigl(A_1(-x) + A_3(-x)\bigr) + A_2(x) \,.
				\label{Eq:ProofLemAuxDistanceBx}
			\end{align} 
			Next, we compute expressions for~$A_1(x) + A_3(x)$ and~$A_2(x)$. Integrating~$A_1(x)$ by parts, we obtain:		
			\isdraft{                                                     
				\begin{equation}                                            
					A_1(x) + A_3(x)					                                  
				=                                                           
					\frac{x}{\sqrt{2\pi}}\exp\Bigl(-\frac{x^2}{2}\Bigr)				
				+
					(1+x^2)\underbrace{\frac{1}{\sqrt{2\pi}}\int_{x}^{+\infty} \exp\Bigl(-\frac{g^2}{2}\Bigr) \,dg}_{:= Q(x)}\,,
				\label{Eq:ProofLemAuxDistanceA1PlusA3}                      
				\end{equation}	                                            
			}{			
			\begin{multline}
					A_1(x) + A_3(x)					
				=
					\frac{x}{\sqrt{2\pi}}\exp\Bigl(-\frac{x^2}{2}\Bigr)
				\\
				+
					(1+x^2)\underbrace{\frac{1}{\sqrt{2\pi}}\int_{x}^{+\infty} \exp\Bigl(-\frac{g^2}{2}\Bigr) \,dg}_{:= Q(x)}\,,
				\label{Eq:ProofLemAuxDistanceA1PlusA3}
			\end{multline}
			}
			where~$Q(x)$ is the $Q$-function. The $Q$-function is not elementary, but the following bounds can be computed, for~$x>0$, and they are sharp for large~$x$~\cite[Eq.\ 2.121]{Wozencraft65-PrinciplesOfCommunicationEngineering}:
			\footnote{
				The lower bound in~\cite[Eq.\ 2.121]{Wozencraft65-PrinciplesOfCommunicationEngineering} is actually $((x^2-1)/x^3)\varphi(x)$. The lower bound in~\eqref{Eq:ProofLemAuxDistanceBoundsQFunction}, however, is tighter and stable near the origin. We found this bound in~\cite{BoundQFunction}. Since we were not able to track it to a published reference, we replicate the proof from~\cite{BoundQFunction} here. For~$x>0$, there holds
				\begin{align*}
						\Big(1 &+ \frac{1}{x^2}\Big)Q(x) 
						= 
						\int_{x}^\infty \Big(1 + \frac{1}{x^2}\Big) \varphi(u)\,du
						\geq
						\int_{x}^\infty \Big(1 + \frac{1}{u^2}\Big) \varphi(u)\,du
					\\
					&=
						-\int_x^\infty \frac{u \,d\varphi(u)/du - \varphi(u)}{u^2}\,du					
					=
						-\int_x^\infty \frac{d}{du}\Big(\frac{\varphi(u)}{u}\Big)\,du
					=
						\frac{\varphi(x)}{x}\,,						
				\end{align*}
				from which the bound follows. In the third step, we used the property $d\varphi(u)/du = -u\varphi(u)$.
			}
			\begin{equation}\label{Eq:ProofLemAuxDistanceBoundsQFunction}
				\frac{x}{1 + x^2}\frac{1}{\sqrt{2\pi}}\exp\Bigl(-\frac{x^2}{2}\Bigr)
				\leq
				Q(x)
				\leq
				\frac{1}{x}\frac{1}{\sqrt{2\pi}}\exp\Bigl(-\frac{x^2}{2}\Bigr)\,.
			\end{equation}				
			The integral in~$A_2(x)$ can be computed in closed-form as
			\begin{equation}\label{Eq:ProofLemAuxDistanceA2}
				A_2(x) 
				=				
				\frac{2x}{\sqrt{2\pi}}\exp\Bigl(-\frac{x^2}{2}\Bigr)\,.				
			\end{equation}
			From~\eqref{Eq:ProofLemAuxDistanceAx}, \eqref{Eq:ProofLemAuxDistanceBx}, \eqref{Eq:ProofLemAuxDistanceA1PlusA3}, and~\eqref{Eq:ProofLemAuxDistanceA2}, we obtain
			\begin{align}
				A(x) &= -\frac{x}{\sqrt{2\pi}}\exp\Bigl(-\frac{x^2}{2}\Bigr) + (1+x^2)Q(x)
				\label{Eq:ProofLemAuxDistanceAClosedForm}
				\\
				B(x) &= \frac{x}{\sqrt{2\pi}}\exp\Bigl(-\frac{x^2}{2}\Bigr) + (1+x^2)Q(-x)\,.
				\label{Eq:ProofLemAuxDistanceBClosedForm}
			\end{align}
			Now we compute bounds for~$A(x)$ and~$B(x)$ based on~\eqref{Eq:ProofLemAuxDistanceBoundsQFunction} and  address the cases $x<0$ and~$x>0$ separately. We will use the property~$Q(x) = 1 - Q(-x)$. We will also make use of the shorthand notation $\varphi(x):=\exp(-x^2/2)/\sqrt{2\pi}$ (cf.\ \eqref{Eq:AuxLemVarPhi}).
			
			Let us start with~$A(x)$. Consider~$x<0$. Then,
			\begin{align}
				  A(x) 
				&=
				  -x\,\varphi(x) + (1+x^2)(1-Q(-x))
				\notag
				\\
				&\leq
				  -x\,\varphi(x) + (1+x^2)\biggl(1+\frac{x}{1 + x^2}\varphi(x)\biggr)
				\notag
				\\
				&=
					1+x^2
				\,,
				\label{Eq:ProofLemAuxDistanceBoundAxNegative}
			\end{align}
			where the inequality is due to the lower bound in~\eqref{Eq:ProofLemAuxDistanceBoundsQFunction}. Now, let~$x>0$. Applying the upper bound in~\eqref{Eq:ProofLemAuxDistanceBoundsQFunction} directly, we obtain
			\begin{equation}\label{Eq:ProofLemAuxDistanceBoundAxPositive}
				  A(x)
				\leq
					-x\varphi(x) + (1 + x^2)\frac{1}{x}\varphi(x)				
				=
					\frac{1}{x}\varphi(x)\,.
			\end{equation}
			
			Now consider~$B(x)$ with~$x<0$. Since~$Q(-x)$ has a positive argument, we use the upper bound in~\eqref{Eq:ProofLemAuxDistanceBoundsQFunction}:
			\begin{equation}\label{Eq:ProofLemAuxDistanceBoundBxNegative}
				  B(x)
				\leq
					x\,\varphi(x) + (1 + x^2)\Bigl(-\frac{1}{x}\varphi(x)\Bigr)				
				=
					\frac{1}{|x|}\varphi(x)\,,
			\end{equation}
			where, in the inequality, we used the fact that~$\varphi(-x) = \varphi(x)$. Assume now~$x>0$. Then,
			\begin{align}
					B(x) 
				&= 
				  x\,\varphi(x) + (1+x^2)(1 - Q(x))
				\notag
				\\
				&\leq
					x\,\varphi(x) + (1+x^2)\biggl(1 - \frac{x}{1 + x^2}\varphi(x)\biggr)
				\notag
				\\
				&=
					1 + x^2\,,
				\label{Eq:ProofLemAuxDistanceBoundBxPositive}
			\end{align}
			where we used the lower bound in~\eqref{Eq:ProofLemAuxDistanceBoundsQFunction}. In sum, \eqref{Eq:ProofLemAuxDistanceBoundAxNegative}, \eqref{Eq:ProofLemAuxDistanceBoundAxPositive}, \eqref{Eq:ProofLemAuxDistanceBoundBxNegative}, and~\eqref{Eq:ProofLemAuxDistanceBoundBxPositive} tell us that
			\begin{align}
					A(x)
				&\leq
					\left\{
						\begin{array}{ll}
							1+x^2  \phantom{s}&,\,\, x<0  \vspace{0.1cm} \\
							\frac{1}{x}\varphi(x) &,\,\, x>0
						\end{array}
					\right.
				\label{Eq:ProofLemAuxDistanceBoundFinalBoundA}
				\\
					B(x)
				&\leq
					\left\{
						\begin{array}{ll}
							\frac{1}{|x|}\varphi(x) &,\,\, x<0 \vspace{0.1cm} \\
								1+x^2 &,\,\, x>0
						\end{array}
					\right.
					\,.
				\label{Eq:ProofLemAuxDistanceBoundFinalBoundB}
			\end{align}
			From~\eqref{Eq:ProofLemAuxDistance1}, \eqref{Eq:ProofLemAuxDistanceBoundFinalBoundA}, and \eqref{Eq:ProofLemAuxDistanceBoundFinalBoundB}, 
			\begin{align*}
				&\mathbb{E}_g \Bigl[\text{dist}\bigl(g,\mathcal{I}(a,b)\bigr)^2\Bigr] 
				=
					A(a+b) + B(a-b)
				\\
				&\leq
					\left\{
						\begin{array}{ll}
							\frac{\varphi(a+b)}{a + b} + \frac{\varphi(a-b)}{|a-b|} &,\,\, |a| < b
							\vspace{0.1cm}
							\\
							1+(a+b)^2 + \frac{\varphi(a-b)}{|a-b|} &,\,\, a+b < 0
							\vspace{0.1cm}
							\\
							\frac{\varphi(a+b)}{a+b} + 1 + (a-b)^2  &,\,\, a-b > 0\,.
						\end{array}
					\right.					
			\end{align*}
			Taking into account that $\varphi(x) = \varphi(-x)$ for any~$x$, this is exactly~\eqref{Eq:LemAuxDistanceToIntervalZeroInside}, \eqref{Eq:LemAuxDistanceToIntervalLeftOfZero}, and~\eqref{Eq:LemAuxDistanceToIntervalRightOfZero}.

	\subsection{Proofs of \text{5)} and \text{6)}:}
		
		Suppose~$a + b = 0$. Since~$a-b < 0$ (recall that~$b > 0$), \eqref{Eq:ProofLemAuxDistanceBoundBxNegative} applies and tells us that
		$
			B(a-b) \leq \varphi(a-b)/(b-a)\,.
		$
		Setting~$x = 0$ in~\eqref{Eq:ProofLemAuxDistanceAClosedForm}, we obtain
		$
			A(a+b) = A(0) = Q(0) = 1/2.
		$
		Therefore, $A(a+b) + B(a-b) = \varphi(a-b)/(b-a) + 1/2$, which is~\eqref{Eq:LemAuxDistanceToIntervalTouchOnTheRight}. The proof of~\eqref{Eq:LemAuxDistanceToIntervalTouchOnTheLeft} is identical.
	\hfill\qed

\section{Proof of Lemma~\ref{lem:AuxLemmaBoundsFunctions}}
\label{App:ProofLemmaBounds}
	Denote~$f(x) := (1-1/x)/\sqrt{\log\,x}$. It can be shown that
	\begin{align*}
		  \frac{d}{dx}f(x)
		&=
		  \frac{2\log\,x + 1 - x}{2x^2 \log^{3/2}x}
		\\
		  \frac{d^2}{dx^2}f(x)
		&=
			\frac{3(x-1)-8\log^2x + 2(x-3)\log\,x}{4x^3\log^{5/2}x}\,.
	\end{align*}
	The stationary points of~$f$ are those for which $\frac{d}{dx}f(x) = 0$, that is, the points that satisfy the equation $2\log\,x = x - 1$. This equation has only one solution, say~$\overline{x}$, for~$x>1$: $\log\,\overline{x} = (\overline{x}-1)/2$. Using this identity, we can conclude that
	\begin{align*}
		  \frac{d^2}{dx^2}f(\overline{x})
		&=
			\frac{3(\overline{x}-1) - 2(\overline{x}-1)^2 + (\overline{x}-3)(\overline{x}-1)}{\frac{1}{\sqrt{2}}\overline{x}^3(\overline{x}-1)^{5/2}}
		\\
		&=
			\frac{2-\overline{x}}{\frac{1}{\sqrt{2}}\overline{x}^3(\overline{x}-1)^{3/2}} < 0\,,
	\end{align*}
	since~$\overline{x} > 2$. This is because $\log2 > 1/2$ and, e.g., $\log11 < 5$ or, in other words, $(x-1)/2$ intersects~$\log\,x$ somewhere in the interval~$2 < x < 11$, that is, $2 < \overline{x}<11$. This means that the only stationary point~$\overline{x}$ is a local maximum. Since $\lim_{x\downarrow 1} f(x) = 0$ and $\lim_{x\xrightarrow{} +\infty}f(x) = 0$ (using for example l'H\^opital's rule), $\overline{x}$ is actually a global maximum. Knowing that~$\overline{x}$ satisfies	$\log\,\overline{x} = (\overline{x}-1)/2$, we have
	$$
		f(\overline{x}) = \frac{\overline{x}-1}{\overline{x}\sqrt{\log\,\overline{x}}}
		=
		\sqrt{2}\frac{\overline{x}-1}{\overline{x}\sqrt{\overline{x}-1}}
		=
		\sqrt{2}\frac{\sqrt{\overline{x}-1}}{\overline{x}}\,.
	$$
	By equating the derivative of the function $\sqrt{x-1}/x$ to zero, we know that it achieves its maximum at~$x = 2$. Therefore,
	$
		f(\overline{x}) \leq \sqrt{2}/2 = 1/\sqrt{2}
	$.
	Dividing by~$1/\sqrt{\pi}$, we obtain~\eqref{Eq:BoundChandrasekaran}. \hfill \qed
	

\section{Proof of Lemma~\ref{lem:ZeroNotInSubGradient}}
\label{App:ProofLemmaZeroNotInSubGradient}	
	
	\subsection{Proof of~\text{1)}}		
	According to~\eqref{Eq:SubgradientL1L1PerComp}, $0 \in \partial f_1^{(i)}(x_i^\star) $ is equivalent to either:
	\begin{subequations}
	\begin{align}
		i &\in IJ\,\, \text{and}\,\, \text{sign}(x_i^\star) + \beta \,\text{sign}(x_i^\star-w_i) = 0\,,
		\text{or}
		\label{Eq:ProofLemmaSubgradients1}
		\\
		i &\in IJ^c\,\, \text{and}\,\,\beta \geq 1\,, \text{or}
		\label{Eq:ProofLemmaSubgradients2}
		\\
		i &\in I^c J\,\, \text{and}\,\,\beta \leq 1\,, \text{or}
		\label{Eq:ProofLemmaSubgradients3}
		\\
		i &\in I^c J^c\,.
		\label{Eq:ProofLemmaSubgradients4}
	\end{align}
	\end{subequations}
	Note that \eqref{Eq:ProofLemmaSubgradients1} cannot be satisfied whenever~$\beta \neq 1$. Hence, conditions~\eqref{Eq:ProofLemmaSubgradients1}-\eqref{Eq:ProofLemmaSubgradients4} can be rewritten as
	\begin{itemize}
		\item $\beta = 1$: $\text{sign}(x_i^\star) + \text{sign}(x_i^\star-w_i) = 0$ for~$i \in IJ$, or $i \in I^cJ$, or $i \in IJ^c$, or~$i \in I^c J^c$.
		\item $\beta > 1$: $i \in IJ^c$ or $i \in I^c J^c$.
		\item $\beta < 1$: $i \in I^c J$ or $i \in I^c J^c$.
	\end{itemize}
				
	We consider two scenarios: $IJ \neq \emptyset$ and~$IJ = \emptyset$. 
	\begin{itemize}		
		\item 
			Let $IJ \neq \emptyset$. When~$\beta = 1$, $0 \not\in \partial f_1(x^\star)$ if and only if there is an~$i \in IJ$ such that~$\text{sign}(x_i^\star) + \text{sign}(x_i^\star-w_i) \neq 0$, i.e., there is at least one bad component: $\overline{h} > 0$. When~$\beta \neq 1$, there is at least one~$i \in IJ$ for which~\eqref{Eq:ProofLemmaSubgradients1} is not satisfied, that is, $0 \not\in \partial f_1(x^\star)$. We thus conclude that part~\text{1)} is true whenever $IJ \neq \emptyset$.
	
		\item
			Let $IJ = \emptyset$ or, equivalently, $x_i^\star = w_i$ for all~$i \in I$. Recall from~\eqref{Eq:LemSetIdentity1} that $|IJ| = h + \overline{h}$. Thus, $IJ = \emptyset$ implies $\overline{h} = 0$. In this case, if~$\beta = 1$, then $0 \in \partial f_1(x^\star)$. On the other hand, for~$\beta > 1$, $0 \not\in \partial f_1(x^\star)$ if and only if~$I^cJ \neq \emptyset$; similarly, for~$\beta < 1$, $0 \not\in \partial f_1(x^\star)$ if and only if~$IJ^c \neq \emptyset$. We next show that~$IJ = \emptyset$, together with~$I \neq \emptyset$ and~$J \neq \emptyset$, implies that both~$I^cJ$ and~$IJ^c$ are nonempty, thus showing that part~\text{1)} is also true whenever $IJ = \emptyset$. In fact, $I\neq \emptyset$ implies $IJ^c \neq \emptyset$, because $I = IJ \cup IJ^c = IJ^c$. Also, $J \neq \emptyset$, that is, $x^\star \neq w$, implies $I^c J \neq \emptyset$. This is because~$IJ = \emptyset$ means that~$x^\star$ and~$w$ coincide on~$I$, and $I^cJ = \{i\,:\, 0 = x_i^\star \neq w_i\}$ is the set of nonzero components of~$w$ outside~$I$. Since~$x^\star$ and~$w$ coincide on~$I$, they have to differ outside~$I$, i.e., $I^cJ \neq \emptyset$.
	\end{itemize}

	\subsection{Proof of \text{2)}}	
	From~\eqref{Eq:SubgradientL1L2PerComp}, $0 \in \partial f_2^{(i)}(x_i^\star)$ is equivalent to either 			
	\begin{align*}
		i &\in IJ\,\, \text{and}\,\, \beta(w_i - x_i^\star) = \text{sign}(x_i^\star)\,, \text{or} 				
 		\\
 		i &\in IJ^c\,, \text{or} 				
 		\\
 		i &\in I^c\,\, \text{and}\,\, \beta \leq 1/|w_i|\,. 	
 	\end{align*} 			
	\hfill \qed

\bibliographystyle{IEEEtran}

{ 
\bibliography{paper}
}

\end{document}